\documentclass[a4paper,10pt]{article}
\usepackage{jheppub}

\usepackage[utf8]{inputenc}
\usepackage{amsmath,amssymb,bm,slashed,braket}
\usepackage{dsfont}
\usepackage{amsfonts}
\usepackage{bm,bbm}
\usepackage{graphicx}
\usepackage{slashed} 
\usepackage[dvipsnames,table]{xcolor}
\usepackage[normalem]{ulem}
\usepackage{soul}
\usepackage{siunitx} 
\usepackage{hyperref}
\hypersetup{colorlinks,citecolor= blue,linkcolor= blue, urlcolor=blue}
\usepackage{ulem}
\usepackage{array}
\usepackage{verbatim}
\usepackage{epsfig}
\usepackage{multirow, booktabs}
\usepackage{bbold}
\usepackage{subcaption}
\usepackage[version=4]{mhchem}
\usepackage[capitalise]{cleveref}
\usepackage{makecell}
\usepackage{orcidlink}

\allowdisplaybreaks[4]
\def\lsim{\mathrel{\raise.3ex\hbox{$<$\kern-.75em\lower1ex\hbox{$\sim$}}}}
\def\gsim{\mathrel{\raise.3ex\hbox{$>$\kern-.75em\lower1ex\hbox{$\sim$}}}}

\usepackage{tikz} 
\usepackage{tkz-euclide}
\usetikzlibrary{backgrounds} 
\usetikzlibrary{decorations.pathmorphing}
\usetikzlibrary{arrows.meta}
\tikzset{
mystyle/.style={line width=1, baseline, scale=0.6, every node/.style={scale=1}},
v/.style={decorate, draw, decoration={snake, segment length=2.mm, amplitude=0.5mm}},
f/.style={draw, decoration={markings,mark=at position #1 with {\arrow[]{Latex[length=1.5mm,width=1.5mm]}}},
    postaction={decorate},node contents=#1},
f/.default=.6,
fb/.style={draw,decoration={markings,mark=at position #1 with {\arrowreversed[]{Latex[length=1.5mm,width=1.5mm]}}},
    postaction={decorate},node contents=#1},
fb/.default=.6,
s/.style={dashed,draw, decoration={markings,mark=at position #1 with {\arrow[]{Latex[length=1.5mm,width=1.5mm]}}},
    postaction={decorate},node contents=#1},
s/.default=.6,    
sb/.style={dashed,draw,decoration={markings,mark=at position #1 with {\arrowreversed[]{Latex[length=1.5mm,width=1.5mm]}}},
    postaction={decorate},node contents=#1},
sb/.default=.4,
snar/.style={dashed,draw,line width =1.25pt},
cross/.style={cross out, draw=black, minimum size=2*(#1-\pgflinewidth), inner sep=0pt, outer sep=0pt}, 
         }

\newcommand{\T}{{\rm T}}

\newcommand{\calL}{ {\cal L} }
\newcommand{\calO}{ {\cal O} }
\newcommand{\calM}{ {\cal M} }
\newcommand{\calS}{ {\cal S} }
\newcommand{\calV}{ {\cal V} }
\newcommand{\calF}{ {\cal F} }
\newcommand{\NR}{{\tt NR}}
\newcommand{\ER}{{\tt ER}}

\newcommand{\pSDM}{\pmb{S}_X}
\newcommand{\pSN}{\pmb{S}_N}
\newcommand{\pq}{\pmb{q}}
\newcommand{\pvn}{{\pmb{v}_N^\perp}}
\newcommand{\mathone}{\mathbb{1}}
\newcommand{\pvt}{{\pmb{v}_T^\perp}}

\title{A systematic investigation on vector dark matter-nucleus scattering in effective field theories}

\author[a,b]{Jin-Han Liang\,\orcidlink{0000-0002-6141-216X},}
\emailAdd{jinhanliang@m.scnu.edu.cn}
\affiliation[a]{Key Laboratory of Atomic and Subatomic Structure and Quantum Control (MOE), 
Guangdong Basic Research Center of Excellence for Structure and Fundamental Interactions of Matter, Institute of Quantum Matter, South China Normal University, Guangzhou 510006, China}
\affiliation[b]{Guangdong-Hong Kong Joint Laboratory of Quantum Matter, 
Guangdong Provincial Key Laboratory of Nuclear Science, 
Southern Nuclear Science Computing Center, South China Normal University, Guangzhou 510006, China}
\author[a,b]{Yi Liao\,\orcidlink{0000-0002-1009-5483},}
\emailAdd{liaoy@m.scnu.edu.cn}
\author[a,b]{Xiao-Dong Ma\,\orcidlink{0000-0001-7207-7793},}
\emailAdd{maxid@scnu.edu.cn}
\author[a,b]{Hao-Lin Wang\,\orcidlink{0000-0002-2803-5657}}
\emailAdd{whaolin@m.scnu.edu.cn}

\abstract{
In this paper, we systematically investigate the general spin-one dark matter-nucleus interactions within the framework of effective field theories (EFT). We consider both the nonrelativistic (NR) and the relativistic EFT descriptions of the DM interactions with nucleons and quarks. In the NREFT framework, we present a complete list of NR
operators for spin-one DM coupling to nucleons and compute their contributions to the DM response functions. 
Next, we consider all possible leading-order relativistic EFT operators between DM and light quarks and the photon, 
and perform NR reductions to match them onto the NREFT. 
We then derive the nuclear scattering rate from these interactions, 
and employ recent DM direct detection data (from both the nuclear recoil and the Migdal effect)
to constrain all these EFT operators and DM electromagnetic properties. 
We find the elastic nuclear recoil data (from PandaX-4T, XENONnT, LZ, and DarkSide-50) set stringent bounds on the EFT coefficients
for a DM mass above a few GeV while the Migdal effect datasets (from PandaX-4T, XENONnT, and DarkSide-50) can probe the DM mass region as small as 20 MeV. 
Lastly, we construct a UV complete model that can provide a complex spin-one DM candidate, and at the same time generate DM-quark/photon operators discussed in this work.
}


\makeatletter
\gdef\@fpheader{}
\makeatother

\begin{document} 

\maketitle
\setcounter{page}{2}

\section{Introduction}

The existence of dark matter (DM) is well established from astrophysical observations, 
including the rotation curves of galaxies, gravitational lensing effects, 
large-scale structure, and the cosmic microwave background radiation, etc~\cite{Bertone:2004pz,Young:2016ala,Arbey:2021gdg}.
But the nature of DM still remains elusive. 
Many particle DM candidates have been proposed in the past decades to account for this missing mass conundrum, 
ranging from the superlight axion or dark photon,
to the weakly interacting massive particles (WIMPs), 
and to the superheavy primordial black hole, to name but a few. A more complete list of candidates can be found in the review papers~\cite{Jungman:1995df,Feng:2010gw,Petraki:2013wwa,Bernal:2017kxu,Escriva:2022duf} and references therein. 
Among them, the most appealing one is the WIMP-like scenario that automatically stabilizes DM from decay due to some residual symmetry in the DM sector. 

For the WIMP-like DM,
the key approach for its detection is through the nuclear recoil signals generated by DM-nucleus elastic scattering in direct detection experiments.
The resulting nuclear recoil energy is eventually converted into detectable signals including light, ionized electrons, and heat.
These nuclear recoil signals are particularly sensitive to DM with a mass comparable to the target nucleus.
The large-volume experiments, such as PandaX~\cite{PandaX:2024qfu}, XENONnT~\cite{XENON:2023cxc}, LUX-ZEPLIN (LZ)~\cite{LZ:2022lsv}, 
and Darkside~\cite{DarkSide-50:2022qzh}, utilize ultrapure liquid xenon or argon targets to maximize sensitivity to these rare events. 
Unfortunately, they have not found any positive DM signal, but they put very stringent bounds on the DM-nucleon scattering cross section for a DM mass around tens of GeV.

For the sub-GeV mass region, the sensitivity to DM-nucleon interactions decreases significantly, because the nuclei cannot
gain sufficient recoil energy from DM-nucleus elastic scattering when the DM mass is much smaller than the nuclear mass.
To overcome this barrier, one of the experimental approaches is to consider the DM-nucleus inelastic scattering through the Migdal effect~\cite{Migdal:1941,Ibe:2017yqa,Dolan:2017xbu}.
Due to the ionization of atomic electrons, this effect leads to an additional electron recoil signal in the keV range that accompanies the nuclear recoil signal, thus enhancing detection sensitivity in this mass range~\cite{Essig:2019xkx,GrillidiCortona:2020owp,Knapen:2020aky,Flambaum:2020xxo,Bell:2021zkr,Bell:2019egg,Li:2022acp,Tomar:2022ofh,Kang:2024kec}.
Recently, PandaX-4T~\cite{PandaX:2022xqx}, XENON1T~\cite{XENON:2019gfn}, and Darkside-50~\cite{DarkSide:2022dhx} 
have performed searches using the Migdal effect and have put meaningful constraints on the DM parameter space below the GeV scale. 

Theoretically, the scalar and fermionic WIMP-like DM scenarios have been extensively  studied in recent years, both in UV models and in effective field theory (EFT) frameworks, 
by leveraging the available direct and indirect detection data to constrain the relevant parameter space~\cite{Fan:2010gt,Harnik:2008uu,Goodman:2010ku,Kumar:2013iva,DelNobile:2011uf,Alanne:2017oqj,Brod:2017bsw,Carpenter:2015xaa,Belyaev:2018pqr}.
In contrast, the exploration of $\mathbb{Z}_2$ protected spin-1 vector DM scenarios has not garnered much attention, primarily due to the complexities associated with model building and the challenges in establishing viable interactions with the SM sector.
In the existing literature, only a limited number of studies have focused on simplified dark gauge models
\cite{Hambye:2008bq,Lebedev:2011iq,Farzan:2012hh,Abe:2012hb,Baek:2013dwa,Hisano:2020qkq,Abe:2020mph,Hu:2021pln}, 
which aim to generate vector DM candidates through the mechanism of spontaneous symmetry breaking of the new gauge symmetry. 
These models offer a promising pathway for understanding the properties and interactions of vector DM, 
yet a comprehensive EFT analysis is still lacking in comparison to the extensive studies conducted on scalar and fermionic counterparts~\cite{Bishara:2017pfq,Liu:2017kmx,Kang:2018rad,Tomar:2022ofh,Liang:2024tef}.
Given the current ignorance of the nature of DM, it is both prudent and necessary to adopt an open-minded approach when exploring various DM scenarios.
Investigating vector DM from the EFT perspective is particularly valuable, especially in light of current direct detection experiments. 
Such investigations could provide crucial insights to model building and deepen our understanding of DM's fundamental characteristics and its role in the cosmos.
 
The purpose of this study is to systematically explore the interactions between spin-1 DM and nucleons/quarks/photons within the EFT framework. 
By considering both nonrelativistic (NR) and relativistic EFT descriptions, 
we aim to provide a comprehensive analysis of the various operators that govern these interactions. 
In the NREFT framework, we compile a complete list of operators for spin-1 DM couplings to nucleons and calculate their contributions to the DM response functions. 
Furthermore, we examine leading-order (LO) relativistic EFT operators involving DM and light quarks or photons, and we
subsequently perform an NR reduction to align them with the NREFT framework. 
Our study derives the nuclear scattering rate from these interactions and utilizes recent direct detection data, 
including both elastic nuclear recoil and the Migdal effect, 
to impose constraints on the EFT operators and the electromagnetic properties of vector DM. 
In our analysis, we will consider two cases in the NREFT framework: one in which the isospin symmetry is imposed, and one in which it is not. 
In the isospin-universal case, a universal Wilson coefficient (WC) is assigned to the operators involving the $p,n$ nucleons.
Conversely, in the isospin-specific case, the contributions from the two nucleons will be considered separately.
Similarly, two cases will be analyzed in the relativistic EFT framework, depending on whether the flavor $\rm SU(3)$ symmetry is imposed. In the flavor-universal case, a universal WC is assigned to the operators involving the $u,d,s$ quarks, while in the flavor-specific case, the contributions from the three quarks will be considered individually.
We find that nuclear recoil data from experiments such as PandaX-4T, XENONnT, LZ, 
and DarkSide-50 provide stringent bounds on the EFT coefficients for a DM mass above a few GeV, 
while Migdal-effect datasets can probe a DM mass as low as 20\,MeV. 
Finally, we propose a UV-complete model that realizes the complex spin-1 DM scenario, thereby contributing to the broader understanding of DM  interactions and properties.

The structure of the paper is organized as follows. In \cref{sec:VDMeft},
we begin by presenting the most general NREFT operators for vector DM-nucleon interactions. 
We then introduce the leading-order (LO) relativistic operators for DM-quark and DM-photon interactions in a low-energy EFT (LEFT)-like framework,
followed by matching them onto the NREFT operators in the NR limit. 
\cref{sec:scattering_rate} is dedicated to the theoretical formalism governing the DM-nucleus scattering due to the involvement of the complete set of NREFT interactions. 
In \cref{sec:results}, we explore the constraints on both NREFT and relativistic EFT interactions from current DM direct detection experiments, 
utilizing datasets from DM-nucleus elastic scattering and the Migdal effect. 
\cref{sec:UVmodel} discusses a UV model of vector DM and its matching to the EFT interactions. 
Finally, our conclusions are summarized in \cref{sec:summary}. 
Additional technical details can be found in \cref{app:NRreduction,app:Msqrd}, 
which cover NR reductions in nuclear cases and the squared matrix elements for each relativistic interaction, respectively.

\section{The nonrelativistic and relativistic EFT interactions for vector DM-nucleus scattering}
\label{sec:VDMeft}

In this section, we focus on EFT descriptions of vector DM scattering with nuclei, 
employing both nonrelativistic and relativistic EFT frameworks. 
We construct LO vector DM-nucleon operators within the NREFT framework to parametrize DM-nucleus scattering. 
Subsequently, we investigate relativistic DM-quark and DM-photon interactions in an extended LEFT framework that can be regarded as part of the dark sector EFT (DSEFT)~\cite{He:2022ljo,Liang:2023yta}, and we establish their connection to NREFT interactions through NR reduction.

\subsection{NR description of vector DM-nucleon interactions}

\begin{table}[t]
	\center
   \resizebox{\linewidth}{!}{
		\renewcommand\arraystretch{1.4}
		\begin{tabular}{| l | c | l | c |}
			\hline
			{\cellcolor{magenta!15}\qquad\quad\, NR operators }  & {\cellcolor{magenta!15}Power counting} & {\cellcolor{magenta!15}\qquad\qquad\, NR operators}   & {\cellcolor{magenta!15}Power counting}
			\\
            \hline
			\cellcolor{gray!15}$\calO_1=\mathone_X \mathone_N$ & $1$ & $\calO_{13} = (\pSDM \cdot \pvn) \left(\frac{i\pq}{m_N}\cdot \pSN  \right)$ & $qv$
            \\
            \hline
            $\calO_3= \mathone_X \left(\frac{i \pq}{m_N}\times \pvn \right) \cdot  \pSN$ & $qv$ &  $\calO_{14} = (\pSDM \cdot \frac{i\pq}{m_N} ) (\pvn \cdot \pSN)$ & $qv$
            \\
            \hline
            $\calO_4 = \pSDM \cdot \pSN$ & $1$ & $\calO_{15} = \pSDM \cdot \frac{\pq}{m_N}  \left[ \frac{\pq}{m_N}  \cdot ( \pvn \times \pSN)\right]$ & $q^2v$
            \\
            \hline
            \cellcolor{gray!15}$\calO_5 = \pSDM \cdot \left(\frac{i\pq}{m_N}\times \pvn\right) \mathone_N$ & $qv$ &  
            \cellcolor{gray!15}$\calO_{17} = \frac{i \pq}{m_N} \cdot \pmb{\tilde\calS}_X \cdot \pvn \,\mathone_N$ & $qv$
            \\
            \hline
            $\calO_6 =  \left(\pSDM\cdot \frac{\pq}{m_N}\right) \left( \frac{\pq}{m_N} \cdot \pSN \right)$ & $q^2$ & $\calO_{18} = \frac{i \pq}{m_N} \cdot \pmb{\tilde\calS}_X \cdot  \pSN$ & $q$
            \\
            \hline
            $\calO_7 = \mathone_X \, \pvn \cdot \pSN$ & $v$ & 
            \cellcolor{gray!15}$\calO_{19} =  \frac{\pq}{m_N} \cdot \pmb{\tilde\calS}_X \cdot  \frac{\pq}{m_N} \mathone_N$ & $q^2$
            \\
            \hline
           \cellcolor{gray!15}$\calO_8 = \pSDM \cdot \pvn \, \mathone_N$ & $v$ & $\calO_{20} =  - \frac{\pq}{m_N}  \cdot \pmb{\tilde\calS}_X \cdot \left(\frac{\pq}{m_N} \times  \pSN\right)$ & $q^2$
            \\
            \hline
            $\calO_9 = -\pSDM \cdot\left(\frac{i\pq}{m_N}\times \pSN \right)$ & $q$ & $\calO_{23} = - \frac{i \pq}{m_N}\cdot\pmb{\tilde\calS}_X \cdot(\pvn \times \pSN)$ & $qv$
            \\
            \hline
            $\calO_{10} =  \mathone_X \, \frac{i \pq}{m_N}\cdot \pSN $ & $q$ & 
            \cellcolor{gray!15}$\calO_{24} = {\pq \over m_N} \cdot \pmb{\tilde\calS}_X \cdot  \left({ \pq \over m_N} \times\pvn\right)$ & $q^2v$
            \\
            \hline
           \cellcolor{gray!15}$\calO_{11} =  \pSDM \cdot \frac{i\pq}{m_N} \mathone_N$ & $q$ &  $\calO_{25} = \left({\pq\over m_N} \cdot \pmb{\tilde\calS}_X \cdot\pvn \right) 
\left({\pq\over m_N} \cdot\pSN \right)$ & $q^2v$
            \\
            \hline
            $\calO_{12} = - \pSDM \cdot (\pvn\times \pSN)$ & $v$ & $\calO_{26} = \left({\pq\over m_N} \cdot \pmb{\tilde\calS}_X \cdot {\pq\over m_N} \right) (\pvn\cdot\pSN)$ & $q^2v$ 
            \\
            \hline
	\end{tabular}
 }
	\caption{The basis of NR operators for vector DM-nucleon interactions up to second order in $\pmb{q}$ and linear order in $\pvn$. Operators $\calO_2$ and $\calO_{16}$, which are quadratic in $\pvn$, are not shown. $\calO_{21}$ and $\calO_{22}$ are excluded, as they do not contribute under the one-nucleon current approximation; however, they become relevant for DM-electron scattering when $\pvn$ is replaced by $\pmb{v}_{\rm el}^\perp$~\cite{Liang:2024lkk,Liang:2024ecw}.} 
	\label{tab:NRop}
\end{table}

The NREFT framework has been widely applied to describe DM interactions with nuclei and atomic electrons in direct detection experiments~\cite{Fan:2010gt,DelNobile:2018dfg,Fitzpatrick:2012ix}. 
Given the nonrelativistic nature of both incoming DM and target particles, 
NREFT can accurately capture the dynamics of DM-nucleus and DM-electron scattering in a model-independent way.
In the literature, Ref.\,\cite{Dent:2015zpa} was the first to enumerate the NREFT operators for the vector DM case. However, the operator set was found to be incomplete in 
Ref.\,\cite{Catena:2019hzw}. Furthermore, the operator basis presented in this latter work is still neither independent nor complete as later pointed out in 
Ref.\,\cite{Gondolo:2020wge}.
In this work, 
we follow a similar convention to that used for the DM-electron scattering in our previous works~\cite{Liang:2024lkk,Liang:2024ecw} to construct a basis of complete and independent LO NREFT operators suitable for DM-nucleon interactions.

For the nucleon $N$ and the vector DM $X$, the relevant NR building blocks in spin space are given by 
$\left\{\mathbbm{1}_N,\,\pmb{S}_N\right\}$ and $\left\{ \mathbbm{1}_X,\, \pmb{S}_X,\,\pmb{\tilde\calS}_X\right\}$, respectively. 
Here, $\mathbbm{1}_N$ and $\pmb{S}_N$ ($\mathbbm{1}_X$ and $\pmb{S}_X$) denote the number and spin operators for the nucleon $N$ (vector DM $X$), respectively. 
$\pmb{\tilde\calS}_X$ is a rank-2 traceless and symmetric tensor operator associated with the vector DM, which is built using the spin operator:
\begin{align}
\tilde\calS_X^{ij} = {1 \over 2} (S_X^i S_X^j+S_X^j S_X^i) - {2 \over 3} \delta^{ij}.
\end{align}
In addition, the two kinetic variables $\left\{i\pmb{q},\, \pmb{v}^\perp_N\right\}$ are also required for the operator construction. 
One is the momentum transfer $\pmb{q}\equiv\pmb{p}-\pmb{p}^\prime$,  with $\pmb{p}$ ($\pmb{p}^\prime$) denoting the initial (final) three-momentum of the vector DM.
The other one, $\pmb{v}_N^\perp$, is the transverse DM-nucleon velocity in the NR limit. 
It relates to the momentum transfer $\pmb{q}$ and the incoming DM-nucleon relative velocity $\pmb{v}$ in the lab frame through the relation  
$\pmb{v}_N^\perp \equiv \pmb{v} -\pmb{q}/2 \mu_{XN}$, where $\mu_{XN}$ is the reduced mass of the DM-nucleon system. 
For elastic scattering, $\pmb{v}_N^\perp \cdot \pmb{q} = 0$. 

Using the aforementioned building blocks, we can construct the LO Hermitian and Galilean-invariant NR operators, 
organized according to the powers of $\pmb{q}$ and $\pmb{v}_N^\perp$. 
All relevant operators up to second order in $\pmb{q}$ and linear order in $\pmb{v}_N^\perp$ are summarized in \cref{tab:NRop}, 
along with the corresponding power counting for each operator.
We highlight the nuclear spin-independent (SI) NREFT operators in gray, while the remaining operators are classified as spin dependent (SD).
Among these, $\calO_{1,3,7,10}$ contribute for DM of any spin, while $\calO_{4-6,8-9,11-15}$ contribute additionally to the fermionic DM case. 
Operators $\calO_{17-25}$ are exclusively relevant to vector DM scattering.
These NR operators encapsulate the most general interactions relevant to vector DM-nucleus scattering, 
with the details of the underlying physics parametrized by their WCs, $c_i^N$. 
To understand the fundamental origin of these operators, an intermediate step involves bridging them to the 
relativistic EFT framework, which can provide insights into possible underlying theory. 
In the next subsection, we will gather the relevant extended LEFT interactions for the vector DM case and match them onto the NREFT interactions in \cref{tab:NRop}. 

\begin{table}[t]
	\center
    \resizebox{\linewidth}{!}{
		\renewcommand\arraystretch{1.48}
		\begin{tabular}{| c | l | c | }
			\hline%
			Dim &\qquad\qquad\quad Relativistic operators & NR reductions 
            \\
            \hline%
            \multicolumn{3}{|c|}{\cellcolor{magenta!15}Vector case A}
		    \\
			\hline%
            \multirow{5}{*}{dim 5}  & 
			$\calO_{q X}^S=(\overline{q} q)(X_\mu^\dagger X^\mu)$ 
            & $-2 F_S^{q/N} m_N \calO_1$
            \\
            \cline{2-3}%
            & $\calO_{q X}^P=(\overline{q}i \gamma_5 q)(X_\mu^\dagger X^\mu)$ 
            & $ 2 F_P^{q/N} m_N \calO_{10}$
            \\
           \cline{2-3}%
& \multirow{2}*{ $\calO_{q X1}^T = {i \over 2} (\overline{q}  \sigma^{\mu\nu} q)
(X_\mu^\dagger X_\nu - X_\nu^\dagger X_\mu) \, (\times) $ } 
& $-4 F_{T,0}^{q/N} m_N \calO_4$
\\
& & {\cellcolor{blue!15} $-2e^2 \Lambda_1 Q_q {m_N^2 \over \bm{q}^2}\left\{2 Q_N \left[ {m_N \over {m_X}}
\left( {\bm{q}^2\over 3 m_N^2 }\calO_1 - \calO_{19}\right)- \calO_5\right]
- g_N \left( { \bm{q}^2 \over m_N^2 } \calO_4 -  \calO_6\right)\right\}$} 
            \\
            \cline{2-3}%
            & \multirow{2}*{$\calO_{q X2}^T = {1\over 2} (\overline{q}\sigma^{\mu\nu}\gamma_5 q) 
           (X_\mu^\dagger X_\nu - X_\nu^\dagger X_\mu)  \, (\times) $} 
            & $4 F_{T,0}^{q/N} m_N\left[-\calO_{12}-{m_N\over m_X }\left(\calO_{18}-\frac{1}{3}\calO_{10}\right)\right]-\left( F_{T,0}^{q/N}-2 F_{T,1}^{q/N}-4 F_{T,2}^{q/N}\right) m_N \calO_{11}$ 
            \\
             &  &  {\cellcolor{blue!15} $+ 4 e^2 \Lambda_1 Q_q Q_N {m_N^2 \over \bm{q}^2}\calO_{11} $}
            \\
            \hline%
             \multirow{12}{*}{dim 6}  & $\calO_{q X1}^V = {1\over 2} [ \overline{q}\gamma_{(\mu} i \overleftrightarrow{D_{\nu)} } q] 
            (X^{\mu \dagger} X^\nu + X^{\nu \dagger} X^\mu  )$ 
            & $A_{20}^{q/N} m_N^2 \calO_1$ 
            \\
            \cline{2-3}%
             & $\calO_{q X2}^V =(\overline{q}\gamma_\mu q)\partial_\nu (X^{\mu \dagger} X^\nu + X^{\nu \dagger} X^\mu  )$ 
            & $-4m_N^2\left[F_1^{q/N}\calO_{17}+\left(F_1^{q/N}+F_2^{q/N}\right)\calO_{20}\right]$
            \\
            \cline{2-3}%
             & $\calO_{q X3}^V =  (\overline{q}\gamma_\mu q)
            ( X_\rho^\dagger \overleftrightarrow{\partial_\nu} X_\sigma )\epsilon^{\mu\nu\rho\sigma}$ 
            & $-4m_X m_N\left[F_1^{q/N}\calO_8-\left(F_1^{q/N}+F_2^{q/N}\right)\calO_9\right]$
            \\
            \cline{2-3}%
             & $\calO_{q X4}^V = (\overline{q}\gamma^\mu q)(X_\nu^\dagger i\overleftrightarrow{\partial_\mu} X^\nu)\,(\times)$ 
            & $-4 F_1^{q/N} m_X m_N \calO_1$
            \\
            \cline{2-3}%
             & $\calO_{q X5}^V = (\overline{q}\gamma_\mu q)i\partial_\nu (X^{\mu \dagger} X^\nu - X^{\nu \dagger}X^\mu)\,(\times)$
            & $2 m_N^2 \left\{ F_1^{q/N} \left[{m_N\over m_X}\left( \frac{\pmb{q}^2}{3m_N^2}\calO_1-\calO_{19}\right) + \calO_5 \right]+ \left(F_1^{q/N} +F_2^{q/N}\right)\left({\pmb{q}^2 \over m_N^2 } \calO_4-\calO_6\right) \right\}$
            \\
            \cline{2-3}%
             & $\calO_{q X6}^V = (\overline{q}\gamma_\mu q) i\partial_\nu (X^\dagger_\rho X_\sigma )\epsilon^{\mu\nu\rho\sigma}\, (\times)$ 
            & $-2 F_{1}^{q/N} m_N^2 \calO_{11}$ 
            \\
            \cline{2-3}%
             & $\calO_{q X1}^A = {1\over 2} [\overline{q}\gamma_{(\mu} \gamma_5 i \overleftrightarrow{D_{\nu)} }  q]
            (X^{\mu \dagger} X^\nu + X^{\nu \dagger} X^\mu  )$ 
            & $ -2 \tilde{A}_{20}^{q/N} m_N^2\left(\frac{4}{3} \calO_{7}+{m_N\over m_X}\calO_9\right) $
            \\
            \cline{2-3}%
             & $\calO_{q X2}^A = (\overline{q}\gamma_\mu \gamma_5 q)\partial_\nu (X^{\mu \dagger} X^\nu + X^{\nu \dagger} X^\mu  )$ 
            & $8 G_A^{q/N} m_N^2 (\calO_{18}-\frac{1}{3}\calO_{10})$
            \\
            \cline{2-3}%
             & $\calO_{q X3}^A = (\overline{q}\gamma_\mu\gamma_5 q) 
            (X_\rho^\dagger \overleftrightarrow{ \partial_\nu} X_\sigma )\epsilon^{\mu\nu\rho\sigma}$ 
            & $8 G_A^{q/N} m_N m_X \calO_4$ 
            \\
            \cline{2-3}%
             & $\calO_{q X4}^A = (\overline{q}\gamma^\mu\gamma_5 q)(X_\nu^\dagger i\overleftrightarrow{\partial_\mu} X^\nu)\,(\times)$ 
            & $8 G_A^{q/N} m_X m_N \calO_7$ 
            \\
            \cline{2-3}%
             & $\calO_{q X5}^A =(\overline{q}\gamma_\mu\gamma_5 q)i \partial_\nu (X^{\mu\dagger} X^\nu - X^{\nu\dagger} X^\mu)\,(\times)$ 
            & $ 4G_A^{q/N} m_N^2 \calO_9$ 
            \\
            \cline{2-3}%
             & $\calO_{q X 6}^A =(\overline{q}\gamma_\mu\gamma_5 q)i\partial_\nu (X^\dagger_\rho X_\sigma)\epsilon^{\mu\nu\rho\sigma}\,(\times)$ 
            & $4G_A^{q/N} m_N^2 \left(\calO_{14}-{m_N\over m_X}\calO_{20}\right)$
            \\
\Xhline{3\arrayrulewidth}
\multirow{2}{*}{dim 4} & {\cellcolor{gray!25}$\calL_{\kappa_\Lambda}=i {{\kappa_\Lambda}\over 2}(X_\mu^\dagger X_\nu-X_\nu^\dagger X_\mu) F^{\mu\nu} \, (\times) $ } 
& $ e \kappa_\Lambda {m_N^2 \over \bm{q}^2} \left\{2  Q_N  \left[ {m_N \over {m_X}}
\left( {\bm{q}^2\over 3 m_N^2 }\calO_1 - \calO_{19}\right)- \calO_5\right]
- g_N \left({\bm{q}^2\over m_N^2 } \calO_4 - \calO_6\right)\right\}$
\\
\cline{2-3}
 & {\cellcolor{gray!25}$\calL_{\tilde\kappa_\Lambda}= i 
 {\Tilde{\kappa}_\Lambda \over 2}(X_\mu^\dagger X_\nu-X_\nu^\dagger X_\mu) \tilde F^{\mu\nu} \, (\times) $} 
& $-2 e Q_N \Tilde{\kappa}_\Lambda m_N^2 {1  \over \bm{q}^2} \calO_{11}$
\\
\hline
\multirow{5}{*}{dim 6} & {\cellcolor{gray!25}$\calO_{X\gamma 1}= \epsilon^{\mu\nu\rho\sigma} \left( X_\rho^\dagger \overleftrightarrow{\partial_\nu}X_\sigma\right)\partial^\lambda F_{\mu\lambda}$ } 
& $ 2 e m_N \left( 2 Q_N m_X \calO_8 - g_N \calO_9 \right)$
\\
\cline{2-3}
 & {\cellcolor{gray!25}$ \calO_{X\gamma 2}= \epsilon^{\mu\nu\rho\sigma} i \partial_\nu\left( X_\rho^\dagger X_\sigma\right)\partial^\lambda F_{\mu\lambda} \,(\times)$ } 
& $2 e Q_N m_N^2 \calO_{11}$
\\
\cline{2-3}
 & {\cellcolor{gray!25}$ \calO_{X\gamma 3}= \left( X_\nu^\dagger i\overleftrightarrow{\partial^\mu} X^\nu\right)\partial^\lambda F_{\mu\lambda}$ } 
& $4 e Q_N m_N m_X \calO_1 $
\\
\cline{2-3}
& {\cellcolor{gray!25}$ \calO_{X\gamma 4}= \partial_\nu(X^{\mu\dagger} X^{\nu} + X^{\nu\dagger} X^{\mu}) \partial^\lambda F_{\mu\lambda} $ } 
& $ 2 e  m_N^2 \left(2 Q_N \calO_{17}+ g_N \calO_{20}\right) $
\\
\cline{2-3}
& {\cellcolor{gray!25}$ \calO_{X\gamma 5}=  i \partial_\nu(X^{\mu\dagger} X^{\nu} - X^{\nu\dagger} X^{\mu}) \partial^\lambda F_{\mu\lambda} \,(\times)$ } 
& $-e  m_N^2 \left\{2 Q_N \left[ {m_N \over m_X} \left({\bm{q}^2  \over 3m_N^2 } \calO_1 - \calO_{19} \right) + \calO_5 \right] + g_N\left({\bm{q}^2 \over m_N^2 } \calO_4 - \calO_6\right) \right\}$
			\\
			\hline
            \multicolumn{3}{|c|}{\cellcolor{magenta!15}Vector case B}
		    \\
			\hline%
			\multirow{6}{*}{dim 7}  & $\tilde \calO_{qX1}^S = (\overline{q}q)X_{\mu\nu}^\dagger  X^{\mu\nu}$ 
            & $4F_S^{q/N} m_N m_X^2 \calO_1$ 
            \\
            \cline{2-3}%
             & $\tilde \calO_{qX2}^S = (\overline{q}q)X_{\mu\nu}^\dagger \tilde X^{ \mu\nu}$ 
            & $4F_S^{q/N} m_N^2 m_X \calO_{11}$ 
            \\
            \cline{2-3}%
            &  $\tilde \calO_{qX1}^P = (\overline{q}i \gamma_5q)X_{\mu\nu}^\dagger X^{ \mu\nu}$ 
            & $-4F_P^{q/N} m_N m_X^2 \calO_{10}$
            \\
            \cline{2-3}%
             & $\tilde \calO_{qX2}^P= (\overline{q}i \gamma_5q)X_{\mu\nu}^\dagger \tilde X^{ \mu\nu}$ 
            & $4F_P^{q/N} m_N^2 m_X \calO_{6}$
\\
\cline{2-3}%
& \multirow{2}*{ $\tilde \calO_{qX1}^T = {i \over 2} (\overline{q}\sigma^{\mu\nu} q)
(X^{\dagger}_{\mu\rho} X^{\rho}_{\,\nu}-X^{\dagger}_{\nu\rho} X^{\rho}_{\,\mu})\,(\times)$ }
& $4F_{T,0}^{q/N} m_N m_X^2 \calO_{4} $
\\%
& & {\cellcolor{blue!15} $ + 2e^2 \Lambda_1 Q_q {m_N^2 m_X^2 \over \bm{q}^2} 
\left\{ 2 Q_N \left[{\bm{q}^2\over 3 m_N^2 } \left(1 + {m_N \over m_X} \right)\calO_1 
- \left( 1 - 2 {m_N \over m_X} \right) \calO_{19} + \calO_5  \right] 
+ g_N \left({\bm{q}^2\over m_N^2 } \calO_4- \calO_6 \right) \right\} $}
\\%
\cline{2-3}
& \multirow{2}*{ $\tilde \calO_{qX2}^T = {1\over 2}(\overline{q} \sigma^{\mu\nu}\gamma_5 q)
(X^{\dagger}_{ \mu\rho} X^{\rho}_{\,\nu}-X^{\dagger}_{ \nu\rho} X^{\rho}_{\,\mu})\,(\times)$ }
& $4F_{T,0}^{q/N} m_N m_X\left(-\frac{2}{3}m_N \calO_{10}+{m_X\over 4}\calO_{11}+m_X\calO_{12}-m_N\calO_{18}\right)$
\\
& & {\cellcolor{blue!15} $-4 e^2 \Lambda_1 Q_q Q_N m_N^2 m_X^2 {1 \over\bm{q}^2} \calO_{11} $ }
\\%
\Xhline{3\arrayrulewidth}
\multirow{2}{*}{dim 6} 
& {\cellcolor{gray!25}$ \Tilde{\calO}_{X\gamma 1} = 
i (X^{\dagger}_{ \mu\rho} X^{\rho}_{\,\nu}-X^{\dagger}_{ \nu\rho} X^{\rho}_{\,\mu}) F^{\mu\nu} \,(\times)$} 
& $ -2e {m_N^2 m_X^2 \over \bm{q}^2} \left\{ 2 Q_N \left[{\bm{q}^2\over 3 m_N^2 }(1 + {m_N \over m_X})\calO_1 
- \left( 1 - 2 {m_N \over m_X} \right) \calO_{19} + \calO_5 \right]
+ g_N \left({\bm{q}^2\over m_N^2 } \calO_4- \calO_6 \right) \right\}$
\\%
\cline{2-3}
& {\cellcolor{gray!25}$ \Tilde{\calO}_{X\gamma 2}=
i (X^{\dagger}_{ \mu\rho} X^{\rho}_{\,\nu}-X^{\dagger}_{ \nu\rho} X^{\rho}_{\,\mu}) \tilde F^{\mu\nu}\,(\times)$} 
&  $ 4eQ_N  m_N^2 m_X^2 {1 \over\bm{q}^2} \calO_{11}$
\\
\hline
\end{tabular}
 }
\caption{ The relevant relativistic vector DM-quark and DM-photon operators 
and their corresponding leading non-vanishing order NR reduction results in terms of the basis in \cref{tab:NRop}. 
Note that the matching results presented here are directly applicable to the case of complex vector DM.  
In the real vector DM case, the relativistic operators marked with `$\times$' vanish due to symmetry restrictions,
and an additional factor of 2 should be included in the NR reduction part of the surviving operators. 
} 
\label{tab:NRmatch}
\end{table}

\subsection{Relativistic description of vector DM-quark/photon interactions and their NR reduction}
\label{sec:NRreduct}

The LEFT is a relativistic EFT designed for physical processes occurring below the electroweak scale $\calO (100~\rm GeV)$. The field contents in LEFT comprise the SM light degrees of freedom after the spontaneous symmetry breaking, and the operators constructed from these fields respect the SM residual symmetry $\rm SU(3)_{c}\otimes U(1)_{em}$. Meanwhile, the heavy degrees of freedom, including the top quark, and the Higgs and $W,Z$ bosons, have been integrated out. 
This LEFT framework is quite general and useful for describing DM direct detection with a small momentum transfer. 
To suit our needs, we extend its field contents to include an additional vector DM particle,  which can be regarded as a special case to the dark sector EFT (DSEFT) framework that we developed previously in~Ref.\,\cite{Liang:2023yta}.   

The DSEFT provides a powerful tool for investigating the low-energy processes participated in by the DM particle. 
This includes applications to DM detection in both nuclear and electron recoils~\cite{Brod:2017bsw,Kang:2018rad,Tomar:2022ofh,Liang:2024lkk, Liang:2024ecw}, 
and to DM production in terrestrial experiments and astronomical environments~\cite{Kundu:2021cmo,Barman:2021hhg}, as well as addressing flavor physics anomalies~\cite{He:2022ljo,He:2023bnk}.
Here, we consider the Lorentz-invariant interactions relevant to vector DM-nucleus scattering. 
Specifically, we investigate the effective interactions between the vector DM and light quarks $(u,d,s)$ and the photon within the DSEFT framework.
For the vector DM-quark interactions, Ref.\,\cite{He:2022ljo} has provided a complete list of operators at LO. These operators are classified into two categories: 
case A and case B. The operators in case A are constructed from the vector DM potential $X^\mu$, 
and there are four independent dimension-5 (dim-5) operators and twelve dim-6 operators.
In case B, the operators are constructed using the vector DM field strength tensor: 
$X^{\mu\nu} \equiv \partial^\mu X^\nu-\partial^\nu X^\mu$, 
and there are six operators at the leading dim-7 order. 
These operators are summarized in the second column of \cref{tab:NRmatch}, 
where a ‘‘$\times$’’ indicates that the corresponding operator vanishes for the real vector DM case, with $X_\mu^\dagger=X_\mu$. 
For the two operators $\calO_{qX1}^{V(A)}$, different from the convention in~Ref.\,\cite{He:2022ljo},
we use the traceless symmetric tensor current 
$\gamma_{(\mu} i \overleftrightarrow{D_{\nu)} } \equiv {1\over 2}[ \gamma_{\mu} i \overleftrightarrow{D_{\nu}} 
+ \mu \leftrightarrow \nu] - {1\over 4} g_{\mu\nu}i\overleftrightarrow{\slashed{D}}$ 
with the covariant derivative convention $A\overleftrightarrow{D_{\mu} }B \equiv A (D_{\mu}B)-(D_{\mu}A)B$, 
which facilitates the NR reduction calculation below.
The electromagnetic operators for vector DM, together with their physical interpretation, have been discussed in previous works~\cite{Hagiwara:1986vm,Gaemers:1978hg,Gounaris:1996rz,Hisano:2020qkq,Chu:2023zbo}. 
However, a complete list of these interactions/operators was only recently provided by us in~Ref.\,\cite{Liang:2024ecw}. 
They are also collected in \cref{tab:NRmatch} and will be used later for a more comprehensive analysis of DM-nucleus scattering. 
In \cref{sec:UVmodel}, we will consider a UV model that naturally gives rise to some of these operators after integrating out the heavy particles, 
thereby underscoring the significance of the EFT analysis in this work.  

With the relativistic EFT interactions in hand, the next step is to match them to the NREFT framework. For each effective DM interaction, 
the basic procedure to obtain its NR counterpart involves performing the NR expansion of the relevant two-by-two scattering amplitude from the insertion of this interaction.
Subsequently, the resulting LO nonvanishing expressions are translated into NR operators along with their corresponding WCs. 
Concretely, for each Lorentz-invariant DM-quark interaction $C_{qX} \calO_{qX}$, its contribution to the scattering amplitude can be expressed as
\begin{align}
\label{eq:ampq}
\mathcal{M}_{XN}^{\tt SD}= 
\langle X(p'), N(k')|C_{qX} \calO_{qX}| X(p), N(k) \rangle 
= C_{qX} \langle X(p')| J_{X}| X(p) \rangle  \langle N(k') | J_{q} | N(k) \rangle,
\end{align}
where $C_{qX}$ denotes the corresponding WC. In the second equality, the amplitude has been factorized into the DM and nucleon parts 
due to the local operator taking a product form of DM and quark bilinears---i.e.,  $\calO_{qX} = J_{X}\times J_{q}$.
For each DM-photon interaction, by integration by parts, it can always be reorganized into the form $ C_{X\gamma} A_\mu {\cal J}_{X}^{\mu}$. 
Then, it contributes to the DM-nucleon scattering process via the long-distance (LD) $t$-channel diagram, mediated by a photon propagator. 
The photon-nucleon vertex in the diagram corresponds to the standard QED interaction, leading to an amplitude that also adopts a similar factorized form,
\begin{align}
\label{eq:ampg}
\mathcal{M}_{XN}^{\rm LD} = - {1 \over q^2} C_{X\gamma} 
\langle X(p') | {\cal J}_{X}^{\mu} | X(p )\rangle  
\langle N(k') |J_{\rm em,\mu} | N (k)\rangle,
\end{align}
where $J_{\rm em}^\mu=-e Q_q \bar q\gamma^\mu q$ represents the electromagnetic currents and the $1/q^2$ factor accounts for the photon propagator.

In \cref{eq:ampq,eq:ampg}, the calculation of the DM part is straightforward, which takes the form 
$\langle X(p') | J_{X}^I | X(p)\rangle 
=  K_{\mu\nu}^I(p,p') \epsilon_X^{s',\mu*}(p') \epsilon_X^{s,\nu}(p)$, 
where $\epsilon_X^{s,\nu}(p)$ [$\epsilon_X^{s',\mu}(p')$] is the polarization vector of the incoming [outgoing] vector DM state, 
while the superscript $I$ is used to denote all the other indices including the Lorentz one, the operator type, etc.
On the other hand, for the nucleon part, one must first parametrize the hadronic matrix elements arising from the quark bilinears 
in terms of independent nucleon spinor bilinears, each multiplied by an appropriate form factor--- namely, 
\begin{align}
\label{eq:ff1}
\langle N(k^\prime)| J_{q}^I | N (k)\rangle = \sum_j F_{I,j}^{q/N} (q^2) \bar u_{N^\prime} \Gamma_j^I u_N ,
\end{align}
where $u_N \equiv u_N(k)$ [$u_{N^\prime}\equiv u_N(k')$] is the spinor of the incoming [outgoing] nucleon state, and
$F_{I,j}^{q/N} (q^2)$ are the hadronic form factors associated with the Lorentz structure $\Gamma_j^I$ resulting from the current $J_{q}^I$. 
The complete parametrization for various quark bilinears, along with numerical inputs for the involved form factors, is collected in \cref{app:NRreduction} for reference. 
Assembling the above results, the resulting amplitude takes the following general form:
\begin{align}
\label{eq:MXN}
\calM_{XN} = C_{X}^I \sum_j  K_{\mu\nu}^I F_{I,j}^{q/N} (q^2) \left(\overline{u}_{N^\prime} \Gamma_j^I u_N \right) \epsilon_X^{s',\mu*}(p') \epsilon_X^{s,\nu}(p).
\end{align}

To connect with the NR operator, the last step is to perform the NR reduction of the amplitude in \cref{eq:MXN}. 
This is achieved by expanding the four-momentum, nucleon spinors, and DM polarization vectors in the NR limit and retaining the leading-order nonvanishing results.
Once the LO amplitude in the NR limit is obtained, 
it can be translated into the building blocks of NR operators by appropriately trading the spinors and polarization vectors with the corresponding spin operators. 
The details of this reduction and translation are provided in~Ref.\,\cite{Liang:2024ecw}, and we do not repeat it here.  
The NR reduction results for each operator are presented in the third column of \cref{tab:NRmatch}.
Given the importance of nonperturbative contributions to DM direct detection for operators involving tensor quark structures
$(\overline q \sigma^{\mu\nu} (\gamma_5) q)$~\cite{Liang:2024tef}, 
we have also included their NR reduction contributions in \cref{tab:NRmatch}, which are highlighted in light blue.
The nine DM-photon interactions contribute to DM scattering via the photon mediator, and we would expect the matching results to receive a $1/\pq^2$ enhancement. However, this LD enhancement is only evident for 
$\calL_{\kappa_\Lambda}$, $\calL_{\tilde\kappa_\Lambda}$, $\Tilde{O}_{X\gamma1}$, and $\Tilde{O}_{X\gamma2}$. In contrast, for the remaining DM-photon operators ($\calO_{X\gamma1,2,3,4,5}$), such LD effects are canceled due to the presence of the structure $\partial^\lambda F_{\mu\lambda}$.

Finally, a remark follows concerning the power of DM mass $m_X$ in the reduction results shown in \cref{tab:NRmatch}.
The factor of $m_X$ is mainly caused by the derivatives acting on the DM fields, where the zeroth component of each nonreducible partial derivative (i.e., those that cannot be reshuffled onto the quark current by integration by parts) contributes a factor of $m_X$ in the NR limit. 
For the parity-odd structure $X^\dagger_{\mu\nu}\Tilde{X}^{\mu\nu}$, only a single factor of $m_X$ appears, because there is only one time derivative due to the involvement of the Levi-Civita tensor $\epsilon$. 
For a total derivative acting on the vector DM part---i.e., $\partial_\nu (X^{\mu\dagger} X^{\nu} \pm X^{\nu\dagger} X^{\mu})$ and $\partial_\nu (X_\rho^\dagger X_\sigma) \epsilon^{\mu\nu\rho\sigma}$---the derivative $\partial_\nu$ could be equivalently transferred to the quark current by integration by parts. Consequently, these terms do not yield an $m_X$ factor. As can be seen from the table, the highest power dependence of $m_X$ is quadractic in the LO operators, and later we will see that the constraints on the WCs strongly depend on the power of $m_X$.

\section{The formalism for DM-nucleus scattering}
\label{sec:scattering_rate}

So far, we have constructed the NREFT operators parametrizing the vector DM-nucleon scattering and matched the LO DSEFT vector DM-quark/photon interactions onto the NREFT building blocks.
In this section, we will concentrate on the detailed calculation of DM-nucleus scattering resulting from the insertion of the general NREFT vector DM-nucleon interactions given in \cref{tab:NRop}.
In the NR limit, the differential cross section of the DM-nucleus scattering with respect to the recoil energy $E_R$ ($E_R=\pmb{q}^2/2m_A$) is expressed as~\cite{DelNobile:2021wmp}
\begin{align}
\label{eq:dsigma}
\frac{d\sigma_T}{d E_R}{=} 
\frac{1}{32 \pi}\frac{1}{m_X^2 m_A}\frac{1}{v^2} \overline{|\mathcal{M}|^2},
\end{align}
where $m_A$ and $m_X$ denote the masses of the target nucleus and vector DM, respectively, and $v$ is the speed of the incoming DM particle in the lab frame. 
The crucial input for the calculation is the spin-summed and -averaged matrix element squared, $\overline{|\mathcal{M}|^2}$, 
which encodes nuclear many-body effects and therefore requires a careful treatment.

With the NREFT operators $\calO_i$ given in \cref{tab:NRop}, the most general form of the nucleon-level interaction in the NR limit is represented as, 
$\calL^N_{\tt NR}=\sum_i c_i\calO_i$, 
where $c_i$ is a dimensionless WC associated with the operator $\calO_i$.
Based on the matching results in \cref{tab:NRmatch}, one can readily identify the contribution to $c_i$ from each DSEFT operator. 
This contribution is simply the WC of the relevant DSEFT operator multiplied by the corresponding factor
associated with the NR operator $\calO_i$ from the NR reduction expression in the table. 
To facilitate the calculation of the nuclear matrix element, the Lagrangian can be reorganized in terms of nucleon building blocks as follows: 
\begin{align}
\calL_{\tt NR}^N & = 
\calS_1^N\,\mathbbm{1}_N
+\calS_2^N\, \pmb{v}_N^\perp \cdot \pmb{S}_N 
+\pmb{\calV}_1^N\cdot \pmb{v}_N^\perp 
+\pmb{\calV}_2^N\cdot \pmb{S}_N 
+\pmb{\calV}_3^N \cdot (\pmb{v}_N^\perp\times\pmb{S}_N),
\label{eq:amp:nucleon}
\end{align}
where $\calS_{1,2}^N $ and $\pmb{\calV}_{1,2,3}^N$ are scalar and vector operators that depend on the momentum transfer $\pmb{q}$ and the DM spin operators.
Their explicit expressions from the complete NR interactions in \cref{tab:NRop} take the forms 
\begin{subequations}
    \label{eq:SN_VN}
    \begin{align}
    \calS_1^N &=c_1^N \mathone_X
    + c_{11}^N \Big(\pSDM \cdot\frac{i\pq}{m_N} \Big)
    + c_{19}^N \Big(\frac{\pq}{m_N}\cdot\pmb{\tilde{\calS}}_X\cdot\frac{\pq}{m_N}\Big),
    \\
    \calS_2^N &= c_7^N \mathone_X 
    + c_{14}^N\Big(\pSDM\cdot\frac{i\pq}{m_N}\Big) 
    + c_{26}^N\Big({\pq\over m_N} \cdot \pmb{\tilde\calS}_X \cdot {\pq\over m_N}\Big),
    \\
    \calV_1^{N,i} &= 
    c_5^N \epsilon^{ijk} S_X^j \frac{i q^k}{m_N}
    + c_8^N S_X^i + c_{17} \frac{iq^j}{m_N}\tilde{\calS}^{ij}_X+c_{24}^N \epsilon^{ijk} \tilde{\calS}^{jl}_X \frac{q^l}{m_N} \frac{q^k}{m_N},
    \\
    \nonumber
    \calV_2^{N,i} &= c_4^N S_X^i + c_6^N \Big(\pSDM\cdot \frac{\pq}{m_N} \Big) \frac{q^i}{m_N} 
    - c_9^N \epsilon^{ijk} S_X^j \frac{i q^k}{m_N}
    + c_{10}^N \frac{i q^i}{m_N} + c_{18}^N \frac{i q^j}{m_N}\tilde{\calS}^{ij}_X
    \\
    & -c_{20}^N \epsilon^{ijk} \tilde{\calS}^{jl}_X \frac{q^l}{m_N} \frac{q^k}{m_N}, 
    \\
    \calV_3^{N,i} &= c_3^N \frac{iq^i}{m_N}- c_{12}^N S_X^i
    +c_{13}^N \epsilon^{ijk} S_X^j \frac{i q^k}{m_N}
      + c_{15}^N \Big(\pSDM\cdot \frac{\pq}{m_N}\Big) \frac{q^i}{m_N} - c_{23}^N \frac{i q^j}{m_N} \Tilde{\calS}^{ij}_X
    \nonumber
     \\
    & +c_{25}^N \epsilon^{ijk} \tilde{\calS}^{jl}_X \frac{q^l}{m_N} \frac{q^k}{m_N}.
\end{align}
\end{subequations}
To derive the expression for $\calV_3^{N,i}$, we have applied the following operator relations concerning $\calO_{13,25}$:
\begin{subequations}
\begin{align}
\calO_{13}& = \Big(\pSDM \times \frac{i\pq}{m_N}\Big) \cdot (\pvn \times \pSN) +(\pSDM\cdot \pSN) \frac{i\pq}{m_N}\cdot\pvn,
\\
\calO_{25}&={\pq\over m_N} \cdot \pmb{\tilde\calS}_X \cdot \Big( {\pq\over m_N}\times (\pvn\times \pSN)\Big)
     + \Big({\pq\over m_N} \cdot \pmb{\tilde\calS}_X \cdot\pvn \Big) {\pq\over m_N} \cdot \pvn,
\end{align}
\end{subequations}
where the two terms proportional to $\pq\cdot\pvn$ have no contribution to the elastic scattering.

Having established the DM$-$single nucleon NR Lagrangian, 
the next step is to properly account for the nuclear many-body effects in order to obtain the DM-nucleus amplitude. 
In doing so, it is particularly beneficial to distinguish the effects related to the overall nuclear motion
and those pertaining to the internal nuclear structure. 
For a given nucleus target with $A$ nucleons, $\pvn$ includes a center-of-mass (c.m.) motion component $(\pmb{v}_T^\perp)$
and an intrinsic nucleon motion component ($\hat{\pmb{v}}_{N}^\perp$) relative to the nuclear c.m.---i.e., $\pmb{v}_N^\perp \equiv \pmb{v}_T^\perp + \hat{\pmb{v}}_{N}^\perp$---where the c.m. component is simply the average of $\pvn$ from all nucleons, 
$\pmb{v}_T^\perp = A^{-1}\sum_{i} \pmb{v}_{i}^\perp$, which depends solely on the nuclear c.m. motion.
With this separation, together with the one-body current domination assumption,
we have a schematic effective Lagrangian describing the $X(p)+T(k)\to X(p')+T(k')$ process at the nuclear level,
\begin{align}
{\cal L}_{XT} = 
\sum_{N=p,n}\sum_{i} \left[S^{N_i}\,\mathbbm{1}_{N_i}
+S_2^{N_i}\, \hat{\pmb{v}}_{N_i}^\perp \cdot \pmb{S}_{N_i}
+\pmb{U}^{N_i}\cdot \pmb{S}_{N_i} 
+\pmb{V}^{N_i} \cdot (\hat{\pmb{v}}_{N_i}^\perp\times\pmb{S}_{N_i}) 
+\pmb{W}^{N_i}\cdot \hat{\pmb{v}}_{N_i}^\perp \right],
\label{eq:LXT}
\end{align}
where $N_i$ denotes the $i$th nucleon (either proton or neutron) in the target nucleus $T$, 
and the coefficients independent of nuclear physics are
\begin{align}
S^N=\calS_1^N+\pmb{\calV}_1^N\cdot\pmb{v}_T^\perp,~
S_2^N = \calS_2^N,~
\pmb{U}^N = \pmb{\calV}_2^N+\calS_2^N \pmb{v}_T^\perp+\pmb{\calV}_3^N\times\pmb{v}_T^\perp,~
\pmb{V}^N = \pmb{\calV}_3^N,~
\pmb{W}^N=\pmb{\calV}_1^N.
\label{eq:struc}
\end{align}
Due to parity and time-reversal selections, the $S_2^N$ term does not contribute to the elastic DM-nucleus scattering~\cite{DelNobile:2021wmp},
and we will not consider it in the following discussion.

Since the nucleus is in an eigenstate of total angular momentum,
the nuclear matrix element from the interaction in \cref{eq:LXT} is typically calculated by performing a multipole expansion,
which accounts for the finite-size effects of the nucleus. For the technical procedures, one can refer to~Ref.\,\cite{DelNobile:2021wmp}.
After a careful treatment of the nuclear part, the final unpolarized DM-nucleus scattering amplitude squared takes the following factorized form:
\begin{align}
    \label{eq:squaredamp}
    \overline{|\mathcal{M}|^2}=\frac{m_A^2}{m_N^2}\sum_{N,N^\prime=p,n}\sum_{X,Y=M,\Delta, \atop \Sigma^\prime,\Sigma^{\prime\prime},\Tilde{\Phi}^\prime,\Phi^{\prime\prime}} 
    F_{XY}^{NN^\prime}(\pmb{q}^2) 
    R_{XY}^{NN^\prime}(\pmb{q}^2,\pmb{v}_T^{\perp2}) .
\end{align}
Here, $F_{XY}^{NN^\prime}(\pmb{q}^2)$ and  $R_{XY}^{NN^\prime}(q^2,\pmb{v}_T^{\perp2})$ are referred to as the nuclear and DM response functions, respectively. 
The nuclear response functions, $F_{XY}^{NN^\prime}(\pmb{q}^2)$, 
exclusively parametrize the nuclear effects and are defined as follows: 
\begin{align}
\label{eq:nuc:res}
F_{XY}^{NN^\prime}(\pmb{q}^2)=\frac{4\pi}{2j+1} \sum_{J}\langle j||\mathcal{T}_J^N[X]||j\rangle \langle j||\mathcal{T}_J^{N^\prime}[Y]||j\rangle,
\end{align}
where $j$ is the spin of the nucleus and $J$ is the summation index. $\langle j||\mathcal{T}_J^N[X]||j\rangle$ is the reduced matrix element of the one-nucleon operator $X^N_J$, with explicit angular momentum $J$ acting on the nucleon $N$.
Following the notations in Refs.\,\cite{Fitzpatrick:2012ix,Anand:2013yka}, the abbreviation $F_{X}^{NN^\prime}\equiv F_{XX}^{NN^\prime}$ is used for diagonal components.
The definition in \cref{eq:nuc:res} implies 
$F_{XY}^{NN^\prime}=F_{YX}^{N^\prime N}$.
For the effective interactions (excluding the $S_2^N$ term) in \cref{eq:LXT}, there are eight nonvanishing and independent nuclear response functions. Among these, six have a diagonal subscript denoted by $X=Y=M,\Sigma^\prime,\Sigma^{\prime\prime},\Phi^{\prime\prime},\Tilde{\Phi}^\prime,\Delta$, while the other two nondiagonal components are  $XY=\Phi^{\prime\prime}M,\Delta\Sigma^\prime$.
The nuclear response functions have been extensively studied in the nuclear community. For instance, they have been calculated for targets such as F, Na, Ge, I, and Xe using the standard shell-model methods in~Refs.\,\cite{Fitzpatrick:2012ix,Anand:2013yka}. 
For later numerical calculations, we adopt their results for the xenon target. 
For the argon target (${}^{40}$Ar), 
there are only three nonvanishing nuclear response functions, 
$F_M^{NN'},\,F_{\Phi^{\prime\prime}}^{NN'},\, F_{M\Phi^{\prime\prime}}^{NN'}$,  
and we use the results provided in~Ref.\,\cite{Catena:2015uha}.

The DM response functions, $R_{XY}^{NN^\prime}(\pmb{q}^2,\pmb{v}_T^{\perp2})$,
are the main goal of our calculation. 
In terms of the coefficients in \cref{eq:struc}, 
they can be calculated through the following compact forms,
\begin{subequations}
\begin{align}
 R^{N N'}_M &= {1\over 2 s_X+1}  S^{N*} S^{N'},
  \\
R^{N N'}_{\Sigma'} &= {1\over 2 s_X+1} {1\over 8}\left[\pmb{U}^{N*}\cdot\pmb{U}^{N'} - { (\pmb{q}\cdot\pmb{U}^{N*}) (\pmb{q}\cdot\pmb{U}^{N'})\over |\pmb{q}|^2} \right],
\\
R^{N N'}_{\Sigma''} &= {1\over 2 s_X+1}  { (\pmb{q}\cdot\pmb{U}^{N*}) (\pmb{q}\cdot\pmb{U}^{N'})\over  4 |\pmb{q}|^2},
 \\
 R^{N N'}_{\tilde\Phi'} &= {1\over 2 s_X+1}  { |\pmb{q}|^2 \over 8 m_N^2}\left[\pmb{V}^{N*}\cdot\pmb{V}^{N'} - { (\pmb{q}\cdot\pmb{V}^{N*}) (\pmb{q}\cdot\pmb{V}^{N'})\over |\pmb{q}|^2} \right],
\\
R^{N N'}_{\Phi''} &= {1\over 2 s_X+1}  { (\pmb{q}\cdot\pmb{V}^{N*}) (\pmb{q}\cdot\pmb{V}^{N'})\over  4m_N^2},
 \\
R^{N N'}_{\Delta} &= {1\over 2 s_X+1}  {\pmb{q}^2 \over 2 m_N^2}\left[\pmb{W}^{N*}\cdot\pmb{W}^{N'} - { (\pmb{q}\cdot\pmb{W}^{N*}) (\pmb{q}\cdot\pmb{W}^{N'})\over |\pmb{q}|^2} \right],
\\
R^{N N'}_{M\Phi''} &= {1\over 2 s_X+1} {S^{N*} (i \pmb{q}\cdot\pmb{V}^{N'}) \over 2 m_N},
 \\
R^{N N'}_{\Sigma'\Delta} &=   {1\over 2 s_X+1}  { - i \pmb{q}\cdot(\pmb{U}^{N*}\times \pmb{W}^{N'}) \over 4 m_N }, 
\end{align}
\end{subequations}
where the prefactor $1/(2 s_X+1)$ accounts for the spin average,
with $s_X=1$ denoting the spin of the vector DM.
By substituting the explicit expressions from \cref{eq:SN_VN,eq:struc} into the above formulas, we obtain the primary results for the DM response functions from the complete list of vector DM operators given in \cref{tab:NRop}:
\begin{subequations}
\label{eq:DMres}
    \begin{align}
    \nonumber
    R_M^{NN^\prime} &= c_1^N c_1^{N^\prime}+\frac{2}{3}\frac{\pmb{q}^2}{m_N^2} \pvt^2 c_5^N c_5^{N^\prime}+\frac{2}{3} \pvt^2 c_8^N c_8^{N^\prime}+ \frac{2}{3}\frac{\pmb{q}^2}{m_N^2} c_{11}^N c_{11}^{N^\prime}+
    \frac{1}{6}\frac{\pmb{q}^2}{m_N^2} \pvt^2 c_{17}^N c_{17}^{N^\prime} 
    \\
    &\qquad+ \frac{2}{9} \frac{\pmb{q}^4}{m_N^4} c_{19}^N c_{19}^{N^\prime} + \frac{1}{6} \frac{\pmb{q}^4}{m_N^4} \pvt^2 c_{24}^N c_{24}^{N^\prime},
    \\
    \nonumber
    R_{\Sigma^\prime}^{NN^\prime} &= \frac{1}{8} \frac{\pmb{q}^2}{m_N^2} \pvt^2 c_{3}^N c_{3}^{N^\prime} +
    \frac{1}{6} c_4^N c_4^{N^\prime} 
    +\frac{1}{8} \pvt^2 c_7^N c_7^{N^\prime} 
    + \frac{1}{6} \frac{\pmb{q}^2}{m_N^2} c_9^N c_9^{N^\prime}
    + \frac{1}{12} \pvt^2 c_{12}^N c_{12}^{N^\prime} 
    \\
    \nonumber
    &
    + \frac{1}{12}\frac{\pmb{q}^2}{m_N^2} \pvt^2 c_{14}^N c_{14}^{N^\prime} + \frac{1}{12} \frac{\pmb{q}^4}{m_N^4} \pvt^2 c_{15}^N c_{15}^{N^\prime}
    + \frac{1}{24}\frac{\pmb{q}^2}{m_N^2} c_{18}^N c_{18}^{N^\prime} 
    + \frac{1}{24}\frac{\pmb{q}^4}{m_N^4} c_{20}^N c_{20}^{N^\prime} 
    \\
    & 
    + \frac{1}{36}\frac{\pmb{q}^2}{m_N^2} \pvt^2 c_{23}^N c_{23}^{N^\prime} 
    + \frac{1}{36}\frac{\pmb{q}^4}{m_N^4} \pvt^2 c_{26}^N c_{26}^{N^\prime}  - \frac{1}{12} \frac{\pmb{q}^2}{m_N^2} \pvt^2 \left( c_{12}^N c_{15}^{N^\prime}
    + c_{15}^N c_{12}^{N^\prime}\right),
    \\
    \nonumber
    R_{\Sigma^{\prime\prime}}^{NN^\prime} &=
    \frac{1}{6}c_4^N c_4^{N^\prime} 
    + \frac{1}{6}\frac{\pmb{q}^4}{m_N^4} c_6^N c_6^{N^\prime} 
    + \frac{1}{4}\frac{\pmb{q}^2}{m_N^2} c_{10}^N c_{10}^{N^\prime} 
   + \frac{1}{6} \pvt^2 c_{12}^N c_{12}^{N^\prime}
    + \frac{1}{6}\frac{\pmb{q}^2}{m_N^2} \pvt^2 c_{13}^N c_{13}^{N^\prime} 
    \\
    \nonumber
    & 
    + \frac{1}{18}\frac{\pmb{q}^2}{ m_N^2} c_{18}^N c_{18}^{N^\prime} 
    + \frac{1}{24}\frac{\pmb{q}^2}{m_N^2} \pvt^2 c_{23}^N c_{23}^{N^\prime} 
    + \frac{1}{24} \frac{\pmb{q}^4}{m_N^4} \pvt^2 c_{25}^N c_{25}^{N^\prime}
    \\
    &
    + \frac{1}{6} \frac{\pmb{q}^2}{m_N^2}\left( c_4^N c_6^{N^\prime}
    + c_6^N c_4^{N^\prime}\right),
    \\
    R_{\Tilde\Phi^{\prime}}^{NN^\prime} &= 
    \frac{1}{6}\frac{\pmb{q}^2}{m_N^2} c_{12}^N c_{12}^{N^\prime} 
    + \frac{1}{6}\frac{\pmb{q}^4}{m_N^4} c_{13}^N c_{13}^{N^\prime} 
    + \frac{1}{24}\frac{\pmb{q}^4}{m_N^4} c_{23}^N c_{23}^{N^\prime}+{\frac{5}{36}} {\pmb{q}
    ^6 \over m_N^6 } c_{25}^N c_{25}^{N^\prime},
    \\
    \nonumber
    R_{\Phi^{\prime\prime}}^{NN^\prime} &= 
    \frac{1}{4}\frac{\pmb{q}^4}{ m_N^4} c_3^N c_3^{N^\prime}
    + \frac{1}{6}\frac{\pmb{q}^2}{m_N^2} c_{12}^N c_{12}^{N^\prime}
    + \frac{1}{6}\frac{\pmb{q}^6}{m_N^6} c_{15}^N c_{15}^{N^\prime} 
    + \frac{1}{18}\frac{\pmb{q}^4}{m_N^4} c_{23}^N c_{23}^{N^\prime} 
    \\    
    & -\frac{1}{6} \frac{\pmb{q}^4}{m_N^4} \left( c_{12}^N c_{15}^{N^\prime} 
    + c_{15}^N c_{12}^{N^\prime}\right),
    \\
    R_{\Delta}^{NN^\prime} &= \frac{2}{3}\frac{\pmb{q}^4}{m_N^4} c_5^N c_5^{N^\prime} + \frac{2}{3}\frac{\pmb{q}^2}{m_N^2} c_8^N c_8^{N^\prime} +  \frac{1}{6}\frac{\pmb{q}^4}{m_N^4} c_{17}^N c_{17}^{N^\prime} + \frac{1}{6} \frac{\pmb{q}^6}{m_N^6} c_{24}^N c_{24}^{N^\prime},
    \\
    R_{M\Phi^{\prime\prime}}^{NN^\prime} &= - \frac{1}{2}\frac{\pmb{q}^2}{m_N^2} c_1^N c_3^{N^\prime} - \frac{1}{3}\frac{\pmb{q}^2}{m_N^2} c_{11}^N c_{12}^{N^\prime} + \frac{1}{3}\frac{\pmb{q}^4}{m_N^4} c_{11}^N c_{15}^{N^\prime} + \frac{1}{9}\frac{\pmb{q}^4}{m_N^4} c_{19}^N c_{23}^{N^\prime},
    \\
    R_{\Sigma^\prime\Delta}^{NN^\prime} &= -\frac{1}{3}\frac{\pmb{q}^2}{m_N^2}c_{4}^N c_{5}^{N^\prime} + \frac{1}{3}\frac{\pmb{q}^2}{m_N^2} c_{9}^N c_{8}^{N^\prime} - \frac{1}{12} \frac{\pmb{q}^4}{m_N^4} c_{20}^N c_{17}^{N^\prime} + \frac{1}{12}\frac{\pmb{q}^4}{m_N^4}c_{18}^N c_{24}^{N^\prime},
\end{align}
\end{subequations}
where $\pvt^2= v^2-{\pmb{q}^2 / (4 \mu_{AX}^2})$ in the NR limit,
with $\mu_{AX}$ being the DM-nucleus reduced mass.
Similarly, the abbreviation $R_{X}^{NN^\prime}\equiv R_{XX}^{NN^\prime}$ is adopted, and the relationships 
$ R_{X}^{NN^\prime}=R_{X}^{N^\prime N}$ and $R_{XY}^{NN^\prime}=R_{YX}^{N^\prime N}$ hold for the elastic scattering case.
As clearly demonstrated above, after neglecting the interference terms, 
the SI NREFT operators are associated with nuclear response functions denoted by the subscripts $M$ and $\Delta$, 
whereas the SD operators correspond to those with subscripts $\Sigma', \Sigma'', \Phi'', \tilde \Phi'$.

The squared matrix element $\overline{|\calM|^2}$ in \cref{eq:squaredamp}, 
formulated in terms of nuclear and DM response functions, 
is quite general and valuable, as it allows for the factorization of nuclear physics and DM dynamics without mutual dependence. 
As previously mentioned, the nuclear response functions have been calculated for various relevant nucleus targets. 
For the DM response functions, our results for the vector DM in \cref{eq:DMres}  derived from the complete list of LO NR operators, 
along with the results for  scalar and fermionic DM cases presented in~Refs.\,\cite{DelNobile:2021wmp,Gorton:2022eed}, 
provide a comprehensive reference for the three primary DM candidates. 
Furthermore, the NR results can be directly linked to the relativistic EFT interactions, 
which encapsulate more fundamental possible origins of DM interactions. 
For a given DSEFT interaction, one can easily identify the corresponding nonvanishing NR WCs $c_i$ via the NR reduction provided in \cref{tab:NRmatch}.
The contribution of these nonvanishing coefficients to the squared matrix element can be readily obtained by incorporating the relevant nuclear and DM response functions.
The explicit forms resulting from the substitution of each DSEFT interaction are summarized in \cref{app:Msqrd}.

\section{Constraints on EFT interactions}
\label{sec:results}

In this section, we calculate the constraints on 
both the nonrelativistic and relativistic EFT operators (\cref{tab:NRop,tab:NRmatch})
from DM-nucleus elastic scattering and the Migdal effect
in DM direct detection experiments, 
including PandaX-4T, XENON1T, XENONnT, LZ, and DarkSide-50.

\subsection{Analysis on DM-nucleus elastic scattering}

For an isotope with mass fraction or abundance $\xi_i$ and mass $m_{A_i}$, the differential event rate per unit time and per target mass for DM-nucleus elastic scattering is given by,
\begin{align}
\frac{dR_{\tt NR}^i}{ dE_R} = 
\frac{\rho_X}{m_X }{\xi_{i} \over m_{A_i} } \int_{v_{\rm min}(E_R)}^{v_{\rm max}} dv F(v) v
{d\sigma_{T_i}\over dE_R} (v,E_R),
\label{eq:NR}
\end{align}
where $v_{\rm min}(E_R) = \sqrt{E_R m_{A_i}}/({\sqrt{2}\mu_{A_i X}})$, $\rho_X$ is the local DM density near the Earth,
and $F(v)$ is the DM velocity distribution in the lab frame,
with the angular dependence having been integrated out.
In the Galaxy rest frame, the DM velocity obeys a normal Maxwell-Boltzmann distribution, which leads to~\cite{Lewin:1995rx}
\begin{align}
F(v)= 
\frac{v}{\sqrt{\pi} v_0 v_E}
\begin{cases}
 e^{-(v-v_E)^2 / v_0^2}-e^{-(v+v_E)^2 / v_0^2}, 
& \text{for}~ 0 \leq v \leq v_{\mathrm{esc}}-v_E 
\vspace{0.1cm}
\\ 
 e^{-(v-v_E)^2 / v_0^2}-e^{-v_{\mathrm{esc}}^2 / v_0^2}, 
& \text{for~}v_{\mathrm{esc}}-v_E<v \leq v_{\mathrm{esc}}+v_E
\end{cases}.
\end{align}
Here, we adopt the averaged Earth-relative velocity $v_E= 244~\rm km/s$ and the suggested standard halo model parameters for the DM given in 
Ref.\,\cite{Baxter:2021pqo}, with the circular velocity $v_0= 238~\rm km/s$,
$\rho_X = 0.3~\si{GeV/cm^3}$, and the escape velocity $v_{\rm esc} = 544~\rm km/s$.

The nuclear target consists of stable isotopes with various spins, leading to different nuclear response functions in \cref{eq:nuc:res}.
To obtain the total event rate, one needs to compute the differential rate for each isotope and sum over their contributions---namely,
\begin{equation}
    \frac{dR_{\tt NR}}{ dE_R} = \sum_{i} \frac{dR_{\tt NR}^i}{ dE_R}.
\end{equation}
For the xenon-target experiments (PandaX-4T, XENON1T, XENONnT, and LZ), we consider the isotopes given in  \cref{tab:xe:abun}.
For the argon-target experiment (DarkSide-50), we focus exclusively on the contribution from $^{40} \rm Ar$, 
which is spinless and accounts for over 99\,\% of the mass fraction.
 
\begin{table}[t]
\center
\resizebox{\linewidth}{!}{
\renewcommand\arraystretch{1.3}
\begin{tabular}{|c|c|c|c|c|c|c|c|}  
\hline  
Isotopes 
& $^{128}$Xe & $^{129}$Xe & $^{130}$Xe & $^{131}$Xe & $^{132}$Xe & $^{134}$Xe & $^{136}$Xe
\\ \hline  
Abundance ($\xi_i$)
&1.9\,\%&26\,\%&4.1\,\%&21\,\%&27\,\%&10\,\%&8.9\,\%
\\ \hline
Spin
& 0& 1/2& 0& 3/2& 0& 0& 0 \\
\hline  
\end{tabular}  
}
\caption{The abundance for xenon isotopes taken from~Ref.\,\cite{DelNobile:2021wmp}.} 
\label{tab:xe:abun}
\end{table}

For the constraints from the elastic DM-nucleus scattering,
we consider the nuclear recoil ($\NR$) data from PandaX-4T~\cite{PandaX:2024qfu}, XENONnT~\cite{XENON:2023cxc}, LZ~\cite{LZ:2022lsv}, 
and DarkSide-50~\cite{DarkSide-50:2022qzh}.
Note the slight font difference for this abbreviation ``$\NR$'' and the ``NR'' used for ``nonrelativistic''.
All these experiments utilize dual-phase time projection chambers, 
which allow us to separate the electron recoil ($\ER$) and $\NR$ events
via different performances in prompt scintillation photon ($S_1$) and drift electron ($S_2$) signals.
Since most SM backgrounds are $\ER$ events~\cite{LZ:2022lsv,XENON:2023cxc,PandaX:2024qfu}, 
distinguishing $\NR$ events from $\ER$ events in the $S_1{\rm-}S_2$ plane is crucial for reducing backgrounds
in the analysis of DM-nucleus elastic scattering.

Among these experiments, LZ excels at separating $\ER$ and $\NR$ signals due to a larger electric field~\cite{LZ:2023poo}. 
This enables us to identify a region in the $S_1{\rm-}S_2$ plane, where approximately 90\,\% of $\NR$ events distribute, 
while simultaneously achieving effective isolation from the $\ER$ backgrounds ~\cite{LZ:2022lsv}.  
In contrast, for the  PandaX-4T and XENONnT experiments, this region is shrunk, 
effectively encompassing only about 50\,\% $\NR$ events because of a smaller electric field.    
Hence, we just take these well-isolated $\NR$ regions as our signal regions (SRs) for setting the constraints.
For the region of interest (ROI) in the LZ experiment, 
we define our SR as the area below the 10\,\% quantile line in the $S_1{\rm-}S_2$ plane---
specifically, the region below the upper dotted red line in Fig.\,1 of~Ref.\,\cite{LZ:2022lsv}.
In the PandaX-4T (XENONnT) experiment,
we designate our SR as the area below the $\NR$ median line 
in the $S_1{\rm-}S_2$ plane, which corresponds to the region below the solid pink (red) line of Fig.\,1 in Ref.\,\cite{PandaX:2024qfu} (Ref.\,\cite{XENON:2023cxc}).
The choice of the SR inside the ROIs given in experimental papers leads to additional efficiency factors $\epsilon_{\rm SR}$ for the calculation of the number of signal events;
$\epsilon_{\rm SR}$ varies as a function of DM mass, because the distribution of signal events in the $S_1{\rm-}S_2$ plane differs for different mass values.
We estimate $\epsilon_{\rm SR} (m_X)$ based on the WIMP signal event simulation conducted by the XENON Collaboration~\cite{XENON:2024xgd}.  We find that for XENONnT, nearly all WIMP $\NR$ events fall below the $\NR$ median line for $m_X \lesssim 6~\rm{GeV}$, whereas only about half of WIMP events fall below this line for $m_X \gtrsim 50~\rm{GeV}$. 
Thus, we adopt a selection efficiency of 100\,\% for $m_X \lesssim 6~\rm{GeV}$ and 
50\,\% for $m_X \gtrsim 50~\rm{GeV}$, applying a linear interpolation for the range $ 6~{\rm GeV} \lesssim  m_X \lesssim  50~{\rm GeV}$.  
Since PandaX-4T has a similar discrimination power for $\NR$ and $\ER$ events, we use the same $\epsilon_{\rm SR} (m_X)$ as that of XENONnT.  
For LZ, a higher efficiency of 90\,\% is adopted for $m_X \gtrsim 50~\rm{GeV}$
due to its superior $\ER$/$\NR$ discrimination capability.

In summary, for the xenon-target experiments, we calculate the number of signal events within the selected signal region using the following formula: 
\begin{align}
N_s = 
 w\times \epsilon_{\rm SR} (m_X) \int_{\rm SR} \epsilon_{\rm ROI} (E_R) \frac{dR_{\tt NR}}{d E_R} dE_R,
\end{align}
where $w$ is the exposure, $\epsilon_{\rm ROI}$ accounts for the combination of efficiency factors of the ROI given by experiment papers,
and $\epsilon_{\rm SR}$ is the additional efficiency factor for the selected signal region in the $S_1{\rm-}S_2$ plane.
For the $\NR$ analyses in PandaX-4T~\cite{PandaX:2024qfu}, XENONnT~\cite{XENON:2023cxc}, and LZ~\cite{LZ:2022lsv}, the observed data are consistent with the background-only hypothesis. 
For simplicity, we adopt the background-only hypothesis to set our constraints. 
We select a signal region that is specifically sensitive to the nuclear recoil, rather than using the entire ROI, in order to obtain more stringent bounds on the parameter space. This is the same strategy employed in~Ref.\,\cite{Bell:2023sdq}.
For a given number of observed events $N_o$ in the signal region,  
the 90\,\% C.L. constraints are determined using the criterion $N_s = N_s^{90}(N_o)$,  
based on the Poisson distribution and the
background-only hypothesis.
The details of how to set constraints for each xenon-target experiment are given as follows:
\begin{itemize}
\item
XENONnT: The exposure is 1.09 ton-yr, and the $\epsilon_{\rm ROI}$ is detailed in Fig.\,2 of~Ref.\,\cite{XENON:2023cxc}.
We choose the signal region 
as the area between the $\NR$ median curve (red solid) and the lower $2\,\sigma$ $\NR$ boundary (lower red dashed curve) in Fig.\,1 of~Ref.\,\cite{XENON:2023cxc} 
to maximize the signal-to-noise ratio, with a maximum nuclear recoil energy of 50~keV. 
Three events are observed within this region, 
resulting in $N_s^{90} = 3.68$.
\item
PandaX-4T: The total exposure is 1.54 ton-yr. The values of $\epsilon_{\rm ROI}$ for Run 0 and Run 1 are provided in Fig.\,15 of~Ref.\,\cite{PandaX:2024med}, with the maximum $\NR$ energy reaching approximately 80 keV.
As shown in Fig.\,1 of~Ref.\,\cite{PandaX:2024qfu}, 
a total of 24 events are observed below the $\NR$ median curve in the combined Run 0 and Run 1 datasets, 
among which 12 events exhibit $\NR$ energies below 20 keV.
For the WIMP with masses below approximately 50 GeV, the expected signal is mostly concentrated in the $\NR$ energy region below 20 keV, as indicated in Fig.\,5 of~Ref.\,\cite{XENON:2024xgd}.
Thus, to enhance the constraining power for the light DM, we consider two signal regions: 
one with $E_R \lesssim  80~\rm keV$ and the other with $E_R \lesssim  20~\rm keV$, corresponding to $N_s^{90} = 5.78$ and $7.58$, respectively.
We calculate the constraints for these two respective regions and choose the better one as our result. 
\item 
LZ: The exposure is 0.9 ton-yr. The efficiency factors for the ROI are given in Fig.\,2 of~Ref.\,\cite{LZ:2022lsv}.
As shown in Fig.\,4 of~\cite{LZ:2022lsv}, below the 10\,\% quantile line, there are 1 (11) events with an $\NR$ energy less than 15 (45) $\rm keV_{\NR}$, corresponding to $N_s^{90} = 2.89$ ($N_s^{90} = 5.6$).
Similarly to the PandaX-4T case, we calculate the constraints for the two respective regions ($E_R \lesssim  15\,\rm keV$ vs $E_R \lesssim  45\,\rm keV$) and choose the stronger one as our result.    
\end{itemize}

For the argon-target experiment, we estimate the constraints from DarkSide-50 using the data provided in~Ref.\,\cite{DarkSide-50:2022qzh} with an exposure of 12.7 ton-days. 
The charge yield for the $\NR$ events, as shown in Fig.\,1 of the paper, allows us to convert $dR_{\tt NR}/dE_R$ into $dR_{\tt NR}/dN_e$, 
where $N_e$ is the number of ionized electrons.
We adopt a global efficiency of 40\,\%~\cite{DarkSide-50:2022qzh}, which accounts for the overall acceptance from both quality and selection cuts. 
We select two signal regions: $3 \leq N_e \leq 5$ and $3 \leq N_e \leq 15$, as depicted in Fig.\,2 of the paper. 
The corresponding $N_{s}^{90}$ values for these signal regions are 42.9 and 101.7, respectively.
Using the simplified methods described above, we can obtain comparable constraints on the SI operators to those shown in each experimental paper, thereby validating the reliability of our analysis.

\subsection{Analysis on the Migdal effect}

In the sub-GeV DM mass range, the sensitivity of $\NR$ signal constraints is significantly reduced, 
because the nucleus is too heavy to effectively acquire enough recoil energy from a light DM. 
Furthermore, most of the $\NR$ energy is converted into heat rather than into detectable photoelectric signals. This limitation can be alleviated by utilizing the Migdal effect~\cite{Migdal:1941,Ibe:2017yqa,Dolan:2017xbu}, 
which involves inelastic effects between DM and the atom, resulting in the direct generation of ionized electrons in addition to the $\NR$ energy. 
This additional $\ER$ signal enables us to extend the constraints into the mass range of tens of MeV.

To obtain the differential event rate for the Migdal effect from a given isotope, 
one simply needs to attach an additional ionization form factor in \cref{eq:NR} and properly account for the transformation of different energies~\cite{Ibe:2017yqa}:  
\begin{align} 
\frac{dR_{\tt Migdal}^i}{dE_{\rm det}} = 
\frac{\rho_X}{m_X}{\xi_i\over m_{A_i}} \int_0^{E_R^{\rm max}} dE_R \int_{v_{\rm min}}^{v_{\rm max}} dv F(v) v {d\sigma_{T_i}\over dE_R} (v,E_R) |Z_{\rm ion}(E_R, \Delta E)|^2, 
\end{align}
where $E_R^{\rm max} = 2 \mu_{A_i X}^2 v_{\rm max}^2/m_{A_i}$.
Here $E_R$, $\Delta E$, and $E_{\rm det}$ are the $\NR$ energy, electron transition energy, 
and the total detected energy generated by the Migdal effect, respectively.
These energies satisfy the relation,  
$E_{\rm det} = \mathcal{L} E_R + \Delta E$, where $\mathcal{L}$ is the quenching factor associated with the $\NR$ signals.
For this analysis, we take $\mathcal{L} = 0$ to establish conservative bounds on the EFT interactions.
The minimum DM velocity for the Migdal effect is determined by
\begin{align} 
v_{\rm min} = \frac{m_{A_i} E_R + \mu_{A_i X} \Delta E}{\mu_{A_i X} \sqrt{2 m_{A_i} E_R }}.
\end{align}
The ionization form factor, $|Z_{\rm ion}|^2$, is defined by 
\begin{align} |Z_{\rm ion}|^2 = \frac{1}{2 \pi} \sum_{n, \ell} \frac{d}{dE_e} p_{q_e}^c(n\ell \rightarrow E_e), 
\end{align}
where $E_e$ denotes the kinetic energy of the ionized electron, calculated as $E_e = \Delta E - |E_{n \ell}|$, 
with $|E_{n\ell}|$ being the binding energy of the electron defined by quantum numbers $n$ and $\ell$. 
The ionization probabilities $p_{q_e}^c$ for the xenon and argon targets are taken from~Ref.\,\cite{Ibe:2017yqa}.
Similarly to the case of DM-nucleus elastic scattering, the total event rate due to the Migdal effect is obtained by summing the differential rates for all isotopes:
\begin{align} 
\frac{dR_{\tt Migdal}}{dE_{\rm det}} = \sum_i \frac{dR_{\tt Migdal}^i}{dE_{\rm det}}. 
\end{align}

We calculate the constraints from the Migdal effect using the $S_2$-only data from 
XENON1T~\cite{XENON:2019gfn}, PandaX-4T~\cite{PandaX:2022xqx}, and DarkSide-50~\cite{DarkSide:2022dhx}.
For the $S_2$-only data,
a complete calculation requires converting the $\ER$ energy into the corresponding number of photoelectrons (PE), as discussed in~Ref.\,\cite{Essig:2019xkx}. 
For the Migdal effect analyses in XENON1T and PandaX-4T,  we adopt a conservative approach by using the total number of events in the signal region rather than performing a spectral analysis. As our results show, this is a sufficiently accurate approximation, since the conversion has only a minor impact on the total event count in the signal region.
The number of signal events for the Migdal effect is then calculated by
\begin{align}
N_s = w \int \epsilon_{\ER}(E_{\rm det}) \frac{dR_{\tt Migdal}}{dE_{\rm det}} dE_{\rm det},
\label{eq:Ns-Migdal}
\end{align}
where $\epsilon_{\ER}$ accounts for the acceptance and efficiency factors for the $\ER$ data in the corresponding DM direct detection experiments.
For DarkSide-50, the data are binned by the number of ionized electrons, $N_e$. Therefore, we convert the differential Migdal event rate from $dR_{\tt Migdal}/dE_{\rm det}$ to $dR_{\tt Migdal}/dN_e$ by using the electron recoil ionization yield $Q_y^{\ER}$ provided in Fig.\,2 of~Ref.\,\cite{DarkSide:2021bnz}. The number of events in each bin is then given by
\begin{align}
N_s = w \int \epsilon_{\ER}(N_e) \frac{dR_{\tt Migdal}}{dN_e} \, dN_e.
\end{align} 
The detailed analysis for each experiment is outlined as follows:
\begin{itemize}
\item XENON1T: We calculate the constraint using the $S_2$-only data set and the efficiency factors given in~Refs.\,\cite{XENON:2019gfn,XENON:2019zpr}, 
with an exposure of $w=$ 22 ton-day. The 90\,\% C.L. constraint is obtained via the criterion of $N_s^{90} < 49$~\cite{Bell:2021zkr,Tomar:2022ofh}
for the region of $0.186~{\rm keV} < E_{\rm det} < 3.9~{\rm keV} $.
\item PandaX-4T: With the exposure of 0.55 ton-yr, 103 events (including 95.8 background events estimated) were observed in the signal region of $ N_{\rm PE}\in [60,200]$~\cite{PandaX:2022xqx}, corresponding to $0.07~{\rm keV} < E_{\rm det} < 0.232~{\rm keV} $.
According to the Poisson distribution, the 90\,\% C.L. constraint requires the number of signal events to be less than 21.5. 
We employ the total efficiency, shown as the red solid line in Fig.\,1 of~Ref.\,\cite{PandaX:2022xqx}.
\item DarkSide-50:
The analysis uses an exposure of 12.3 ton-day.
The overall acceptance, which is nearly flat with respect to recoil energy, ranges from 38\,\% at $N_e = 4$ to 40\,\% for $N_e \geq 15$.
As shown in Fig.\,3 of~Ref.\,\cite{DarkSide:2022dhx}, the Migdal effect at the DarkSide-50 provides leading constraints for a DM mass below 0.5 GeV.
For DM with a mass below 0.5 GeV, the signals generated by the Migdal effect are primarily distributed in the range with $N_e \leq 15$.
Given the $S_2$ threshold of $N_e = 4$ in~Ref.\,\cite{DarkSide:2022dhx}, we define the SR as $4 \leq N_e \leq 15$, which is uniformly divided into 11 bins for our analysis.
For each bin, we impose the requirement that the number of signal events not exceed the 90\,\% C.L. limit based on the Poisson distribution.    
\end{itemize}

With the above methods, 
we can obtain constraints on 
the SI DM-nucleon cross section that are comparable to those presented in the experiment papers, thus reinforcing the
robustness of our analysis. 
Before we embark on detailed analyses of experimental constraints in the following subsections, let us outline the main conclusions we will achieve. 
Generally, for $m_X > 1~\mathrm{GeV}$, the most stringent constraints come from the $\NR$ data on DM-nucleus elastic scattering,
while in the sub-GeV region, the constraints are dominated by $\ER$ data due to the Migdal effect.  
Specifically, LZ $\NR$ data provide the strongest limits for $m_X \gtrsim 10~\mathrm{GeV}$, 
while PandaX-4T $\NR$ data are more restrictive in the intermediate range of $1~\mathrm{GeV} \lesssim m_X \lesssim 10~\mathrm{GeV}$ 
due to its higher efficiency for the low-$\NR$ energy data.  
Notably, for SI operators and certain SD operators, DarkSide-50 $\NR$ data surpass PandaX-4T in sensitivity at $m_X \simeq 2~\mathrm{GeV}$, 
although the exposure of DarkSide-50 is much smaller than that of PandaX-4T.
This is because the lower atomic mass of the argon target
enhances recoil energy capture from low-mass DM particles, 
giving DarkSide-50 superior sensitivity compared to xenon-target experiments.
In the sub-GeV region, PandaX-4T $\ER$ data provide the most stringent constraints via the Migdal effect, 
with sensitivity extending down to $\sim 5~\mathrm{MeV}$.

\subsection{Constraints on NREFT interactions}

Based on the numerical approach described in previous subsections, we will present the constraints on the NREFT interactions in this subsection by considering two cases: the isospin-universal case where the WCs $c_i^p =c_i^n \equiv c_i$ are assumed for each operator, and the isospin-specific case where the proton and neutron interactions are treated independently. 
Additionally, we will examine one interaction a time to derive the constraints on its coupling constant using all of the datasets discussed above.

\subsubsection{Isospin-universal case}

\begin{figure}
\centering
\includegraphics[width=0.99 \textwidth]{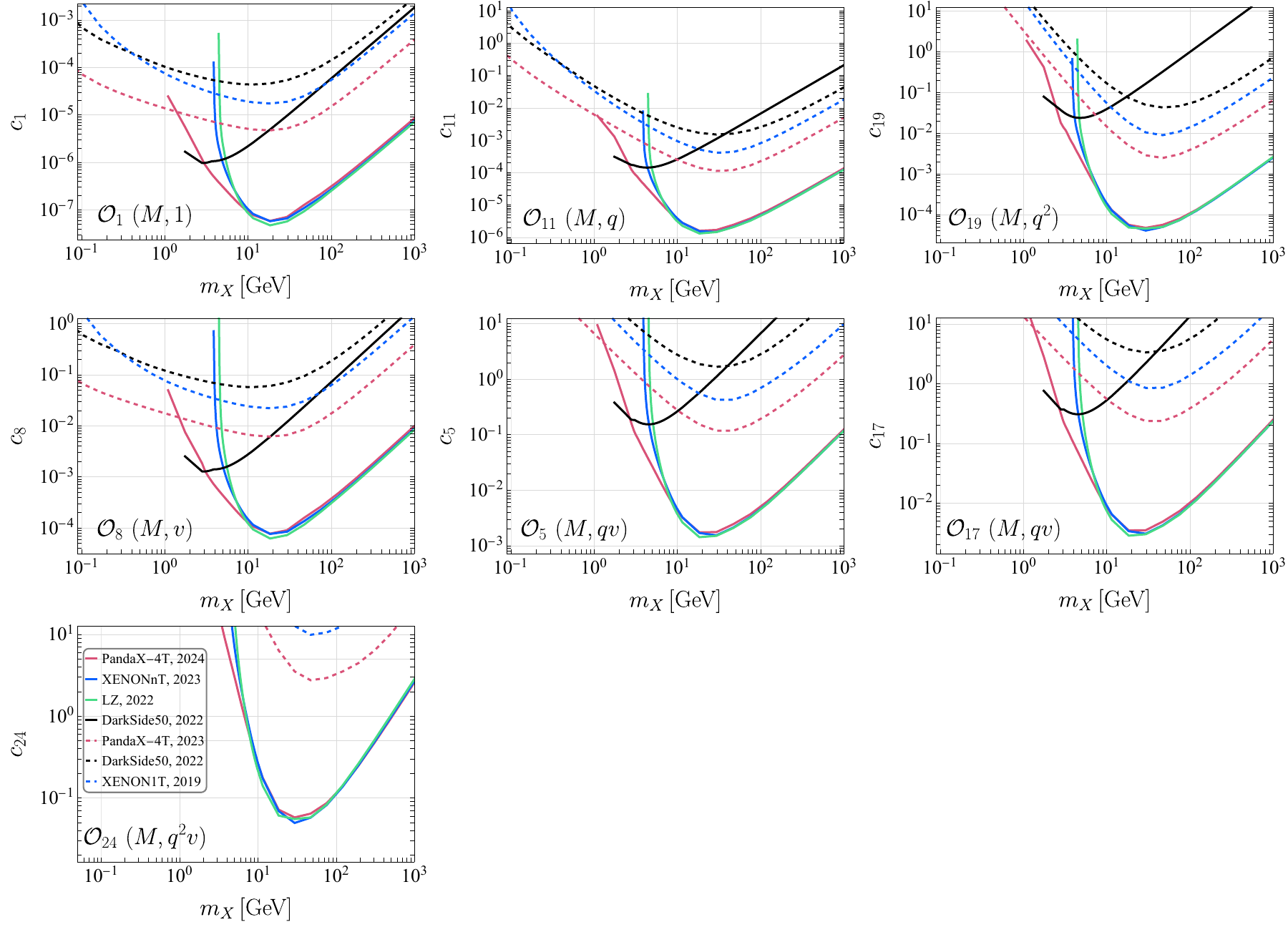}
\caption{Constraints on the dimensionless WCs of SI NREFT operators
from DM-nucleus elastic scattering (solid curves) and the Migdal effect (dashed curves).
The panels are organized according to the $(q,v)$ power counting, which is indicated in the parentheses following the operator label, 
along with the corresponding dominant nuclear response function.
} 
\label{fig:con-NR-SI}
\end{figure}

\begin{figure}
\centering
\includegraphics[width=0.99 \textwidth]{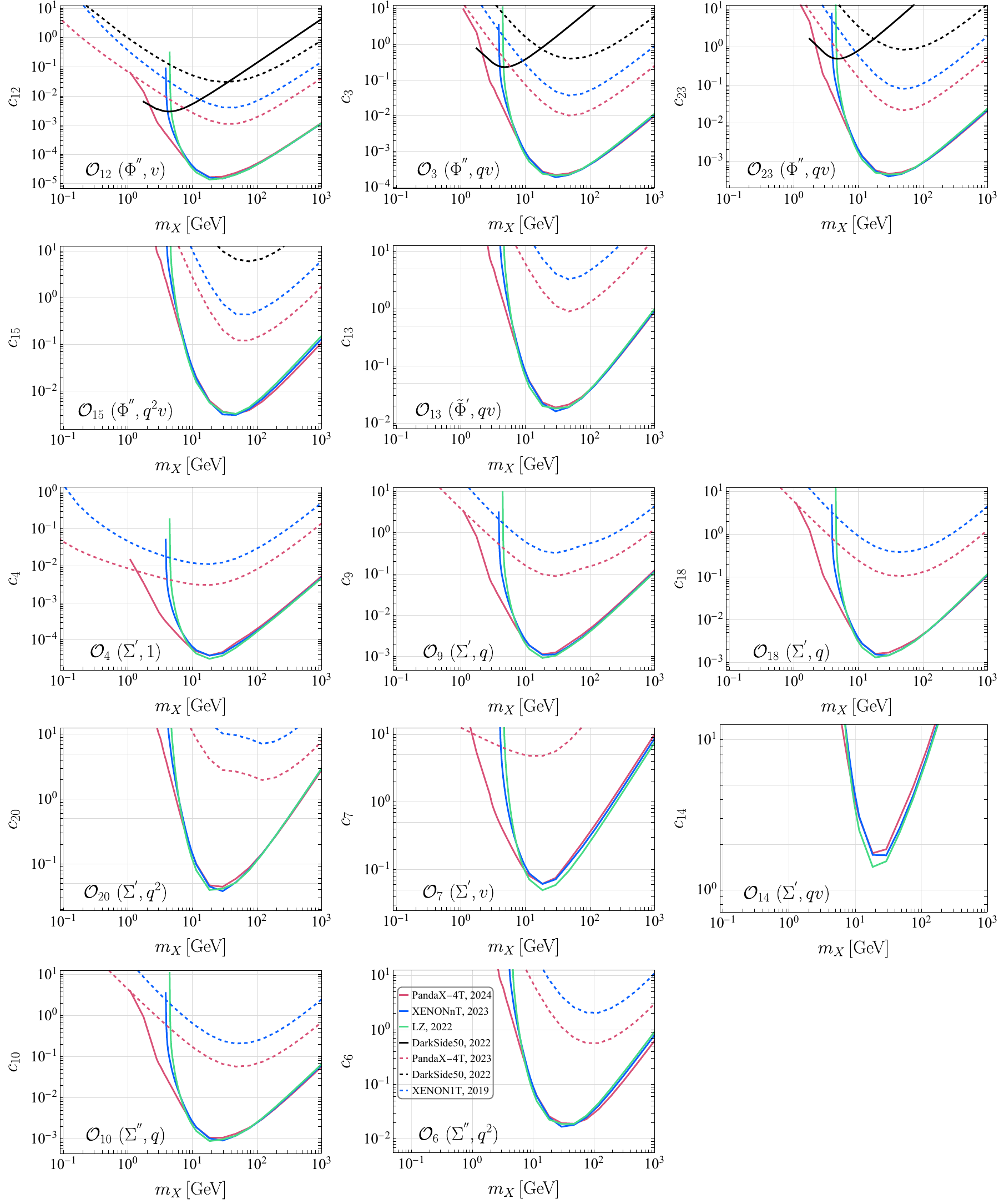}
\caption{
Same as \cref{fig:con-NR-SI}, but for SD NREFT operators.
Notice that we have further categorized the operators based on their  dominant nuclear response functions. 
Constraints on $\calO_{25}$ and $\calO_{26}$ are not shown, as they are too weak to be considered meaningful.
} 
\label{fig:con-NR-SD}
\end{figure}

Both DM-nucleus elastic scattering and the Midgal effect 
exhibit strong dependence on the nucleon spin structure of the effective operators,
as reflected in the magnitude of the nuclear response functions. 
For instance, for $^{131} \mathrm{Xe}$ at zero momentum transfer, 
$\{F_{M}^{nn}(0), F_{\Delta}^{nn}(0) \}= \{ 5928, 0.56 \}$ for the SI case, and 
$\{ F_{\Phi^{''}}^{nn},F_{\tilde{\Phi}^{'} }^{nn}, F_{ \Sigma^{'} }^{nn},F_{\Sigma^{''} }^{nn} \} = \{ 202, 0.84,  0.18, 0.088\}$ for the SD case, respectively. 
It is thus illuminating to discuss the SI and SD interactions separately.

{\it SI case}. In \cref{fig:con-NR-SI}, 
we present the constraints on the WCs ($c_i$) of the seven SI NREFT operators, assuming isospin symmetry with $c_i \equiv c_i^p = c_i^n$.    
We have reorganized the order of these operators based on their power counting in $q$ and $v$, as indicated in each plot.
This arrangement facilitates comparison among the constraints on different operators and provides a cross-check for our results. 
For all SI operators, the dominant nuclear response function is $F_M^{NN'}$, 
which is also indicated in each plot by the  label $M$ in parentheses. 
A comparison between the constraint on $c_8$ or $c_{11}$ and that on $c_1$ from the same experiment shows that the power of $v$ results in stronger suppression than the power of $q$. 
We observe that, approximately, the constraints on the WCs are weakened by about 3 orders of magnitude with each additional power of $v$ for the operators, 
in contrast to a reduction of about 1 to 2 orders of magnitude, with each additional power of $q$ for the operators.  
The strongest constraint for SI operators occurs at $m_X \simeq 20~\rm{GeV}$ for $c_1$, with the constraint from the LZ $\NR$ data reaching a value of $\simeq 4 \times 10^{-8}$.

{\it SD case}. The results for the SD operators are shown in \cref{fig:con-NR-SD}. 
Based on their dominant nuclear response functions in direct detection, 
we categorize them into three groups: 
\{$\calO_3$, $\calO_{12}$, $\calO_{15}$, $\calO_{23}$\}, \{$\calO_{13}$\}, and  
\{$\calO_4$, $\calO_6$, $\calO_7$, $\calO_9$, $\calO_{10}$, $\calO_{14}$, $\calO_{18}$, $\calO_{20}$\}, 
with the corresponding dominant nuclear response functions being $F_{\Phi^{''}}^{NN'}$, 
$F_{\tilde{\Phi}^{'}}^{NN'}$, and $F_{\Sigma^{'},\Sigma^{''}}^{NN'}$, respectively. Similarly to the SI case, 
these operators are also reorganized based on the $(q,v)$ power counting and the dominant nuclear response functions, as indicated in each plot. 
As the spin of the Xe isotopes increases, 
these functions first appear for spin-0 ($^{128,130,132,134,136}\rm Xe$), spin-1/2 ($^{129}\rm Xe$), and spin-3/2 ($^{131}\rm Xe$) nuclei, respectively.  
For operators with the same dominant nuclear response function, the constraints generally follow the same pattern of power counting in $q$ and $v$ as the SI operators. 
The strongest constraint for SD operators occurs at $m_X \simeq 20~\mathrm{GeV}$ for $c_{12}$, with the LZ $\NR$ data reaching a value of $\simeq 2 \times 10^{-5}$.  
Although $^{40}\mathrm{Ar}$ is spin-0, it exhibits $\Phi^{''}$ nuclear response functions.  
Therefore, DarkSide-50, using a spin-0 argon target, can also probe NR SD operators, including $\calO_3$, $\calO_{12}$, $\calO_{13}$, $\calO_{15}$, and $\calO_{23}$.

\subsubsection{Isospin-specific case}

\begin{figure}
\centering
\includegraphics[width=0.99 \textwidth]{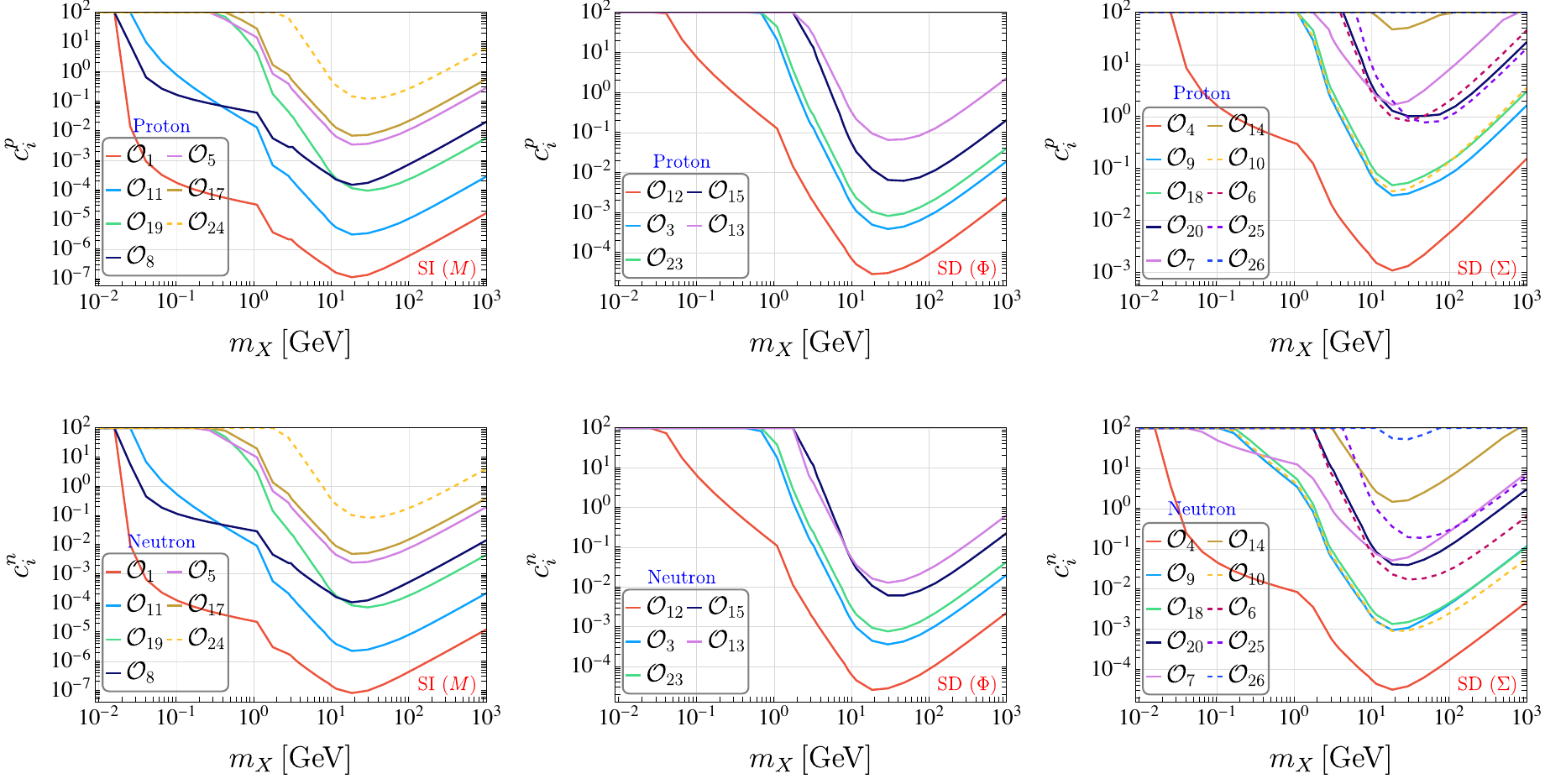}
\caption{
Constraints on the WCs, $c_i$, associated with each NREFT
operator for the isospin-specific case, featuring contributions from proton-only (upper) and neutron-only
(lower) interactions. Each curve (either solid or dashed) represents the strongest constraint obtained by combining results from all considered experiments. 
} 
\label{fig:con-NR-IV}
\end{figure}

There have been considerable discussions regarding the isospin-specific DM scenario~\cite{Kurylov:2003ra, Giuliani:2005my, Chang:2010yk, Feng:2011vu}.
To provide a comprehensive analysis, we derive constraints for scenarios with proton-only and neutron-only couplings, respectively. In contrast to the isospin-universal case, we only consider the combined strongest bound on the WC $c_i^{p(n)}$ of each NREFT operator $\calO_i^{p(n)}$ at each mass point from all experimental data. 
In \cref{fig:con-NR-IV}, the upper (lower) panels show the combined constraints on proton-only (neutron-only) couplings $c_i^p$ ($c_i^n$).
For clarity, we categorize the NR operators into three groups based on their dominant nuclear response functions, represented by their labels $M,\Phi, \Sigma$ in each panel. For all SI operators, the dominant response function is $M$, 
while for SD operators, they are further divided into two groups: 
one dominated by $\Phi\equiv \Phi'',~\tilde{\Phi}'$, 
and the other by $\Sigma\equiv\Sigma',~\Sigma''$.  

For a given NREFT interaction, the difference in constraints between the proton-only and neutron-only cases in \cref{fig:con-NR-IV} arises from the variations in the nuclear response functions, $F_{X}^{pp}$ and $F_{X}^{nn}$.
This behavior is consistent with the dependence of the scattering rate on the nuclear structure, which includes factors such as the neutron-to-proton ratio and the total spin of the target nucleus.
In general, each $c_i^n$ receives a stronger constraint than that on $c_i^p$ due to the larger number of neutrons in a xenon nucleus.
For the SI operators, the ratio of the constraints can be approximately estimated as $c_i^p / c_i^n = N_n / N_p \simeq 1.4$,
where $N_p$ ($N_n$) is the number of protons (neutrons) in the xenon nucleus.
For the $\Phi$-dominated SD operators, the isospin violation effect becomes smaller, with $c_i^p / c_i^n \simeq 1.15$, except for $\calO_{13}$, where $c_i^n / c_i^p \simeq 6.5$. This is because the dominant response function for  
$\calO_{13}$ is $F_{\tilde{\Phi}'}^{NN}$, which differs significantly between the two nucleons.      
For $\Sigma$-dominated SD operators, the isospin violation effect varies  significantly between different operators and DM mass values, 
with the ratio $c_i^p / c_i^n$ ranging from 1 to 70. This variation is primarily due to the total nuclear spin being predominantly contributed by the unpaired neutrons (the number of protons in argon and xenon nuclei is an even number).

\subsection{Constraints on relativistic EFT interactions}

To derive a constraint on each DSEFT operator, we perform calculations using the matching relations provided in \cref{tab:NRmatch} and evaluate the contribution of a single DSEFT operator at a time. 
For a DSEFT operator that matches to multiple NR operators,  interference effects among different NR contributions must be taken into account.
Except for the two dim-4 interactions $\calL_{\kappa_\Lambda}$ and $\calL_{\tilde\kappa_\Lambda}$, 
we impose a constraint on the effective scale $\Lambda \equiv |C_i|^{1/(n-4)}$ associated with each dim-$n$ ($n>4$) operator $\calO_i$, with $C_i$ denoting its corresponding WC.
For the DM-quark interactions, we further consider two scenarios: the flavor-universal case, in which a universal WC is assigned to  an operator involving any of the
$u$, $d$, $s$ quarks, such that $C_i^u = C_i^d = C_i^s \equiv C_i$ for the operator; and the flavor-specific case, where the contributions from the $u$, $d$, and $s$ quarks are calculated independently. 
In the following part, we will present constraints for the general complex DM case. When restricting to the real DM scenario, one simply ignores the operators marked with a ``$\times$'' in \cref{tab:NRmatch} and scales the bound on the effective scale associated with a surviving dim-$n$ operator by a factor of $2^{1/(n-4)}$.

\begin{figure}[t]
\centering
\includegraphics[width=0.99 \textwidth]{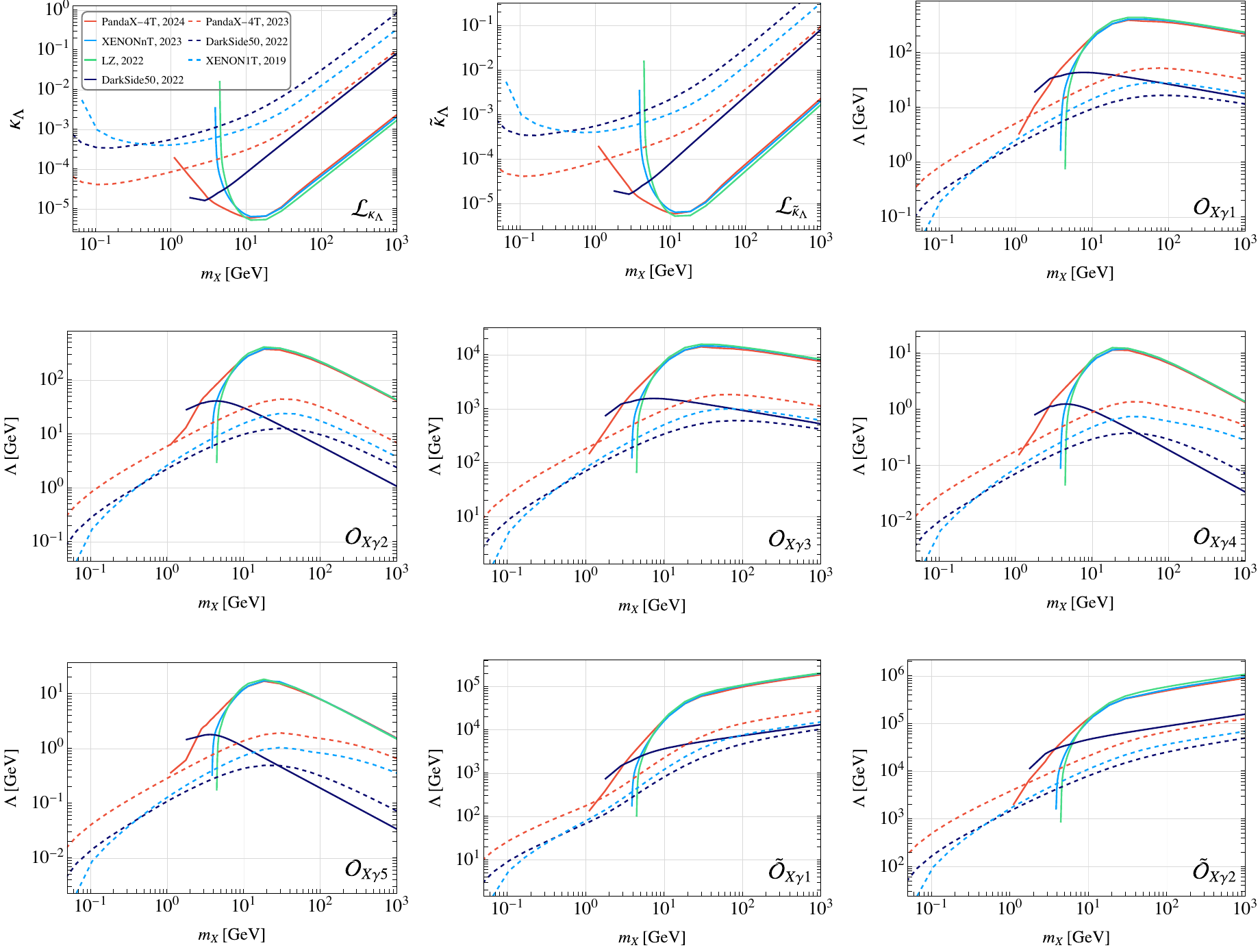}
\caption{ Constraints on the coupling or effective scale related to each relativistic vector DM-photon operator
from both the DM-nucleus elastic scattering (solid curves) and the Migdal effect (dashed curves). 
}
\label{fig:con-Rel-EM}
\end{figure}

\subsubsection{DM-photon interactions}   

In \cref{fig:con-Rel-EM}, 
we show the constraints on DM-photon interactions from various datasets. 
For the two dim-4 interaction terms, $\calL_{\kappa_\Lambda}$ and $\calL_{\tilde{\kappa}_\Lambda}$,
the behavior of the constraints on their dimensionless couplings follows a similar pattern to those on the WCs of NREFT operators shown in \cref{fig:con-NR-SI,fig:con-NR-SD}. 
For the remaining dim-6 DM-photon operators, the constraints are imposed on the effective energy scale, 
exhibiting different behavior for each operator.
These operators contribute through the exchange of a photon.
However, only $\calL_{\kappa_\Lambda}$ and $\calL_{\tilde{\kappa}_\Lambda}$ in case A, 
and the operators $\tilde{\calO}_{X\gamma 1}$ and $\tilde{\calO}_{X\gamma 2}$ in case B, exhibit an LD $1/\pmb{q}^2$ factor enhancement in their matching relations, as shown in \cref{tab:NRmatch}. 
This is because the remaining operators (all dim-6 ones in case A) involve a partial derivative, which compensates their LD enhancement factor, as discussed in \cref{sec:NRreduct}.
The constraints on the two terms ($\calL_{\kappa_\Lambda}$ and $\calL_{\tilde{\kappa}_\Lambda}$) are nearly identical, 
with $\kappa_{\Lambda}$ and $\tilde{\kappa}_{\Lambda}$ being constrained to $\lesssim 10^{-5}$ at $m_X \simeq 10~\rm GeV$.  
Due to the $1/\pmb{q}^2$-factor enhancement, they also receive strong constraints in the sub-GeV region from the Migdal effect, reaching approximately $\sim 10^{-4}$.  
For $\tilde{\calO}_{X\gamma 1}$ and $\tilde{\calO}_{X\gamma 2}$, in addition to the $1/\pmb{q}^2$-factor enhancement, 
their contribution to the amplitude has a global $m_X^2$ dependence. This makes them subject to the most stringent constraints, which become progressively stronger as $m_X$ increases.

\subsubsection{DM-quark interactions: Flavor-universal case}

\begin{figure}[t]
\centering
\includegraphics[width=0.995 \textwidth]{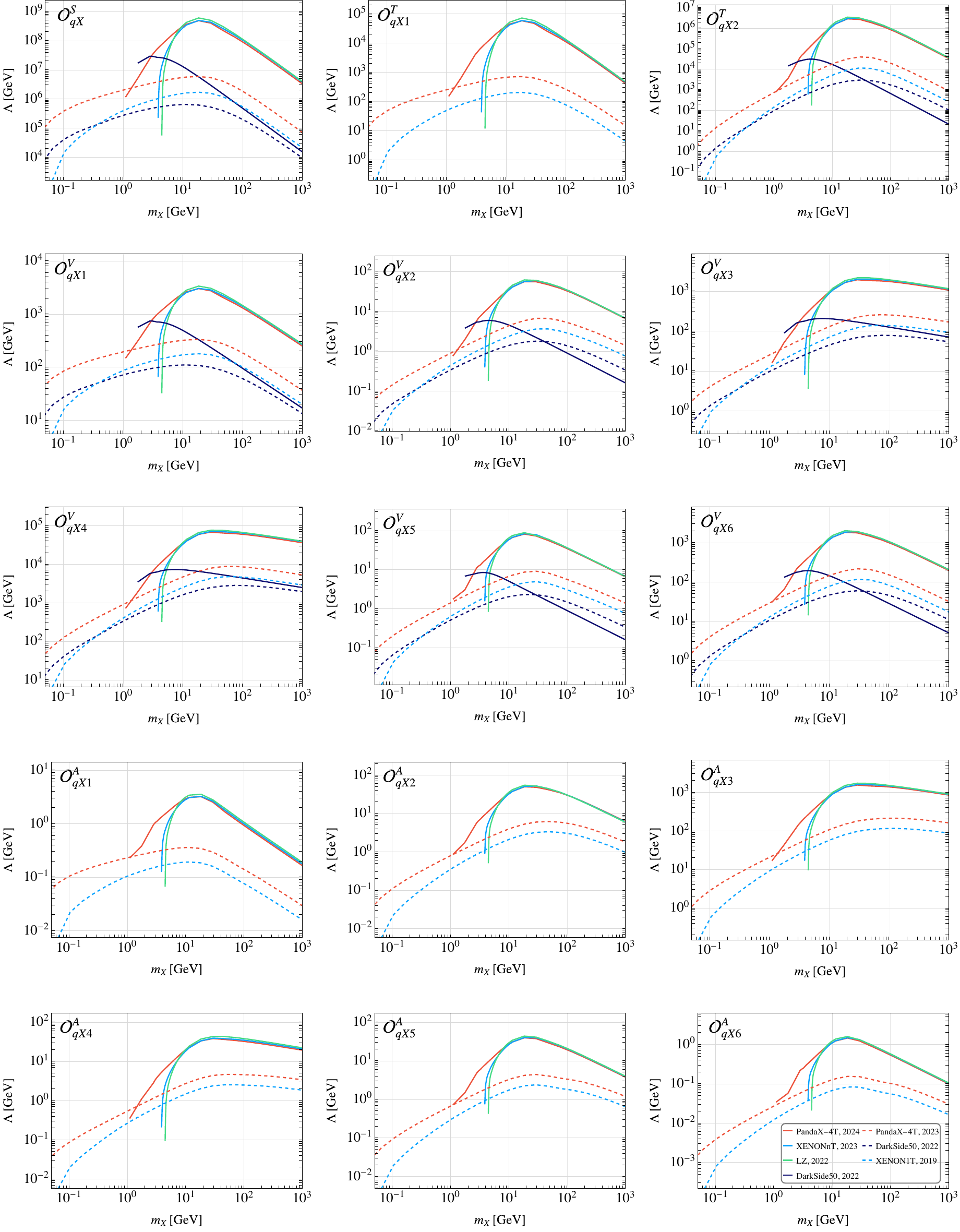}
\caption{ Constraints on the effective scale associated with each relativistic vector DM-quark operator in case A 
under the flavor-universal assumption with a universal coupling for $u$, $d$, and $s$ quarks.
The solid curves represent results from the DM-nucleus elastic scattering, while the dashed curves correspond to the Migdal effect.
}
\label{fig:con-Rel-A}
\end{figure}

\begin{figure}[t]
\centering
\includegraphics[width=0.99 \textwidth]{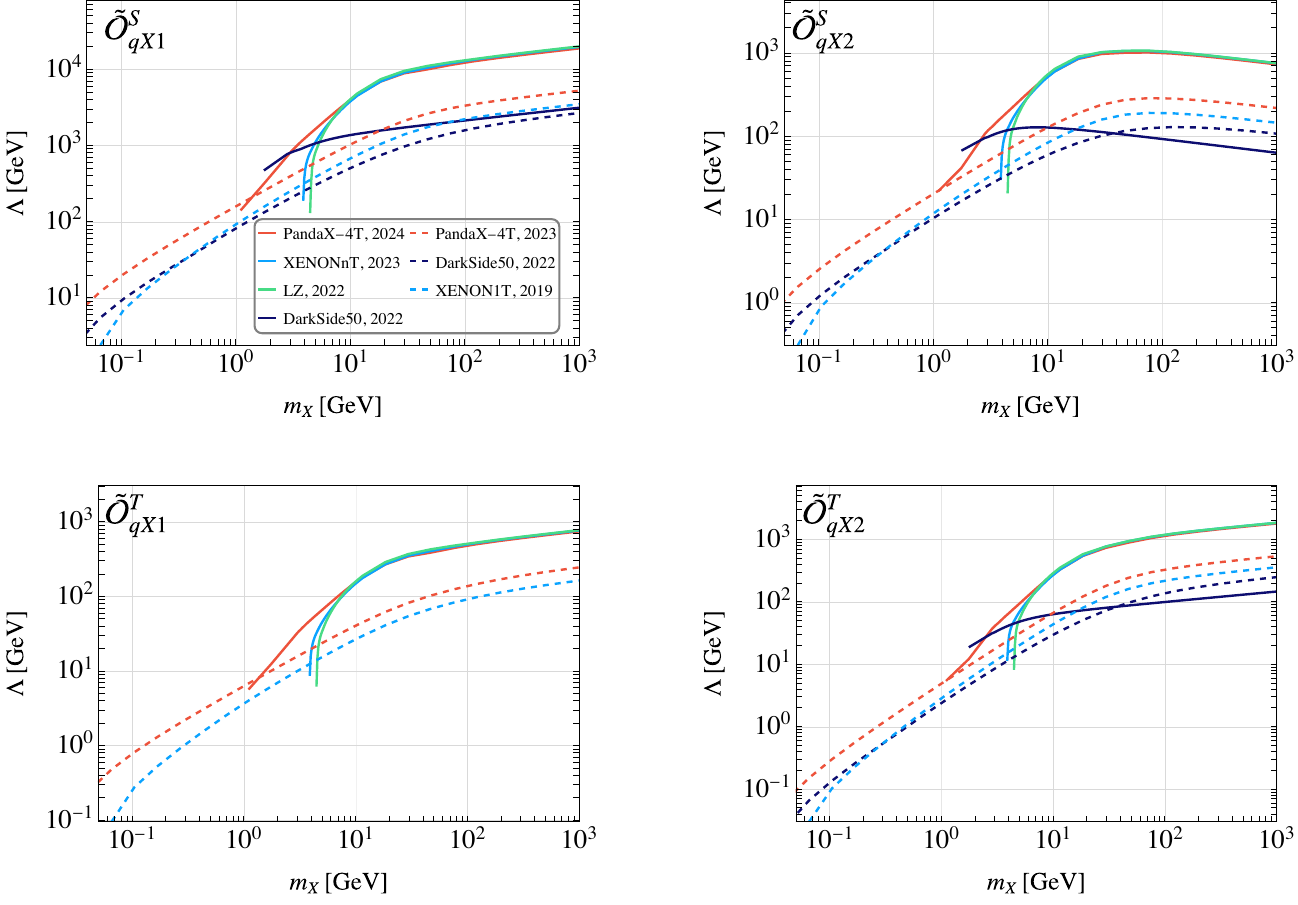}
\caption{Same as \cref{fig:con-Rel-A}, but for case B.
}
\label{fig:con-Rel-B}
\end{figure}

In \cref{fig:con-Rel-A,fig:con-Rel-B}, we show constraints on DM-quark operators in case A and case B, in which we assume flavor symmetry for the $u$, $d$, and $s$ quarks.
Under this assumption, for the operators $\calO_{qX}^P$, $\tilde{\calO}_{qX1}^P$, and $\tilde{\calO}_{qX2}^P$, the contributions from $u$, $d$, and $s$ quarks cancel out at the leading chiral order, 
and therefore, they are not shown in the two figures. 
This is because these operators involve a pseudoscalar quark bilinear $\bar q i\gamma_5 q$, and in the large-$N_c$ and chiral limits~\cite{Cheng:1988im}, we have $\langle N'|\bar u i \gamma_5 u +\bar d i \gamma_5 d +\bar s i \gamma_5 s |N\rangle=0$.
The constraints on DM-quark operators generally follow the power counting based on the dimensions of the operators, 
with higher-dimensional operators yielding weaker constraints.  
The most stringent constraint is $\Lambda \simeq 6 \times 10^8~\mathrm{GeV}$ for $\calO_{qX}^S$,  
a dim-5 operator that directly yields $\calO_1$ in the NR limit, which receives the strongest constraints among all NR operators. 
However, there are following points that can enhance the constraints, causing deviations from the simple dimension-based power counting:

\begin{itemize}
\item DM mass factors in the NR reduction:
The constraints on the effective scale in the large-$m_X$ region (i.e., $m_X \gtrsim 10~\rm GeV$) exhibit three distinct types of scaling with the increase of $m_X$: decreasing, nearly flat, and increasing. They reflect three different powers of $m_X$ in the dominant terms of the NR reduction: $m_X^0$ for $(\calO_{qX}^{S(P)},\calO_{qX1,2}^T,\calO_{qX1,2,5,6}^{V(A)})$, $m_X^1$ for $(\calO_{qX3,4}^{V(A)},\tilde\calO_{qX2}^{S,P})$, and $m_X^2 $ for $(\tilde\calO_{qX1}^{S(P)},\tilde\calO_{qX1,2}^{T})$, respectively.  
This can be understood as follows: For a given dim-$(n+4)$ DSEFT operator with an NR reduction relation of $c_i \propto m_X^m / \Lambda^n$, the event rate is proportional to $m_X^{2m-3} / \Lambda^{(2n)}$, as determined by the $m_X$ dependence in \cref{eq:NR,eq:dsigma}. This leads to the scaling relation $\Lambda \propto m_X^{(2m-3)/(2n)}$ in the large-$m_X$ region.
Consequently, when $m=0$, $m=1$, and $m=2$, the scaling behavior in the large-$m_X$ region corresponds to decreasing, nearly flat, and increasing, respectively. 
\item Matching to SI versus SD NR operators: Operators that primarily match to SI NR operators tend to have stronger constraints than those matching to SD NR operators. For instance, the constraints on $\calO^V_{qX}$ operators are generally stronger than those on $\calO^A_{qX}$, as the former mainly matches to SI operators, and the latter to SD operators in the NR limit.
\item Additional LD contributions arising from nonperturbative effects: For tensor operators ($\calO^T_{qX1,2}$ and $\tilde{\calO}^T_{qX1,2}$), 
in addition to the short-distance contributions, QCD nonperturbative effects can induce DM dipole moments, 
resulting in significant LD contributions~\cite{Liang:2024tef}. These LD effects can enhance the event rate, particularly for the Migdal effect, where the momentum transfer is relatively small.
\end{itemize}

\subsubsection{DM-quark interactions: Flavor-specific case}

\begin{figure}
\centering
\includegraphics[width=0.99 \textwidth]{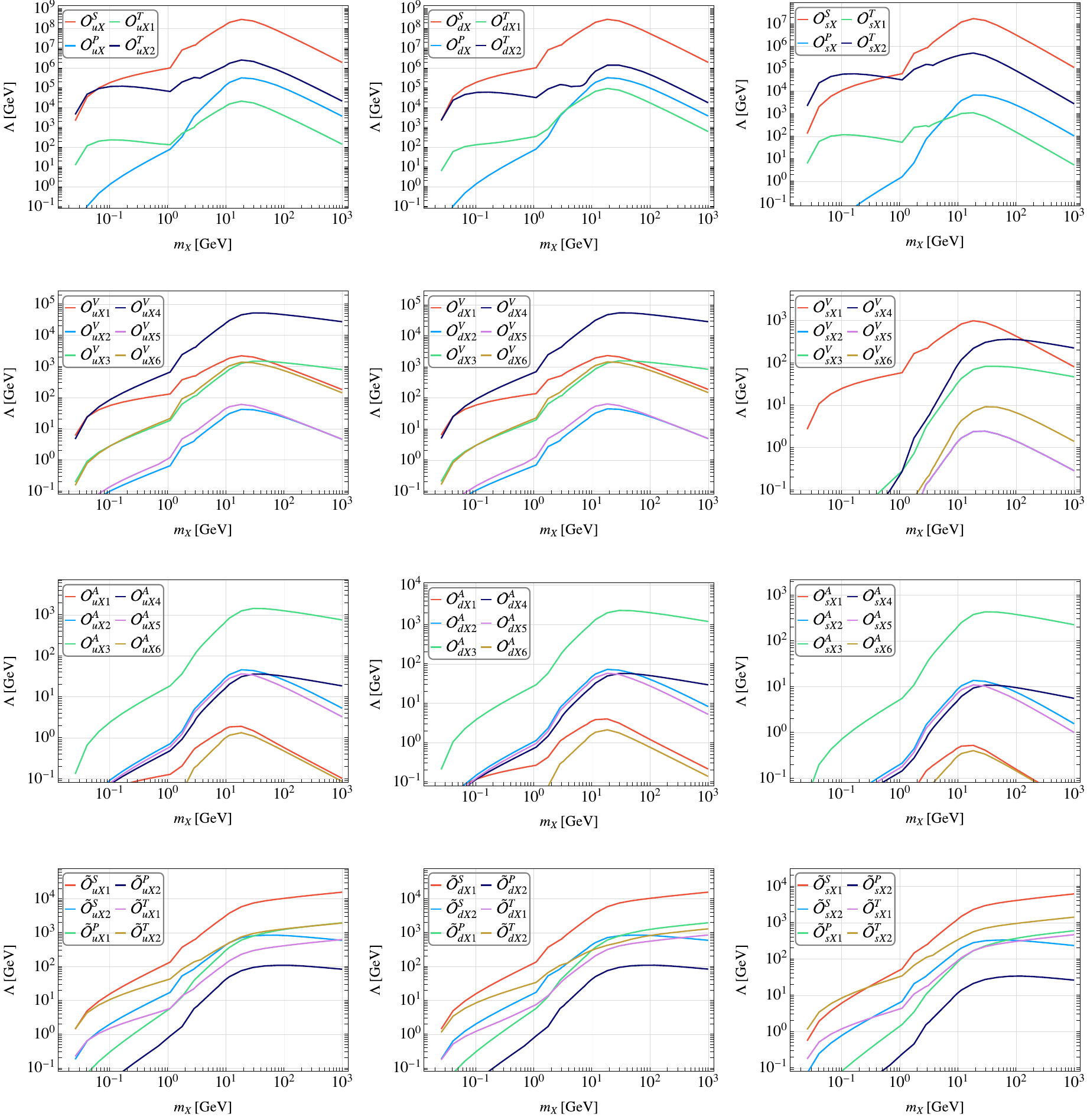}
\caption{
Constraints on the effective scale associated with each relativistic vector DM-quark operator are shown for the flavor-specific case, 
featuring contributions from  $u$-only (left), $d$-only (middle), and $s$-only (right) interactions.
Each solid curve represents the strongest constraint obtained by combining results from all considered experiments.
} 
\label{fig:con-Rel-uds}
\end{figure}

For many UV models, such as the one provided in the next section, the assumption of flavor symmetry for quarks does not hold.
Hence, in \cref{fig:con-Rel-uds}, we further present the constraints on the effective scale associated with each specific quark flavor,
as indicated by the subscript $u$, $d$, or $s$ attached to the operator, corresponding to the three columns, respectively. 
Similarly to the isopin-specific case presented in \cref{fig:con-NR-IV}, we only 
display the combined limit curve for each operator for brevity. 
These limits are derived by combining results from various DM direct detection experiments and reserving the most stringent limit in each mass region. 
The constraints for the $u$ and $d$ quarks are similar due to the isospin symmetry of the nucleon form factors.
In general, since $u$ and $d$ quarks are the valence quarks of the nucleon, the constraints on $u$ and $d$ cases are much stronger than those on the $s$ case, 
except for the tensor operators $\calO^T_{qX1}$, $\calO^T_{qX2}$, $\tilde{\calO}^T_{qX1}$, and $\tilde{\calO}^T_{qX2}$.
This exception arises because these tensor operators induce DM electromagnetic dipole moments, 
which are proportional solely to the charge of the quark.
As a result, the constraints on the $s$ case become comparable to those on the $u$ and $d$ cases in the sub-GeV region, 
where the LD interactions induced by the dipole moments dominate~\cite{Liang:2024tef}.

\section{ A UV model of vector DM }
\label{sec:UVmodel}

In this section we provide a UV-complete model that can lead to a single complex vector DM candidate, and at the same time, 
generate some of those relativistic effective operators at low energy after integrating out the heavy states. 
The model extends the SM gauge symmetry by an additional dark $\rm SU(2)_D$ symmetry with the gauge field $V^I\,(I=1,2,3)$. 
In addition, a real scalar triplet $\phi=(\phi_1, \phi_2, \phi_3)^\T$ and a vector-like fermion doublet $\Psi=(\psi, \chi)^\T$ under the $\rm SU(2)_D$ are introduced, 
where the new fermion $\Psi$ is also charged under the SM $\rm U(1)_Y$ group with hypercharge $Y=1$, though any other value of hypercharge is also possible.
The new physics Lagrangian takes the form
\begin{align}
{\cal L}_{\tt NP} & = 
 - {1\over 4}V^I_{\mu\nu} V^{I,\mu\nu} + \bar \Psi i \slashed{D}\Psi 
 - \left[ m_\Psi \bar \Psi_L \Psi_R 
 + y_\Psi \bar\Psi_L \sigma^I  \Psi_R \phi^I + \rm H.c.,\right]
\nonumber
\\
&
  + {1\over 2} (D_\mu \phi)^\T (D^\mu \phi) - 
  \left[ - {1\over 2} \mu_\phi^2 \phi^\T \phi 
  + {1\over 4} \lambda_\phi (\phi^\T \phi)^2
  + \lambda_{\phi h} (H^\dagger H) (\phi^\T \phi)
  \right],
\end{align}
where $y_\Psi$ is the new Yukawa coupling for the fermion field whose bare mass is $m_\Psi$,  
$\mu_\phi^2$ ($\lambda_\phi$) is the $\phi$ quadratic (quartic) coupling, 
and $\lambda_{\phi h}$ is its coupling to the SM Higgs. 
Due to the phase freedom of $\Psi_L$ and $\Psi_R$, $m_\Psi$ and $y_\Psi$ both can be taken to be real parameters. 
The covariant derivatives for the fermion and scalar fields are defined by 
\begin{align}
D_\mu \Psi  = \left(\partial_\mu - i g_D {\sigma^I \over 2} V_\mu^I + i g_1 B_\mu \right)\Psi, \quad 
D_\mu \phi = \partial_\mu  \phi - i g_D T^I V_\mu^I \phi,
\end{align}
where $g_D~[g_1]$ is the $\rm SU(2)_D~[\rm U(1)_Y]$ coupling constant with the gauge field $V^I_\mu~[B_\mu]$.
$\sigma^I$ are the Pauli matrices, while $T^I$ are adjoint representation matrices of
$\rm SU(2)_D$ with elements $(T^I)_{JK}= i \epsilon^{JIK}$.

We assume that this $\rm SU(2)_D$ is spontaneously broken to a dark $\rm U(1)_D$ by the scalar field $\phi$, which develops a nonvanishing expectation value (VEV) $v_\phi$.  
Denoting the physical scalar field by $\tilde S$---i.e., $\phi  = (0, \, 0,\, v_\phi+\tilde S)$ in the unitary gauge---the two gauge components $V_\mu^{1,2}$ become massive particles that are identified as the DM particles, 
while $V_\mu^3$ is still massless and serves as a massless dark photon. 
Due to the unbroken $\rm U(1)_D$ symmetry, one will find that the DM carries $\rm U(1)_D$ charge and thus naturally becomes a stable particle without any prior assumption of $\mathbb{Z}_2$ parity. 
To see it more clearly, we reorganize the DM and dark photon fields:  
\begin{align}
X_\mu \equiv { V_\mu^1 - i V_\mu^2 \over \sqrt{2} }, \quad 
X_\mu^* \equiv { V_\mu^1 + i V_\mu^2 \over \sqrt{2} }, \quad 
A_\mu' \equiv V_\mu^3.
\end{align}
Therefore, the scalar kinetic term in the unitary gauge becomes, 
\begin{align}
\label{eq:dmint} 
{1\over 2} (D_\mu \phi)^\T  (D^\mu \phi) = 
{1\over 2} \partial_\mu \tilde S \partial^\mu \tilde S 
+ g_D^2 (v_\phi + \tilde S)^2 X_\mu X^{*\mu}.
\end{align}
This implies that the DM mass $m_X \equiv g_D v_\phi$.
Furthermore, the dark gauge kinetic term becomes
\begin{align}
\label{eq:Vkin}
  - {1 \over 4} V^I_{\mu\nu} V^{I\mu\nu} = 
  - {1\over 4} A'_{\mu\nu} A'^{\mu\nu}
  - {1\over 2} X_{\mu\nu} X^{*\mu\nu}  
  + i g_D A'^{\mu\nu} X_\mu X^*_\nu 
  + {g_D^2 \over 2} (X_\mu X^*_\nu - X_\nu X^*_\mu) X^{\mu} X^{*\nu},
\end{align}
where
$A'_{\mu\nu} \equiv \partial_\mu A'_\nu - \partial_\nu A'_\mu$,
$X_{\mu\nu} \equiv D_\mu X_\nu - D_\nu X_\mu$, and the DM covariant derivative is 
$D_\mu X_\nu \equiv \partial_\mu X_\nu - i g_D A'_\mu X_\nu$.

Now, we go to the fermion part. After taking into account the above definitions of DM and dark photon fields, 
and the relation of $B_\mu$ in terms of the photon $A$ and $Z$ boson, the fermion part becomes
\begin{align}
\calL_{\Psi} & = 
\bar \psi \left( i \slashed{\partial} - m_\psi + { g_D\over2}  \slashed{A'} - e \slashed{A}- g_1 s_W \slashed{Z}  
\right)\psi 
+ \bar \chi \left( i \slashed{\partial} - m_\chi - {g_D\over2} \slashed{A'} - e \slashed{A}- g_1 s_W \slashed{Z}  \right)\chi
\nonumber
\\
& + { g_D\over \sqrt{2}} (\bar\psi \slashed{X}\chi+\bar\chi \slashed{X^*}\psi)
 - {y_\Psi\over 2} \tilde S \left( \bar\psi \psi - \bar\chi \chi \right),
\end{align}
where $e$ [and later $g_2$] is the SM $\rm U(1)_{em}$ [$\rm SU(2)_L$] gauge coupling, and $s_W=\sin\theta_W$ (and later $c_W=\cos\theta_W$), with $\theta$ being the weak mixing angle.
Due to the new Yukawa terms, the charged fermion doublet splits into two nondegenerate states $\psi$ and $\chi$, with masses  
$m_\psi \equiv m_\Psi + {1\over2}y_\Psi v_\phi$ and 
$m_\chi \equiv m_\Psi - {1\over2}y_\Psi v_\phi$.
This mass splitting will lead to the interesting kinetic mixing between the photon and the dark photon via the one-loop vacuum polarization diagrams. The details of this mixing and its phenomenological implications are postponed to our future work.  

Due to the Higgs portal interaction $\lambda_{\phi h}$, the scalar state $\tilde S$ will mix with the SM Higgs. To find the physical scalars, we need to analyze the full scalar potential including the SM part:
\begin{align}
 V(H,\phi) = - \mu_H^2 H^\dagger H + \lambda (H^\dagger H)^2 
 - {1\over 2} \mu_\phi^2 \phi^\T \phi + {1\over 8} \lambda_\phi (\phi^\T \phi)^2 + \lambda_{\phi h}(H^\dagger H) (\phi^\T \phi).
\end{align}
In the unitary gauge, we take $H \to {1\over \sqrt{2} }(v_h + \tilde h)$ and $\phi \to v_\phi + \tilde S $, with $v_h\approx 246~\rm GeV$ being the VEV of the SM Higgs doublet. The minimization conditions of the vacuum are 
\begin{subequations}
\begin{align}
 {\partial  V(v_h,v_\phi) \over \partial v_h}  = 0, \quad 
 {\partial  V(v_h,v_\phi) \over \partial v_\phi} = 0. 
\end{align}
\end{subequations}
Solving the above equations, we obtain 
\begin{subequations}
\begin{align}
 \mu_H^2 = \lambda v_h^2 + \lambda_{\phi h} v_\phi^2, \quad
 \mu_\phi^2 = {1\over 2}\lambda_\phi  v_\phi^2 + \lambda_{\phi h} v_h^2.
\end{align}
\end{subequations}
Thus, after the symmetry breaking, the scalar sector becomes, 
\begin{align}
V &= 
 \lambda v_h^2 \tilde h^2 
+ 2 \lambda_{\phi h} v_h v_\phi \tilde h \tilde S 
+ {1\over 2}\lambda_\phi v_\phi^2 \tilde S^2 
\nonumber
\\
&
+  \lambda v_h h^3 
+ \lambda_{\phi h} v_\phi \tilde h^2 \tilde S 
+ \lambda_{\phi h} v_h \tilde h \tilde S^2 
+ {1 \over 2} \lambda_\phi v_\phi \tilde S^3
\nonumber
\\
& + { 1\over 4}\lambda   \tilde h^4 
+ { 1 \over 2}\lambda_{\phi h} \tilde h^2 \tilde S^2 
+ { 1 \over 8}\lambda_\phi \tilde S^4. 
\end{align}
The mixing between $\tilde S$ and $\tilde h$ is eliminated by performing a rotation,  
\begin{align}
\begin{pmatrix}
\tilde h \\ \tilde  S  
\end{pmatrix} 
\equiv
\begin{pmatrix}
c_\theta & s_\theta \\
- s_\theta & c_\theta 
\end{pmatrix}
\begin{pmatrix}
h \\  S  
\end{pmatrix},
\end{align}
where $c_\theta \equiv \cos\theta$ and $s_\theta \equiv \sin\theta$. $h$ is identified as the observed scalar boson while $S$ is a new scalar particle. 
The rotation angle is fixed by the scalar VEVs and the scalar couplings, 
\begin{align}
  \tan(2\theta) = {4 \lambda_{\phi h} v_h v_\phi \over \lambda_\phi v_\phi^2 - 2 \lambda v_h^2}.
\end{align}
In terms of the rotation angle and scalar potential parameters, the expressions for the masses of 
$h$ and $S$ are
\begin{align}
m_{h,S}^2
={1\over 2} \left[ 
2 \lambda v_h^2 + \lambda_\phi v_\phi^2 
\mp {\lambda_\phi v_\phi^2 -2\lambda v_h^2 \over  1 - 2 s_\theta^2}
\right]. 
\end{align}
In the following, in addition to the SM parameters, we choose $\{g_D,\, s_\theta, m_S, m_X, m_\psi, m_\chi \}$ 
as 
input parameters for performing the matching calculation onto the DSEFT interactions given \cref{tab:NRmatch}. 

\begin{figure}[t]
\centering
\begin{tikzpicture}[mystyle,scale=0.7]
\begin{scope}[shift={(1,1)}]
\draw[v,blue] (-2.5, 0.866025) node[left]{$X^*$} -- (-0.5,0.866025);
\draw[v, blue] (-2.5,-0.866025) node[left]{$X^{~}$}--  (-0.5,-0.866025);
\draw[orange] (1,0) arc[start angle=0, end angle=120, radius =1cm] ;
\draw[purple] (-0.5,0.866025) arc[start angle=120, end angle=240, radius =1cm] ;
\draw[orange] (-0.5,-0.866025) arc[start angle=240, end angle=360, radius =1cm] ; 
\node[purple] at (-0.5,0) {$\chi$};
\node[orange] at (0.65,0) {$\psi$};
\draw[-stealth,purple](-1,0.1) -- (-1,-0.1);
\draw[-stealth,orange](0.80711,0.60711) -- (0.70711,0.80711);
\draw[-stealth,orange](0.60711,-0.80711) -- (0.80711,-0.60711);
\draw[v,black] (1,0) -- (3,0) node[right]{$\gamma$};
\end{scope}
\end{tikzpicture} \,\,
\begin{tikzpicture}[mystyle,scale=0.7]
\begin{scope}[shift={(1,1)}]
\draw[v,blue] (-2.5, 0.866025) node[left]{$X^*$} -- (-0.5,0.866025);
\draw[v, blue] (-2.5,-0.866025) node[left]{$X^{~}$}--  (-0.5,-0.866025);
\draw[purple] (1,0) arc[start angle=0, end angle=120, radius =1cm] ;
\draw[orange] (-0.5,0.866025) arc[start angle=120, end angle=240, radius =1cm] ;
\draw[purple] (-0.5,-0.866025) arc[start angle=240, end angle=360, radius =1cm] ; 
\node[orange] at (-0.5,0) {$\psi$};
\node[purple] at (0.65,0) {$\chi$};
\draw[-stealth,orange](-1,-0.1) -- (-1,0.1);
\draw[-stealth,purple](0.60711,0.80711) -- (0.80711,0.60711);
\draw[-stealth,purple](0.80711,-0.60711) -- (0.60711,-0.80711);
\draw[-stealth,orange](-1,-0.1) -- (-1,0.1);
\draw[v,black] (1,0) -- (3,0) node[right]{$\gamma$};
\end{scope}
\end{tikzpicture}
\caption{Diagrams generating the DM electromagnetic properties.}
\label{fig:diagDMEM}
\end{figure}
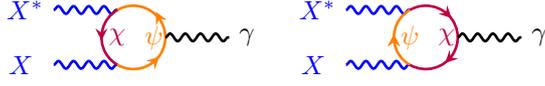

{\it Dark matter EM properties:~} 
Due to the heavy fermions that carry both $\rm U(1)_D$ and $\rm U(1)_{\rm em}$ charges, 
this model will generate electromagnetic properties of the vector DM through the diagrams shown in \cref{fig:diagDMEM}. 
We work on the assumption that $m_X \ll m_\chi\sim m_\psi$ and $m_X \ll m_{S,h}$. 
By direct calculation together with a low-energy expansion, we obtain the following dim-4 DM-photon interactions given in \cref{tab:NRmatch} 
\begin{align}
{\calL_{\kappa_\Lambda}} = i \kappa_\Lambda 
X^*_\mu X_\nu F^{\mu\nu}, \quad 
\kappa_\Lambda  = - {e g_D^2 \over 16\pi^2}\Big\{ F_1(r) + {m_X^2 \over m_\psi^2} [ F_2(r) + F_3(r)] \Big\},
\label{eq:kappaLam}
\end{align}
where $r\equiv m_\chi/m_\psi$ is the heavy fermion mass ratio. 
The loop functions are defined as 
\begin{subequations}
\begin{align}
F_1(r) & = {1 - 4 r + 4 r^3 - r^4 
- 2r(1-r + r^2) \ln r^2 \over (1- r^2)^2},
\\
F_2(r) & = { 5 \over 9} 
{1 + 9 r^2 - 9 r^4 - r^6 
+ 6 r^2 (1 + r^2) \ln r^2  \over (1-r^2)^4 },
\\
F_3(r) & = - 2 r 
{ 3 - 3 r^4  + (1 + 4 r^2 + r^4) \ln r^2 \over (1-r^2)^4 },
\\
F_4(r) & = - 
{ 2 - 2 r^2 + (1 + r^2) \ln r^2 \over 3(1-r^2)^2 },
\end{align}
\end{subequations}
where $F_4(r)$ will be relevant to the following discussion.
Furthermore, expanding to next-to-leading order with an additional pair of derivatives, 
we obtain three more DM EM operators:
$\calO_{X\gamma3}$, $\calO_{X\gamma5}$, 
and $\tilde\calO_{X\gamma1}$, given in \cref{tab:NRmatch}. 
Their WCs are 
\begin{align}
C_{X\gamma3} = - {e g_D^2 \over 16 \pi^2 m_\psi^2} G_2(r), \quad 
C_{X\gamma5} = - {e g_D^2 \over 16 \pi^2 m_\psi^2} G_3(r), \quad
\tilde C_{X\gamma1} =-{e g_D^2 \over 16 \pi^2 m_\psi^2}G_4(r),
\end{align}
with the loop functions
\begin{align}
G_2(r) = {F_2(r)\over 10}  + {F_3(r)\over 6}  + F_4(r), \quad 
G_3(r) ={F_1(r)\over 3 r}  - {F_2(r)\over 10}  - {F_3(r)\over 6} , \quad 
G_4(r) = - {F_2(r)\over 10}.
\end{align}

\begin{figure}[t]
\centering
\begin{tikzpicture}[mystyle,scale=0.7]
\begin{scope}[shift={(1,1)}]
\draw[v,blue] (-2.5, 0.866025) node[left]{$X^*$} -- (-0.5,0.866025);
\draw[v, blue] (-2.5,-0.866025) node[left]{$X^{~}$}--  (-0.5,-0.866025);
\draw[orange] (1,0) arc[start angle=0, end angle=120, radius =1cm] ;
\draw[purple] (-0.5,0.866025) arc[start angle=120, end angle=240, radius =1cm] ;
\draw[orange] (-0.5,-0.866025) arc[start angle=240, end angle=360, radius =1cm] ;
\node[purple] at (-0.5,0) {$\chi$};
\node[orange] at (0.65,0) {$\psi$};
\draw[-stealth,purple](-1,0.1) -- (-1,-0.1);
\draw[-stealth,orange](0.80711,0.60711) -- (0.70711,0.80711);
\draw[-stealth,orange](0.60711,-0.80711) -- (0.80711,-0.60711);
\draw[v,black] (1,0) -- (2.5,0) node[midway,yshift = 8 pt]{$Z$};
\draw[f] (2.5, 0) -- (4.5,1.5) node[right]{$q$};
\draw[fb] (2.5, 0) -- (4.5,-1.5) node[right]{$\bar q$};
\end{scope}
\end{tikzpicture}
\begin{tikzpicture}[mystyle,scale=0.7]
\begin{scope}[shift={(1,1)}]
\draw[v,blue] (-2.5, 0.866025) node[left]{$X^*$} -- (-0.5,0.866025);
\draw[v, blue] (-2.5,-0.866025) node[left]{$X^{~}$}--  (-0.5,-0.866025);
\draw[purple] (1,0) arc[start angle=0, end angle=120, radius =1cm] ;
\draw[orange] (-0.5,0.866025) arc[start angle=120, end angle=240, radius =1cm] ;
\draw[purple] (-0.5,-0.866025) arc[start angle=240, end angle=360, radius =1cm] ;
\node[orange] at (-0.5,0) {$\psi$};
\node[purple] at (0.65,0) {$\chi$};
\draw[-stealth,orange](-1,-0.1) -- (-1,0.1);
\draw[-stealth,purple](0.60711,0.80711) -- (0.80711,0.60711);
\draw[-stealth,purple](0.80711,-0.60711) -- (0.60711,-0.80711);
\draw[v,black] (1,0) -- (2.5,0) node[midway,yshift = 8 pt]{$Z$};
\draw[f] (2.5, 0) -- (4.5,1.5) node[right]{$q$};
\draw[fb] (2.5, 0) -- (4.5,-1.5) node[right]{$\bar q$};
\end{scope}
\end{tikzpicture}
\begin{tikzpicture}[mystyle,scale=0.7]
\begin{scope}[shift={(1,1)}]
\draw[v,blue] (-1.5,1.5) node[left]{$X^*$}  -- (0,0);
\draw[v, blue] (-1.5,-1.5) node[left]{$X^{~}$}-- (0,0);
\draw[snar, black] (0,0) -- (2,0) node[midway,yshift = 8 pt]{$h,S$};
\draw[f] (2, 0) -- (3.5,1.5) node[right]{$q$};
\draw[fb] (2, 0) -- (3.5,-1.5) node[right]{$\bar q$};
\end{scope}
\end{tikzpicture}
\caption{Diagrams generating the DM-quark interactions mediated by $Z$ and scalar particles.}
\label{fig:diagramXXqq}
\end{figure}
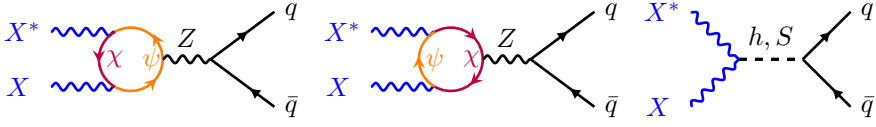

{\it Dark matter-quark interactions:~}
Now, we turn to the interactions of the vector DM with quarks. 
This is mediated by one-loop diagrams attached with a $Z$ boson or
tree-level diagrams via the scalar mediators $h,~S$, as shown in \cref{fig:diagramXXqq}. 
After integrating out the heavy particles, we match out three operators $\calO_{qX}^S, \calO_{qX5}^V, \calO_{qX5}^A$ at leading order, with the corresponding WCs 
\begin{align}
C_{qX}^{S,q} = 2 s_\theta c_\theta 
g_D m_X {m_q \over v_h} {m_S^2 - m_h^2 \over m_h^2 m_S^2}, \quad 
C_{qX5}^{V,q} = - {g_2 g_V^q\over c_W m_Z^2} \kappa_\Lambda, \quad 
C_{qX5}^{A,q} = - {g_2 g_A^q\over c_W m_Z^2} \kappa_\Lambda,
\end{align}
where $\kappa_\Lambda$ is given in \cref{eq:kappaLam}
and the SM couplings for
the three light quarks ($u,d,s$) are 
\begin{align}
g_V^u =  {1\over 4} - {2\over 3}s_W^2, \quad 
g_V^{d,s} = - {1\over 4} + {1\over 3} s_W^2,\quad  
g_A^u = - {1\over 4}, \quad 
g_A^{d,s} = {1\over 4}. 
\end{align}

As illustrated above, this model generates seven operators in \cref{tab:NRmatch}, 
thereby underscoring the significance of the EFT study in this work. 
Moreover, alternative UV models can be readily constructed by replacing the vector-like fermion doublet with other fields that possess nontrivial representations of both the dark $\rm SU(2)_D$ and the standard model groups. 
For instance, a scalar doublet with a nonzero hypercharge can yield a similar set of operators, 
albeit with a richer scalar phenomenology. Additionally, models featuring heavy mediators that carry SM $\rm SU(2)_L$ quantum numbers can produce operators with different structures. 
Conversely, one could replace the scalar triplet $\phi$ with a scalar doublet, allowing all three vector components to acquire mass following symmetry breaking. In this scenario, the vector DM  becomes a real particle, and several studies have explored this possibility~\cite{Hambye:2008bq,Baouche:2021wwa}. 
While a comprehensive phenomenological analysis of the current model is beyond the scope of this work, we intend to address it in future research.

\section{Summary}
\label{sec:summary}

In this paper, we have systematically investigated the interactions between spin-1 dark matter (DM) and nuclei within the framework of effective field theories (EFTs). 
By considering both nonrelativistic (NR) and relativistic EFT descriptions, 
we have provided a comprehensive analysis of the various operators that govern these interactions with nucleons and quarks/photons. 
Within the NR EFT framework, we compiled a complete list of NR operators for spin-1 DM coupling to nucleons and calculated their contributions to the DM response functions, 
thereby enhancing our understanding on how these interactions manifest in detection experiments. 
Furthermore, to gain deeper insights into potential fundamental origins of the NR interactions, we investigated all possible leading-order relativistic EFT operators that involve DM and light quarks or photon. 
By conducting the NR reduction of these relativistic interactions, we were able to establish clear relationships between the relativistic interactions and their NR counterparts. 
This analysis not only enhances our understanding of how these interactions are connected, but also provides a framework for
translating the experimental bounds on NR interactions into those on relativistic interactions, thereby enriching our overall comprehension of DM interactions.

Our derivation of the nuclear scattering rate from these interactions allowed us to employ recent direct detection data, 
including both DM-nucleus elastic scattering and the Migdal effect, to impose stringent constraints on the EFT operators and the electromagnetic properties of vector DM. 
Notably, we found that nuclear recoil data from experiments such as PandaX-4T, XENONnT, LZ, and DarkSide-50
set stringent bounds on the EFT Wilson coefficients for a DM mass above a few GeV. 
In contrast, the Migdal effect datasets from PandaX-4T, XENON1T, and DarkSide-50 enable probing of the DM mass region as low as 20 MeV, 
highlighting the sensitivity of these methods to lighter dark matter candidates.

Finally, we proposed a simple UV-complete model that effectively realizes the complex spin-1 dark matter scenario. 
This model extends the standard model by incorporating a dark $\rm SU(2)_D$ gauge group 
and introduces a scalar triplet $\phi$ along with a charged vector-like fermion doublet $\Psi$ that transforms under this new symmetry. 
We analyzed the fundamental structure of the model and performed a matching onto the relativistic EFT interactions, 
revealing the emergence of several relativistic DM-quark operators and associated electromagnetic properties. 
Additionally, variations of this model, achieved by
modifying its components,
can lead to a richer phenomenology and a more diverse operator structure. 
Such models not only provide a solid theoretical foundation for our findings but also pave the way for future research, 
potentially guiding experimental efforts aimed at detecting spin-1 DM and enhancing our understanding of its role in the Universe. 
Overall, our work contributes to the expanding body of knowledge regarding dark matter interactions 
and underscores the significance of effective field theories in elucidating the properties of this enigmatic component of the cosmos.

\acknowledgments
We thank Meng-Chao Gao and Xing-Bo Yuan for helping us correct some of the errors in Eq.\,(3.10).
This work was supported in part by Grants 
No.\,NSFC-12035008, 
No.\,NSFC-12247151, 
No.\,NSFC-12305110,
and No.\,NSFC-12347121,
and by Guangdong Major Project of Basic
and Applied Basic Research No.\,2020B0301030008.

\appendix
\section{NR reduction in nuclear case}
\label{app:NRreduction}
 
In this appendix, we provide the complete parametrization of quark bilinears in \cref{eq:ff1}, 
and we collect the numerical inputs for the involved form factors.
Due to the Lorentz and discrete $C,P,T$ symmetries of the QCD dynamics, 
the hadronic matrix elements of quark currents between nucleon states have the following general parametrizations~\cite{DelNobile:2021wmp,Adler:1975he,Ji:1998pc} 
{\small
\begin{subequations}
\label{eq:hadff}
\begin{align}
\label{eq:hadff:scalar}
 \langle N(k')| \bar q q | N(k)\rangle & = 
 F_S^{q/N}\, \bar u_{N'}u_N,
 \\
 \langle N(k')| \bar q i\gamma_5 q | N(k)\rangle & =
 F_P^{q/N}\, \bar u_{N'}i \gamma_5 u_N,
 \\
 \langle N(k')| \bar q \gamma_\mu q | N(k)\rangle & =
\overline{u}_{N'} \left[
F_{1}^{q/N}\, \gamma_\mu  + F_{2}^{q/N}\,  \frac{i\sigma_{\mu\nu} q^\nu}{2m_{N}}  
\right] u_{N} , 
\\
\langle N(k') |  \bar q \gamma_{\mu}\gamma_5 q | N(k) \rangle
&=
\overline{u}_{N'} \bigg[ 
G_A^{q/N}\, \gamma_\mu\gamma_5 + 
G_P^{q/N}\, \frac{ q_\mu}{2m_{N}} \gamma_5
\bigg] u_{N}, 
\\
\langle N(k') | \bar q \sigma_{\mu\nu} q | N(k) \rangle
&  \Rightarrow
\overline{u}_{N'} \left[ 
 F_{T,0}^{q/N}\, \sigma_{\mu\nu} 
+ F_{T,1}^{q/N}\, \frac{i \gamma_{[\mu}q_{\nu]} }{m_{N}}
+ F_{T,2}^{q/N}\, \frac{i K_{[\mu} q_{\nu]} }{m_{N}^2} \right] u_{N},
\\
\label{eq:hadff:tensorL}
\langle N(k') | \bar q \sigma_{\mu\nu} q | N(k) \rangle
& \overset{\rm LD}{\Rightarrow} 
d_q {i q_\mu \over q^2}
\left[ 
 Q_N ( 2 m_N g_\nu^0 \mathbbm{1}_N - g_\nu^i K^i )
 + g_N g_\nu^i (i \pmb{q}\times\pmb{S}_N)^i
\right] -\mu \leftrightarrow \nu,
\\
\langle N(k')| ( \overline{q}\gamma_{(\mu} i \overleftrightarrow{D}_{\nu)} q) | N(k) \rangle
& = \overline{u}_{N'} \left[  
A_{20}^{q/N}\, \gamma_{(\mu} K_{\nu)} + B_{20}^{q/N}\, {i \sigma_{(\mu\rho} K_{\nu)} q^{\rho} \over 2 m_N}
+ C_{20}^{q/N}\, { 2 q_{(\mu}q_{\nu)} \over m_N}\right] u_{N},
\label{eq:qDgammaq}
\\
\langle N(k')| ( \overline{q}\gamma_{(\mu}\gamma_5 i \overleftrightarrow{D}_{\nu)} q) | N(k) \rangle
& = \overline{u}_{N'} \left[  
\tilde A_{20}^{q/N}\, \gamma_{(\mu}\gamma_5 K_{\nu)} 
+ \tilde B_{20}^{q/N}\, {K_{(\mu}q_{\nu)} \over 2 m_N}\gamma_5  \right] u_{N},
\label{eq:qDgamma5q}
\end{align}
\end{subequations} }%
where the momentum transfer is $q^\mu\equiv k'^\mu-k^\mu$, and $K^\mu\equiv k'^\mu+k^\mu$. 
The symbols $u_{N}\equiv u_N(k)$ and $\bar{u}_{N'} \equiv \bar{u}_N(k')$ are the Dirac spinors of initial- and final-state nucleons.
The $q^2$ dependence of the form factors is implied---i.e., 
$I^{q/N}_i\equiv I^{q/N}_i(q^2)$. 
The symmetrization notation without a trace for the last two currents is defined by 
$ A_{(\mu } B_{\nu)} = {1\over 2}(A_{\mu}B_{\nu} + A_{\nu}B_{\mu}) -  {1\over 4}g_{\mu\nu} A_\rho B^\rho$. 
For the LD part of the tensor current, 
$d_q \equiv Q_q e^2 c_T \Lambda_\chi /(8\pi^2)$~\cite{Liang:2024tef}.
From the matrix element results given above, we can extract the following LO NR reduction correspondence:
\begin{subequations}
\begin{align}
\bar q q & \Rightarrow 2 m_N F_S^{q/N} \mathbbm{1}_N,
\\
\bar q i\gamma_5 q  &  \Rightarrow  - 2 F_P^{q/N} i \pmb{q}\cdot \pmb{S}_N, 
\\
\bar q \gamma_\mu q & \Rightarrow  F_{1}^{q/N} \left( 2 m_N g_\mu^0 \mathbbm{1}_N 
-  g_\mu^i K^i \right)
+ 2 \left(F_{1}^{q/N} +F_{2}^{q/N} \right) g_\mu^i (i \pmb{q}\times\pmb{S}_N)^i, 
\\
\bar q \gamma_{\mu}\gamma_5 q & \Rightarrow  
- 4 m_N G_A^{q/N} g_\mu^i S_N^i 
+ 2 G_A^{q/N} g_\mu^0 \pmb{K}\cdot \pmb{S}_N
- 2 G_P^{q/N} {q_\mu \over m_N} \pmb{q}\cdot \pmb{S}_N, 
\\
 \bar q \sigma_{\mu\nu} q &  \Rightarrow   
4 m_N F_{T,0}^{q/N} g_\mu^i g_\nu^j \epsilon^{ijk} S_N^k 
+ \left( F_{T,0}^{q/N} -2 F_{T,1}^{q/N} - 4 F_{T,2}^{q/N} \right) (g_\mu^0 g_\nu^i - g_\nu^0 g_\mu^i ) i q^i \mathbbm{1}_N
\nonumber
\\
& + 2 F_{T,0}^{q/N} (g_\mu^0 g_\nu^i - g_\nu^0 g_\mu^i ) (\pmb{K}\times \pmb{S}_N)^i, 
\\
\overline{q}\gamma_{(\mu} i \overleftrightarrow{D}_{\nu)} q
&\Rightarrow 
4 m_N^2 A_{20}^{q/N}\left(g_\mu^0 g_\nu^0 - {1\over 4 } g_{\mu\nu}\right)  \mathbbm{1}_N, 
\\
\overline{q}\gamma_{(\mu}\gamma_5 i \overleftrightarrow{D}_{\nu)} q
&\Rightarrow 
2 m_N \tilde A_{20}^{q/N} \left[ - 2 m_N (g_\mu^0 g_\nu^i + g_\mu^i g_\nu^0 )
 +(2 g_\mu^0 g_\nu^0 \delta^{ij} + g_\mu^i g_\nu^j+g_\mu^j g_\nu^i) K^j \right] S_N^i.
\end{align}
\end{subequations}%

\begin{table}[t]
	\center
    \resizebox{\linewidth}{!}{
		\renewcommand\arraystretch{1.2}
		\begin{tabular}{| c | c  c  c | c  c  c | c  c  c |}
            \hline
            \multicolumn{10}{|c|}{\cellcolor{cyan!15}$\calF_1^{q/p}(q^2)$~\cite{Bishara:2017pfq} }
            \\
            \hline
            & $a_{F,\pi}^{u/p}(0)$ & $a_{F,\pi}^{d/p}(0)$ & $a_{F,\pi}^{s/p}(0)$ & $a_{F,\eta}^{u/p}(0)$ & $a_{F,\eta}^{d/p}(0)$ & $a_{F,\eta}^{s/p}(0)$ & $b_{F}^{u/p}(0)$ & $b_{F}^{d/p}(0)$ & $b_{F}^{s/p}(0)$ 
            \\
            \hline
            $G_P^{q/N} (q^2)$  & $2.5446(46)$ & $-2.5446(46)$ & $0$ & $0.388667$ & $0.388667$ & $-0.777334$  & $-4.65(25)$ & $3.28(25)$ & $0.32(18)$
            \\
            $F_P^{q/N} (q^2)$  & $3.616(78)$ & $-3.616(78)$ & $0$ & $0.552(39)$ & $0.552(39)$ & $-1.105(78)$ & $-$ & $-$ & $-$
            \\
            \hline
            \multicolumn{7}{|c|}{\cellcolor{blue!15}$\calF_2^{q/p}(q^2)$~\cite{Bishara:2017pfq} }
            & \multicolumn{3}{c|}{\cellcolor{purple!15}$A_{20}^{q/p}(0)$ ~\cite{Alexandrou:2020sml}}
			\\
            \hline
			 & $\calF_2^{u/p}(0)$ & $\calF_2^{d/p}(0)$ & $\calF_2^{s/p}(0)$ & $\calF_2'^{\,u/p}(0)$ & $\calF_2'^{\,d/p}(0)$ & $\calF_2'^{\,s/p}(0)$ &
             $A_{20}^{u/p}(0)$ & $A_{20}^{d/p}(0)$ & $A_{20}^{s/p}(0)$
			\\
            \cline{1-7}
            $F_1^{q/N}(q^2)$ & $2$ & $1$ & $0$ & $5.57(9)$ & $2.84(5)$ & $-0.018(9)$ &
            $0.359(30)$ & $0.188(19)$ & $0.052(12)$
			\\
           \cline{8-10}
            $F_2^{q/N}(q^2)$ & $1.609(17)$ & $-2.097(17)$ & $-0.064(17)$ & $-$ & $-$ & $-$ 
            & \multicolumn{3}{c|}{\cellcolor{purple!15}$\tilde A_{20}^{q/p}(0)$~\cite{Blumlein:2010rn}}
			\\
            \hline
             $G_A^{q/N}(q^2)$ & $0.90$ & $-0.38$ & $-0.031(5)$ & $-$ & $-$ & $-$ & 
             $\tilde A_{20}^{u/p}(0)$ & $\tilde A_{20}^{d/p}(0)$ & $\tilde A_{20}^{s/p}(0)$
			\\
            \cline{1-7}
             $F_S^{q/N}(q^2)$ & $7.73$ & $6.81$ & $0.435$ & $-$  & $-$ & $-$ &
             $0.151(4)$ & $-0.039(7)$ & $0.0025(12)$
			\\ 
            \hline
             $F_{T,0}^{q/N}(q^2)$ & $0.784(28)(10)$ & $-0.204(11)(10)$ & $-0.0027(16)$ & $0.6272$  & $-0.1428$ & $-0.0027$ & & &
			\\
            $F_{T,1}^{q/N}(q^2)$ & $-1.5(1.0)$ & $0.5(3)$ & $0.009(5)$ & $-1.5$  & $-0.05$ & $-$ & & &
            \\
            $F_{T,2}^{q/N}(q^2)$ & $0.1(2)$ & $-0.6(3)$ & $-0.004(3)$ & $0.12$  & $-0.6$ & $-$ & & & 
			\\
            \hline
	\end{tabular}
 } 
\caption{Summary of the adopted numerical values for the parameters related to the proton form factors. The results for the neutron case can be obtained from isospin symmetry by swapping the superscripts $p\leftrightarrow n$ and $u\leftrightarrow d$. } 
	\label{tab:fftype2}
\end{table}

We will now discuss in more detail the treatment of the hadronic form factors involved in \cref{eq:hadff}. 
First, all the form factors appearing in the usual Lorentz structures from  \cref{eq:hadff:scalar} to \cref{eq:hadff:tensorL} have been consistently determined within the framework of chiral perturbation theory (ChPT) under the single-nucleon current approximation~\cite{Bishara:2017pfq,Liang:2024tef}.
In ChPT, they describe the contributions to DM-nucleon scattering stemming from both local DM-nucleon interactions and $t$-channel diagrams that involve intermediate mesonic degrees of freedom. Notably, the tensor bilinear also receives contributions through the LD photon exchange diagrams at higher chiral orders.

The two form factors, $F_P$ and $G_P$, corresponding to the pseudoscalar and axial-vector quark bilinears, 
receive LO contributions from $\pi$ and $\eta$ exchange diagrams within the ChPT framework, based on  the chiral power counting. 
Meanwhile, the contact DM-nucleon interactions via the one-body current provide NLO contributions. 
Collectively denoting the two form factors as $\mathcal{F}_1^{q/N}(q^2)$, we express them in terms of the three individual components as follows: 
\begin{align}
    \mathcal{F}_1^{q/N}(q^2) &= \frac{m_N^2}{m_\pi^2-q^2} a_{{\mathcal{F}_1},\pi}^{q/N} + \frac{m_N^2}{m_\eta^2-q^2} a_{{\mathcal{F}_1},\eta}^{q/N} + b_{\mathcal{F}_1}^{q/N},\quad\text{for}~ \mathcal{F}_1= G_P, F_P.
\end{align}
Note that there are no calculations on the NLO terms $b_{F_P}^{q/N}$, so we have neglected their contributions in our numerical analysis.
The other form factors, $F_{1,2}^{q/N}$, $G_A^{q/N}$, $F_S^{q/N}$, and $F_{T, \{0,1,2\}}^{q/N}$, 
parametrize the direct contributions from local DM-nucleon interactions. 
By representing them collectively as $\mathcal{F}_2^{q/N}(q^2)$, 
we can expand the LO and NLO contributions from the ChPT power counting as follows:
\begin{align}
    \mathcal{F}_2^{q/N}(q^2) &= \mathcal{F}_2^{q/N}(0) +\mathcal{F}_2'^{q/N}(0) { q^2 \over {\rm GeV}^2 }.
\end{align}
In \cref{tab:fftype2}, we list the numerical values for the relevant parameters that parametrize $\mathcal{F}_1^{q/p}(q^2)$ and $\mathcal{F}_2^{q/p}(q^2)$ related to the proton, 
as summarized in~Ref.\,\cite{Bishara:2017pfq}. The corresponding values for the neutron case are obtained by employing isospin symmetry, 
with the exchange of superscripts, $p\leftrightarrow n$ and $u\leftrightarrow d$. 

For the two bilinears involving a derivative, 
$\overline{q}\gamma_{(\mu} i \overleftrightarrow{D}_{\nu)} q$ in \cref{eq:qDgammaq} and $\overline{q}\gamma_{(\mu}\gamma_5 i \overleftrightarrow{D}_{\nu)} q$ in \cref{eq:qDgamma5q}, 
only the form factors $A_{20}^{q/N}$ and $\tilde A_{20}^{q/N}$ are relevant to our calculation, as shown in \cref{tab:NRmatch}. 
There is no ChPT analysis for these currents, but fortunately they have been calculated by the lattice QCD or QCD-based methods.  
Due to the momentum transfer $|q^2|\lesssim 0.01~\rm GeV^2$ in the DM direct detection, 
we can safely use their values at the forward limit, $A_{20}^{q/N}(0)$ and $\tilde A_{20}^{q/N}(0)$. 
For our numerical analysis, we adopt the lattice QCD calculation results for $A_{20}^{q/N}(0)$ from~Ref.\,\cite{Alexandrou:2020sml} and the NLO QCD analysis for $\tilde A_{20}^{q/N}(0)$ in~Ref.\,\cite{Blumlein:2010rn}. Their values are also collected in \cref{tab:fftype2}. 
In addition, the numerical values for the other parameters are taken from~Refs.\,\cite{Mateu:2007tr,DelNobile:2021wmp}:
$ g_n=-3.83, \,
g_p=+5.59, \,
Q_n=0, \,
Q_p= +1,\,
\Lambda_\chi=4\pi F_\pi, \,
c_T=-3.2,\, 
F_\pi=93\,{\rm MeV}$.

\section{Squared matrix element for each relativistic interaction}
\label{app:Msqrd}
For simplicity, we denote
\begin{align}
F_{12}^{i/I} \equiv F_1^{i/I} + F_2^{i/I}, \quad 
F_{T,012}^{i/I} \equiv F_{T,0}^{i/I}
-2F_{T,1}^{i/I}-4F_{T,2}^{i/I}.
\end{align}
Then, for the vector DM case A, the spin-summed and spin-averaged matrix elements squared  in the NR limit are
{\small
\begin{subequations}
\begin{align}
\overline{\left|{\cal M}_{qX}^{\tt S}\right|^2} & = 
4 m_{N}^2 F_{S}^{i/I} F_{S}^{j/J}  
F_{M}^{IJ} C_{qX}^{{\tt S},i}C_{qX}^{{\tt S},j} ,
\\
\overline{\left|{\cal M}_{qX}^{\tt P}\right|^2} & = 
\pmb{q}^2 F_{P}^{i/I} F_{P}^{j/J}  
F_{\Sigma''}^{IJ} C_{qX}^{ {\tt P},i} C_{qX}^{{\tt P},j},  
\\
\overline{\left|{\cal M}_{qX1}^{\tt T}\right|^2} & = \frac{4}{3} \left\{ 
d_{i} d_{j}  \left[
 Q_I Q_J 
\left( 2 F_{\Delta }^{IJ} + \left(\frac{m_{N}^2 }{m_X^2} +2 \frac{m_{N}^2 |\pmb{v}_T^\perp|^2 }{\pmb{q}^2} \right) F_{M}^{IJ}   \right)
+{1\over 8 }g_I  g_J  F_{ \Sigma'}^{IJ}
- g_I Q_J F_{\Sigma'\Delta}^{IJ}  \right]
\right.
\nonumber
\\
&
\left.
+  2 m_{N}^2 F_{T,0}^{i/I} F_{T,0}^{j/J} \left(F_{\Sigma'}^{IJ}+F_{\Sigma''}^{IJ}\right) 
- m_{N} d_{i}
 F_{T,0}^{j/I} \left( g_J F_{\Sigma'}^{IJ} - 4 Q_J F_{\Sigma'\Delta}^{IJ} \right)
\right\} C_{qX1}^{{\tt T},i} C_{qX1}^{{\tt T},j},
\\
\overline{\left|{\cal M}_{qX2}^{\tt T}\right|^2} & = 
\frac{2}{3} m_{N}  
\left[
{\pmb{q}^2 \over m_N} \left(  F_{T,012}^{i,I} F_{T,012}^{j,J} F_{M}^{IJ}
+ 4 F_{T,0}^{i,I} F_{T,0}^{j,J} \left(F_{\Phi'}^{IJ}+F_{\Phi''}^{IJ}\right)
- 4 F_{T,0}^{j,J} F_{T,012}^{i,I} F_{M\Phi''}^{IJ}  \right)
\right.
\nonumber
\\
& \left. + 4 d_i  Q_I \left( 2  F_{T,0}^{j/J} F_{M\Phi''}^{IJ} -F_{T,012}^{j/J} F_{M}^{IJ} \right)
+ 4 { m_N \over \pmb{q}^2 }
 d_i d_j Q_I Q_J  F_{M}^{IJ} \right] 
 C_{qX2}^{{\tt T},i}C_{qX2}^{{\tt T},j},
\\
\overline{\left|{\cal M}_{qX1}^{{\tt V}}\right|^2} & = 
 m_{N}^4 A_{20}^{i/I} A_{20}^{j/J} F_{M}^{IJ} C_{qX1}^{{\tt V},i} C_{qX1}^{{\tt V},j}, 
\\
\overline{\left|{\cal M}_{qX2}^{{\tt V}}\right|^2} & = \frac{2}{3}  \pmb{q}^4 
\left[
 4 F_1^{i/I} F_1^{j/J}\left(F_{\Delta }^{IJ} 
 + \frac{m_{N}^2 |\pmb{v}_T^\perp|^2|}{\pmb{q}^2}  F_{M}^{IJ} \right)
 \right.
\nonumber
\\
& \left.
 + F_{12}^{i/I} F_{12}^{j/J}  F_{\Sigma'}^{IJ}
- 4  F_1^{i/I} F_{12}^{j/J} F_{\Sigma'\Delta}^{JI}
\right]
C_{qX2}^{{\tt V},i}C_{qX2}^{{\tt V},j},
\\
\overline{\left|{\cal M}_{qX3}^{{\tt V}}\right|^2} & = \frac{8}{3}  m_X^2 \pmb{q}^2
\left[
4  F_1^{i/I} F_1^{j/J}\left(F_{\Delta }^{IJ}  
+  \frac{m_{N}^2 |\pmb{v}_T^\perp|^2  }{\pmb{q}^2} F_{M}^{IJ} \right) 
\right.
\nonumber
\\
& \left.
+  F_{12}^{i/I} F_{12}^{j/J} F_{\Sigma'}^{IJ}
-4 F_1^{i/I} F_{12}^{j/J} F_{\Sigma'\Delta}^{JI}
\right]
C_{qX3}^{{\tt V},i}C_{qX3}^{{\tt V},j},
\\
\overline{\left|{\cal M}_{qX4}^{{\tt V}}\right|^2} & = 
16  m_{N}^2 m_X^2 F_1^{i/I} F_1^{j/J} 
F_{M}^{IJ} C_{qX4}^{{\tt V},i} C_{qX4}^{{\tt V},j}, 
\\
\overline{\left|{\cal M}_{qX5}^{{\tt V}}\right|^2} & = \frac{2}{3}\pmb{q}^4  \left\{
2 F_1^{i/I} F_1^{j/J}  \left[ 2 F_{\Delta }^{IJ}  
+ \left( \frac{m_{N}^2}{m_X^2 }+2\frac{m_{N}^2 |\pmb{v}_T^\perp|^2 }{\pmb{q}^2} \right) F_{M}^{IJ} \right]
\right.
\nonumber
\\
& \left.
+  F_{12}^{i/I} F_{12}^{j/J} F_{\Sigma'}^{IJ} 
-4  F_1^{i/I} F_{12}^{j/J} F_{\Sigma'\Delta}^{JI}
\right\} C_{qX5}^{{\tt V},i}C_{qX5}^{{\tt V},j},
\\
\overline{\left|{\cal M}_{qX6}^{{\tt V}}\right|^2} & = 
\frac{8}{3} \pmb{q}^2 m_{N}^2 F_1^{i/I} F_1^{j/J} 
F_{M}^{IJ} C_{qX6}^{{\tt V},i}C_{qX6}^{{\tt V},j},
\\
\overline{\left|{\cal M}_{qX1}^{{\tt A}}\right|^2} & = 
\frac{2}{3} \left( \frac{\pmb{q}^2}{m_X^2}+\frac{4}{3} |\pmb{v}_T^\perp|^2 \right) 
m_{N}^4 \tilde A_{20}^{i/I} \tilde A_{20}^{j/J} F_{\Sigma'}^{IJ} 
C_{qX1}^{{\tt A},i}C_{qX1}^{{\tt A},j},
\\
\overline{\left|{\cal M}_{qX2}^{{\tt A}}\right|^2} & = 
\frac{8}{3} \pmb{q}^2 m_{N}^2 G_{A}^{i/I} G_{A}^{j/J} 
\left(F_{\Sigma'}^{IJ}+2 F_{\Sigma''}^{IJ}\right)C_{qX2}^{{\tt A},i}C_{qX2}^{{\tt A},j},
\\
\overline{\left|{\cal M}_{qX3}^{{\tt A}}\right|^2} & =
\frac{32}{3} m_{N}^2 m_X^2 G_{A}^{i/I} G_{A}^{j/J} 
\left(F_{\Sigma'}^{IJ}+F_{\Sigma''}^{IJ}\right) C_{qX3}^{{\tt A},i} C_{qX3}^{{\tt A},j},
\\
\overline{\left|{\cal M}_{qX4}^{{\tt A}}\right|^2} & = 
8 |\pmb{v}_T^\perp|^2 m_{N}^2 m_X^2 G_{A}^{i/I} G_{A}^{j/J} 
F_{\Sigma'}^{IJ} C_{qX4}^{{\tt A},i}C_{qX4}^{{\tt A},j},      
\\
\overline{\left|{\cal M}_{qX5}^{{\tt A}}\right|^2} & = \frac{8}{3} \pmb{q}^2 m_{N}^2
G_{A}^{i/I} G_{A}^{j/J} F_{\Sigma'}^{IJ} C_{qX5}^{{\tt A},i} C_{qX5}^{{\tt A},j},      
\\
\overline{\left|{\cal M}_{qX6}^{{\tt A}}\right|^2} & = 
\frac{2}{3}\left(\frac{\pmb{q}^2}{m_X^2}+2 |\pmb{v}_T^\perp|^2 \right) 
\pmb{q}^2 m_{N}^2  G_{A}^{i/I} G_{A}^{j/J} 
F_{\Sigma'}^{IJ} C_{qX6}^{{\tt A},i} C_{qX6}^{{\tt A},j}. 
\end{align}
\end{subequations} }%
where the quark flavor indices $i,j=u,d,s$, while the nucleon indices $I,J=p,n$, both of which have been summed over in each matrix element squared. 

For the vector DM case B, 
{\small
\begin{subequations}
\begin{align}          
\overline{\left|\tilde{\cal M}_{qX1}^{\tt S}\right|^2}  & =  16 m_{N}^2 m_X^4 F_{S}^{i/I} F_{S}^{j/J}  F_{M}^{IJ} 
{\tilde C}_{qX1}^{{\tt S},i} {\tilde C}_{qX1}^{{\tt S},j},  
\\%
\overline{\left|\tilde{\cal M}_{qX2}^{\tt S}\right|^2}  & =  \frac{32}{3} {\pmb q}^2 m_{N}^2 m_X^2  F_{S}^{i/I} F_{S}^{j/J} F_{M}^{IJ} 
{\tilde C}_{qX2}^{{\tt S},i} {\tilde C}_{qX2}^{{\tt S},j},  
\\%
\overline{\left|\tilde{\cal M}_{qX1}^{\tt P}\right|^2}  & =  4 {\pmb q}^2 m_X^4 F_{P}^{i/I} F_{P}^{j/J} F_{\Sigma''}^{IJ} 
{\tilde C}_{qX1}^{{\tt P},i} {\tilde C}_{qX1}^{{\tt P},j},   
\\%
\overline{\left|\tilde{\cal M}_{qX2}^{\tt P}\right|^2}  & =  \frac{8}{3} {\pmb q}^4 m_X^2 F_{P}^{i/I} F_{P}^{j/J} F_{\Sigma''}^{IJ} 
{\tilde C}_{qX2}^{{\tt P},i} {\tilde C}_{qX2}^{{\tt P},j},   
\\ 
\overline{\left|\tilde{\cal M}_{qX1}^{\tt T}\right|^2}  & =  \frac{4}{3}  m_X^4 
\left\{ 
d_{i}  d_{j} \left[ 
2 Q_I Q_J\left(F_{\Delta }^{IJ} + 
\left({m_N^2 \over m_X^2} +\frac{m_N^2  \pmb{v}_T^2 }{\pmb{q}^2} \right) F_{M}^{IJ}\right)
+{1\over 8}g_I g_J F_{\Sigma'}^{IJ}
- g_I Q_J F_{\Sigma'\Delta}^{IJ} 
\right]
\right.
\nonumber
\\
& \left.
+ 2 m_{N}^2  F_{T,0}^{i/I} F_{T,0}^{j/J} \left(F_{\Sigma'}^{IJ}+F_{\Sigma''}^{IJ}\right)
-  m_N d_{i} F_{T,0}^{j/I}\left( g_J F_{\Sigma'}^{IJ} -4 Q_J F_{\Sigma'\Delta}^{IJ} \right)
\right\} 
{\tilde C}_{qX1}^{{\tt T},i} {\tilde C}_{qX1}^{{\tt T},j},
\\
\overline{\left|\tilde{\cal M}_{qX2}^{\tt T}\right|^2}  & =  \frac{4}{3}m_N m_X^2\left\{
{ m_X^2 \pmb{q}^2 \over 2m_N} \left(  F_{T,012}^{i,I} F_{T,012}^{j,J} F_{M}^{IJ}
+ 4 F_{T,0}^{i,I} F_{T,0}^{j,J} \left(F_{\Phi'}^{IJ}+F_{\Phi''}^{IJ}\right)
- 4 F_{T,0}^{j,J} F_{T,012}^{i,I} F_{M\Phi''}^{IJ} \right)
\right.
\nonumber
\\
&\left. +2 m_X^2  d_{i} Q_I \left( 2 F_{T,0}^{j/J} F_{ M\Phi''}^{IJ} -F_{T,012}^{j/J} F_{M}^{IJ} \right)
+2 m_{N} { m_X^2 \over \pmb{q}^2} 
d_{i} d_{j} Q_I  Q_J  F_{M}^{IJ} 
\right\} {\tilde C}_{qX2}^{{\tt T},i} {\tilde C}_{qX2}^{{\tt T},j}. 
\end{align}
\end{subequations} }%

Finally, for the DM-photon interactions,
{\small
\begin{subequations}
\begin{align}
\overline{\left|{\cal M}_{\kappa_\Lambda}\right|^2}  & =  \frac{4}{3} e^2 {\tilde \kappa}_\Lambda^2 \left[
Q_I Q_J 
\left( 2 F_{\Delta }^{IJ}  + \left(\frac{m_{N}^2 }{m_X^2} +2 \frac{m_{N}^2 |\pmb{v}_T^\perp|^2 }{\pmb{q}^2} \right) F_{M}^{IJ}  \right)
+{1\over 8 }g_I  g_J  F_{ \Sigma'}^{IJ}
- g_I Q_J F_{\Sigma'\Delta}^{IJ}  \right] 
\\%
\overline{\left|{\cal M}_{\tilde \kappa_\Lambda}\right|^2}  & = \frac{8}{3} e^2 {\tilde \kappa}_\Lambda^2 
{ m_{N}^2 \over {\pmb q}^2 } Q_I Q_J F_{M}^{IJ}  , 
\\
\overline{\left|{\cal M}_{X\gamma1}\right|^2}  & = \frac{8}{3} e^2 m_X^2 \pmb{q}^2 
\left[
4  Q_I Q_J\left(F_{\Delta }^{IJ}  
+  \frac{m_{N}^2 |\pmb{v}_T^\perp|^2  }{\pmb{q}^2} F_{M}^{IJ} \right) 
+  {g_I g_J \over 4 m_X^2} F_{\Sigma'}^{IJ}
-2  {Q_I g_J \over m_X} F_{\Sigma'\Delta}^{JI}
\right]C_{X\gamma 1}^2, 
\\%
\overline{\left|{\cal M}_{X\gamma2}\right|^2}  & =  \frac{8}{3} e^2  {\pmb q}^2 m_{N}^2 Q_I Q_J F_{M}^{IJ}  C_{X\gamma 2}^2, 
\\%
\overline{\left|{\cal M}_{X\gamma3}\right|^2}  & =  16  e^2 m_{N}^2 m_X^2 Q_I Q_J F_{M}^{IJ}  C_{X\gamma 3}^2, 
\\%
\overline{\left|{\cal M}_{X\gamma4}\right|^2}  & = \frac{2}{3} e^2 \pmb{q}^4 
\left[
 4 Q_I Q_J \left(F_{\Delta }^{IJ} 
 + \frac{m_{N}^2 |\pmb{v}_T^\perp|^2|}{\pmb{q}^2}  F_{M}^{IJ} \right)
 + {g_I g_J \over 4}  F_{\Sigma'}^{IJ}
 -  {2 Q_I g_J \over 2} F_{\Sigma'\Delta}^{JI}
\right]
C_{X\gamma 4}^2, 
\\%
\overline{\left|{\cal M}_{X\gamma5}\right|^2}  & =
\frac{4}{3} e^2  \pmb{q}^4 
\left[ Q_I Q_J\left(2 F_{\Delta }^{IJ} + 
\left({m_N^2 \over m_X^2} +2\frac{m_N^2  \pmb{v}_T^2 }{\pmb{q}^2} \right) F_{M}^{IJ}\right)
+{1\over 8}g_I g_J F_{\Sigma'}^{IJ}
- g_I Q_J F_{\Sigma'\Delta}^{IJ} 
\right] C_{X\gamma 5}^2, 
\\%
\overline{\left|\tilde{\cal M}_{X\gamma1}\right|^2}  & = \frac{16}{3} e^2 m_X^4  
 \left[ 
2 Q_I Q_J\left(F_{\Delta }^{IJ} + 
\left({m_N^2 \over m_X^2} +\frac{m_N^2  \pmb{v}_T^2 }{\pmb{q}^2} \right) F_{M}^{IJ}\right)
+{1\over 8}g_I g_J F_{\Sigma'}^{IJ}
- g_I Q_J F_{\Sigma'\Delta}^{IJ} 
\right] {\tilde C}_{X\gamma 1}^2, 
\\%
\overline{\left|\tilde{\cal M}_{X\gamma2}\right|^2}  & = \frac{32}{3} e^2  { m_{N}^2 m_X^4 \over {\pmb q}^2 } 
 Q_I Q_J F_{M}^{IJ}  {\tilde C}_{X\gamma 2}^2, 
\end{align}
\end{subequations}
}

\bibliography{refs_paper.bib}{}

@article{Bell:2023sdq,
    author = "Bell, Nicole F. and Newstead, Jayden L. and Shaukat-Ali, Iman",
    title = "{Cosmic-ray boosted dark matter confronted by constraints on new light mediators}",
    eprint = "2309.11003",
    archivePrefix = "arXiv",
    primaryClass = "hep-ph",
    doi = "10.1103/PhysRevD.109.063034",
    journal = "Phys. Rev. D",
    volume = "109",
    number = "6",
    pages = "063034",
    year = "2024"
}

@article{He:2023bnk,
    author = "He, Xiao-Gang and Ma, Xiao-Dong and Valencia, German",
    title = "{Revisiting models that enhance B+\textrightarrow{}K+\ensuremath{\nu}\ensuremath{\nu}\textasciimacron{} in light of the new Belle II measurement}",
    eprint = "2309.12741",
    archivePrefix = "arXiv",
    primaryClass = "hep-ph",
    doi = "10.1103/PhysRevD.109.075019",
    journal = "Phys. Rev. D",
    volume = "109",
    number = "7",
    pages = "075019",
    year = "2024"
}

@article{Feng:2011vu,
    author = "Feng, Jonathan L. and Kumar, Jason and Marfatia, Danny and Sanford, David",
    title = "{Isospin-Violating Dark Matter}",
    eprint = "1102.4331",
    archivePrefix = "arXiv",
    primaryClass = "hep-ph",
    reportNumber = "UCI-TR-2011-03, UH511-1157-2011",
    doi = "10.1016/j.physletb.2011.07.083",
    journal = "Phys. Lett. B",
    volume = "703",
    pages = "124--127",
    year = "2011"
}

@article{Bertone:2004pz,
    author = "Bertone, Gianfranco and Hooper, Dan and Silk, Joseph",
    title = "{Particle dark matter: Evidence, candidates and constraints}",
    eprint = "hep-ph/0404175",
    archivePrefix = "arXiv",
    reportNumber = "FERMILAB-PUB-04-047-A",
    doi = "10.1016/j.physrep.2004.08.031",
    journal = "Phys. Rept.",
    volume = "405",
    pages = "279--390",
    year = "2005"
}

@article{Young:2016ala,
    author = "Young, Bing-Lin",
    title = "{A survey of dark matter and related topics in cosmology}",
    doi = "10.1007/s11467-016-0583-4",
    journal = "Front. Phys. (Beijing)",
    volume = "12",
    number = "2",
    pages = "121201",
    year = "2017",
    note = "[Erratum: Front.Phys.(Beijing) 12, 121202 (2017)]"
}

@article{Arbey:2021gdg,
    author = "Arbey, A. and Mahmoudi, F.",
    title = "{Dark matter and the early Universe: a review}",
    eprint = "2104.11488",
    archivePrefix = "arXiv",
    primaryClass = "hep-ph",
    reportNumber = "CERN-TH-2021-066",
    doi = "10.1016/j.ppnp.2021.103865",
    journal = "Prog. Part. Nucl. Phys.",
    volume = "119",
    pages = "103865",
    year = "2021"
}

@article{Feng:2010gw,
    author = "Feng, Jonathan L.",
    title = "{Dark Matter Candidates from Particle Physics and Methods of Detection}",
    eprint = "1003.0904",
    archivePrefix = "arXiv",
    primaryClass = "astro-ph.CO",
    reportNumber = "UCI-TR-2009-13",
    doi = "10.1146/annurev-astro-082708-101659",
    journal = "Ann. Rev. Astron. Astrophys.",
    volume = "48",
    pages = "495--545",
    year = "2010"
}

@article{Escriva:2022duf,
author = "Escriv\`a, Albert and Kuhnel, Florian and Tada, Yuichiro",
title = "{Primordial Black Holes}",
eprint = "2211.05767",
archivePrefix = "arXiv",
primaryClass = "astro-ph.CO",
month = "11",
year = "2022"
}

@article{Jungman:1995df,
    author = "Jungman, Gerard and Kamionkowski, Marc and Griest, Kim",
    title = "{Supersymmetric dark matter}",
    eprint = "hep-ph/9506380",
    archivePrefix = "arXiv",
    reportNumber = "SU-4240-605, UCSD-PTH-95-02, IASSNS-HEP-95-14, CU-TP-677",
    doi = "10.1016/0370-1573(95)00058-5",
    journal = "Phys. Rept.",
    volume = "267",
    pages = "195--373",
    year = "1996"
}

@article{Bernal:2017kxu,
    author = "Bernal, Nicol\'as and Heikinheimo, Matti and Tenkanen, Tommi and Tuominen, Kimmo and Vaskonen, Ville",
    title = "{The Dawn of FIMP Dark Matter: A Review of Models and Constraints}",
    eprint = "1706.07442",
    archivePrefix = "arXiv",
    primaryClass = "hep-ph",
    reportNumber = "PI-UAN-2017-602FT, HIP-2017-08-TH, PI-UAN--2017--602FT, HIP--2017--08-TH",
    doi = "10.1142/S0217751X1730023X",
    journal = "Int. J. Mod. Phys. A",
    volume = "32",
    number = "27",
    pages = "1730023",
    year = "2017"
}

@article{Petraki:2013wwa,
    author = "Petraki, Kalliopi and Volkas, Raymond R.",
    title = "{Review of asymmetric dark matter}",
    eprint = "1305.4939",
    archivePrefix = "arXiv",
    primaryClass = "hep-ph",
    reportNumber = "NIKHEF-2013-016",
    doi = "10.1142/S0217751X13300287",
    journal = "Int. J. Mod. Phys. A",
    volume = "28",
    pages = "1330028",
    year = "2013"
}

@article{Goodman:2010ku,
    author = "Goodman, Jessica and Ibe, Masahiro and Rajaraman, Arvind and Shepherd, William and Tait, Tim M. P. and Yu, Hai-Bo",
    title = "{Constraints on Dark Matter from Colliders}",
    eprint = "1008.1783",
    archivePrefix = "arXiv",
    primaryClass = "hep-ph",
    reportNumber = "UCI-HEP-TR-2010-15",
    doi = "10.1103/PhysRevD.82.116010",
    journal = "Phys. Rev. D",
    volume = "82",
    pages = "116010",
    year = "2010"
}

@article{Alanne:2017oqj,
    author = "Alanne, Tommi and Goertz, Florian",
    title = "{Extended Dark Matter EFT}",
    eprint = "1712.07626",
    archivePrefix = "arXiv",
    primaryClass = "hep-ph",
    doi = "10.1140/epjc/s10052-020-7999-2",
    journal = "Eur. Phys. J. C",
    volume = "80",
    number = "5",
    pages = "446",
    year = "2020"
}

@article{Farzan:2012hh,
    author = "Farzan, Yasaman and Akbarieh, Amin Rezaei",
    title = "{VDM: A model for Vector Dark Matter}",
    eprint = "1207.4272",
    archivePrefix = "arXiv",
    primaryClass = "hep-ph",
    doi = "10.1088/1475-7516/2012/10/026",
    journal = "JCAP",
    volume = "10",
    pages = "026",
    year = "2012"
}

@article{Dent:2015zpa,
    author = "Dent, James B. and Krauss, Lawrence M. and Newstead, Jayden L. and Sabharwal, Subir",
    title = "{General analysis of direct dark matter detection: From microphysics to observational signatures}",
    eprint = "1505.03117",
    archivePrefix = "arXiv",
    primaryClass = "hep-ph",
    doi = "10.1103/PhysRevD.92.063515",
    journal = "Phys. Rev. D",
    volume = "92",
    number = "6",
    pages = "063515",
    year = "2015"
}

@article{Fan:2010gt,
    author = "Fan, JiJi and Reece, Matthew and Wang, Lian-Tao",
    title = "{Non-relativistic effective theory of dark matter direct detection}",
    eprint = "1008.1591",
    archivePrefix = "arXiv",
    primaryClass = "hep-ph",
    doi = "10.1088/1475-7516/2010/11/042",
    journal = "JCAP",
    volume = "11",
    pages = "042",
    year = "2010"
}

@article{Blumlein:2010rn,
    author = "Blumlein, Johannes and Bottcher, Helmut",
    title = "{QCD Analysis of Polarized Deep Inelastic Scattering Data}",
    eprint = "1005.3113",
    archivePrefix = "arXiv",
    primaryClass = "hep-ph",
    reportNumber = "DESY-09-131, SFB-CPP-10-032",
    doi = "10.1016/j.nuclphysb.2010.08.005",
    journal = "Nucl. Phys. B",
    volume = "841",
    pages = "205--230",
    year = "2010"
}

@article{Harnik:2008uu,
    author = "Harnik, Roni and Kribs, Graham D.",
    title = "{An Effective Theory of Dirac Dark Matter}",
    eprint = "0810.5557",
    archivePrefix = "arXiv",
    primaryClass = "hep-ph",
    reportNumber = "SLAC-PUB-13482",
    doi = "10.1103/PhysRevD.79.095007",
    journal = "Phys. Rev. D",
    volume = "79",
    pages = "095007",
    year = "2009"
}

@article{DelNobile:2011uf,
    author = "Del Nobile, Eugenio and Sannino, Francesco",
    title = "{Dark Matter Effective Theory}",
    eprint = "1102.3116",
    archivePrefix = "arXiv",
    primaryClass = "hep-ph",
    reportNumber = "CP3-ORIGINS-2011-05",
    doi = "10.1142/S0217751X12500650",
    journal = "Int. J. Mod. Phys. A",
    volume = "27",
    pages = "1250065",
    year = "2012"
}

@article{Kumar:2013iva,
    author = "Kumar, Jason and Marfatia, Danny",
    title = "{Matrix element analyses of dark matter scattering and annihilation}",
    eprint = "1305.1611",
    archivePrefix = "arXiv",
    primaryClass = "hep-ph",
    doi = "10.1103/PhysRevD.88.014035",
    journal = "Phys. Rev. D",
    volume = "88",
    number = "1",
    pages = "014035",
    year = "2013"
}

@article{Brod:2017bsw,
	author = "Brod, Joachim and Gootjes-Dreesbach, Aaron and Tammaro, Michele and Zupan, Jure",
	title = "{Effective Field Theory for Dark Matter Direct Detection up to Dimension Seven}",
	eprint = "1710.10218",
	archivePrefix = "arXiv",
	primaryClass = "hep-ph",
	reportNumber = "DO-TH-17-21, DO-TH 17/21",
	doi = "10.1007/JHEP10(2018)065",
	journal = "JHEP",
	volume = "10",
	pages = "065",
	year = "2018",
	note = "[Erratum: JHEP 07, 012 (2023)]"
}

@article{Catena:2019hzw,
    author = "Catena, Riccardo and Fridell, K\r{a}re and Krauss, Martin B.",
    title = "{Non-relativistic Effective Interactions of Spin 1 Dark Matter}",
    eprint = "1907.02910",
    archivePrefix = "arXiv",
    primaryClass = "hep-ph",
    doi = "10.1007/JHEP08(2019)030",
    journal = "JHEP",
    volume = "08",
    pages = "030",
    year = "2019"
}

@article{Liang:2023yta,
    author = "Liang, Jin-Han and Liao, Yi and Ma, Xiao-Dong and Wang, Hao-Lin",
    title = "{Dark sector effective field theory}",
    eprint = "2309.12166",
    archivePrefix = "arXiv",
    primaryClass = "hep-ph",
    doi = "10.1007/JHEP12(2023)172",
    journal = "JHEP",
    volume = "12",
    pages = "172",
    year = "2023"
}

@article{Fitzpatrick:2012ix,
    author = "Fitzpatrick, A. Liam and Haxton, Wick and Katz, Emanuel and Lubbers, Nicholas and Xu, Yiming",
    title = "{The Effective Field Theory of Dark Matter Direct Detection}",
    eprint = "1203.3542",
    archivePrefix = "arXiv",
    primaryClass = "hep-ph",
    doi = "10.1088/1475-7516/2013/02/004",
    journal = "JCAP",
    volume = "02",
    pages = "004",
    year = "2013"
}

@article{DelNobile:2021wmp,
    author = "Del Nobile, Eugenio",
    title = "{The Theory of Direct Dark Matter Detection: A Guide to Computations}",
    eprint = "2104.12785",
    archivePrefix = "arXiv",
    primaryClass = "hep-ph",
    doi = "10.1007/978-3-030-95228-0",
    month = "4",
    year = "2021"
}

@article{Anand:2013yka,
    author = "Anand, Nikhil and Fitzpatrick, A. Liam and Haxton, W. C.",
    title = "{Weakly interacting massive particle-nucleus elastic scattering response}",
    eprint = "1308.6288",
    archivePrefix = "arXiv",
    primaryClass = "hep-ph",
    doi = "10.1103/PhysRevC.89.065501",
    journal = "Phys. Rev. C",
    volume = "89",
    number = "6",
    pages = "065501",
    year = "2014"
}

@article{Adler:1975he,
    author = "Adler, Stephen L. and Colglazier, Jr., E. W. and Healy, J. B. and Karliner, Inga and Lieberman, Judy and Ng, Yee Jack and Tsao, Hung-Sheng",
    title = "{Renormalization Constants for Scalar, Pseudoscalar, and Tensor Currents}",
    reportNumber = "COO-2220-38",
    doi = "10.1103/PhysRevD.11.3309",
    journal = "Phys. Rev. D",
    volume = "11",
    pages = "3309",
    year = "1975"
}

@article{Ji:1998pc,
    author = "Ji, Xiang-Dong",
    title = "{Off forward parton distributions}",
    eprint = "hep-ph/9807358",
    archivePrefix = "arXiv",
    reportNumber = "UMD-PP-98-092, DOE-ER-40762-144",
    doi = "10.1088/0954-3899/24/7/002",
    journal = "J. Phys. G",
    volume = "24",
    pages = "1181--1205",
    year = "1998"
}

@article{Liang:2024lkk,
    author = "Liang, Jin-Han and Liao, Yi and Ma, Xiao-Dong and Wang, Hao-Lin",
    title = "{Revisiting general dark matter-bound-electron interactions}",
    eprint = "2405.04855",
    archivePrefix = "arXiv",
    primaryClass = "hep-ph",
    doi = "10.1103/PhysRevD.110.L091701",
    journal = "Phys. Rev. D",
    volume = "110",
    number = "9",
    pages = "L091701",
    year = "2024"
}

@article{Liang:2024tef,
    author = "Liang, Jin-Han and Liao, Yi and Ma, Xiao-Dong and Wang, Hao-Lin",
    title = "{Comprehensive constraints on fermionic dark matter-quark tensor interactions in direct detection experiments}",
    eprint = "2401.05005",
    archivePrefix = "arXiv",
    primaryClass = "hep-ph",
    doi = "10.1088/1674-1137/ad77b3",
    journal = "Chin. Phys. C",
    volume = "48",
    number = "12",
    pages = "123103",
    year = "2024"
}

@article{Gondolo:2020wge,
    author = "Gondolo, Paolo and Kang, Sunghyun and Scopel, Stefano and Tomar, Gaurav",
    title = "{Effective theory of nuclear scattering for a WIMP of arbitrary spin}",
    eprint = "2008.05120",
    archivePrefix = "arXiv",
    primaryClass = "hep-ph",
    reportNumber = "CQUeST-2020-0648",
    doi = "10.1103/PhysRevD.104.063017",
    journal = "Phys. Rev. D",
    volume = "104",
    number = "6",
    pages = "063017",
    year = "2021"
}

@article{He:2022ljo,
    author = "He, Xiao-Gang and Ma, Xiao-Dong and Valencia, German",
    title = "{FCNC B and K meson decays with light bosonic Dark Matter}",
    eprint = "2209.05223",
    archivePrefix = "arXiv",
    primaryClass = "hep-ph",
    doi = "10.1007/JHEP03(2023)037",
    journal = "JHEP",
    volume = "03",
    pages = "037",
    year = "2023"
}

@article{Liang:2024ecw,
    author = "Liang, Jin-Han and Liao, Yi and Ma, Xiao-Dong and Wang, Hao-Lin",
    title = "{A systematic investigation on dark matter-electron scattering in effective field theories}",
    eprint = "2406.10912",
    archivePrefix = "arXiv",
    primaryClass = "hep-ph",
    doi = "10.1007/JHEP07(2024)279",
    journal = "JHEP",
    volume = "07",
    pages = "279",
    year = "2024"
}

@article{XENON:2023cxc,
    author = "Aprile, E. and others",
    collaboration = "XENON",
    title = "{First Dark Matter Search with Nuclear Recoils from the XENONnT Experiment}",
    eprint = "2303.14729",
    archivePrefix = "arXiv",
    primaryClass = "hep-ex",
    doi = "10.1103/PhysRevLett.131.041003",
    journal = "Phys. Rev. Lett.",
    volume = "131",
    number = "4",
    pages = "041003",
    year = "2023"
}

@article{LZ:2022lsv,
    author = "Aalbers, J. and others",
    collaboration = "LZ",
    title = "{First Dark Matter Search Results from the LUX-ZEPLIN (LZ) Experiment}",
    eprint = "2207.03764",
    archivePrefix = "arXiv",
    primaryClass = "hep-ex",
    doi = "10.1103/PhysRevLett.131.041002",
    journal = "Phys. Rev. Lett.",
    volume = "131",
    number = "4",
    pages = "041002",
    year = "2023"
}

@article{XENON:2019gfn,
    author = "Aprile, E. and others",
    collaboration = "XENON",
    title = "{Light Dark Matter Search with Ionization Signals in XENON1T}",
    eprint = "1907.11485",
    archivePrefix = "arXiv",
    primaryClass = "hep-ex",
    doi = "10.1103/PhysRevLett.123.251801",
    journal = "Phys. Rev. Lett.",
    volume = "123",
    number = "25",
    pages = "251801",
    year = "2019"
}

@article{PandaX:2022xqx,
    author = "Li, Shuaijie and others",
    collaboration = "PandaX",
    title = "{Search for Light Dark Matter with Ionization Signals in the PandaX-4T Experiment}",
    eprint = "2212.10067",
    archivePrefix = "arXiv",
    primaryClass = "hep-ex",
    doi = "10.1103/PhysRevLett.130.261001",
    journal = "Phys. Rev. Lett.",
    volume = "130",
    number = "26",
    pages = "261001",
    year = "2023"
}

@article{PandaX:2024qfu,
    author = "Bo, Zihao and others",
    collaboration = "PandaX",
    title = "{Dark Matter Search Results from 1.54 Tonne$\cdot$Year Exposure of PandaX-4T}",
    eprint = "2408.00664",
    archivePrefix = "arXiv",
    primaryClass = "hep-ex",
    month = "8",
    year = "2024"
}

@article{PandaX:2024med,
    author = "Luo, Yunyang and others",
    collaboration = "PandaX",
    title = "{Signal response model in PandaX-4T}",
    eprint = "2403.04239",
    archivePrefix = "arXiv",
    primaryClass = "physics.ins-det",
    doi = "10.1103/PhysRevD.110.023029",
    journal = "Phys. Rev. D",
    volume = "110",
    number = "2",
    pages = "023029",
    year = "2024"
}

@article{Bell:2021zkr,
    author = "Bell, Nicole F. and Dent, James B. and Dutta, Bhaskar and Ghosh, Sumit and Kumar, Jason and Newstead, Jayden L.",
    title = "{Low-mass inelastic dark matter direct detection via the Migdal effect}",
    eprint = "2103.05890",
    archivePrefix = "arXiv",
    primaryClass = "hep-ph",
    reportNumber = "MI-TH-215",
    doi = "10.1103/PhysRevD.104.076013",
    journal = "Phys. Rev. D",
    volume = "104",
    number = "7",
    pages = "076013",
    year = "2021"
}

@article{Tomar:2022ofh,
    author = "Tomar, Gaurav and Kang, Sunghyun and Scopel, Stefano",
    title = "{Low-mass extension of direct detection bounds on WIMP-quark and WIMP-gluon effective interactions using the Migdal effect}",
    eprint = "2210.00199",
    archivePrefix = "arXiv",
    primaryClass = "hep-ph",
    reportNumber = "CQUeST-2022-0698",
    doi = "10.1016/j.astropartphys.2023.102851",
    journal = "Astropart. Phys.",
    volume = "150",
    pages = "102851",
    year = "2023"
}

@article{Ibe:2017yqa,
    author = "Ibe, Masahiro and Nakano, Wakutaka and Shoji, Yutaro and Suzuki, Kazumine",
    title = "{Migdal Effect in Dark Matter Direct Detection Experiments}",
    eprint = "1707.07258",
    archivePrefix = "arXiv",
    primaryClass = "hep-ph",
    reportNumber = "IPMU17-0100",
    doi = "10.1007/JHEP03(2018)194",
    journal = "JHEP",
    volume = "03",
    pages = "194",
    year = "2018"
}

@article{DarkSide-50:2022qzh,
    author = "Agnes, P. and others",
    collaboration = "DarkSide-50",
    title = "{Search for low-mass dark matter WIMPs with 12~ton-day exposure of DarkSide-50}",
    eprint = "2207.11966",
    archivePrefix = "arXiv",
    primaryClass = "hep-ex",
    reportNumber = "FERMILAB-PUB-22-589-ND-PPD-SCD",
    doi = "10.1103/PhysRevD.107.063001",
    journal = "Phys. Rev. D",
    volume = "107",
    number = "6",
    pages = "063001",
    year = "2023"
}

@article{XENON:2019zpr,
    author = "Aprile, E. and others",
    collaboration = "XENON",
    title = "{Search for Light Dark Matter Interactions Enhanced by the Migdal Effect or Bremsstrahlung in XENON1T}",
    eprint = "1907.12771",
    archivePrefix = "arXiv",
    primaryClass = "hep-ex",
    doi = "10.1103/PhysRevLett.123.241803",
    journal = "Phys. Rev. Lett.",
    volume = "123",
    number = "24",
    pages = "241803",
    year = "2019"
}

@article{DarkSide:2022dhx,
    author = "Agnes, P. and others",
    collaboration = "DarkSide",
    title = "{Search for Dark-Matter\textendash{}Nucleon Interactions via Migdal Effect with DarkSide-50}",
    eprint = "2207.11967",
    archivePrefix = "arXiv",
    primaryClass = "hep-ex",
    doi = "10.1103/PhysRevLett.130.101001",
    journal = "Phys. Rev. Lett.",
    volume = "130",
    number = "10",
    pages = "101001",
    year = "2023"
}

@article{Lewin:1995rx,
    author = "Lewin, J. D. and Smith, P. F.",
    title = "{Review of mathematics, numerical factors, and corrections for dark matter experiments based on elastic nuclear recoil}",
    reportNumber = "RAL-TR-95-024",
    doi = "10.1016/S0927-6505(96)00047-3",
    journal = "Astropart. Phys.",
    volume = "6",
    pages = "87--112",
    year = "1996"
}

@article{Bishara:2017pfq,
    author = "Bishara, Fady and Brod, Joachim and Grinstein, Benjamin and Zupan, Jure",
    title = "{From quarks to nucleons in dark matter direct detection}",
    eprint = "1707.06998",
    archivePrefix = "arXiv",
    primaryClass = "hep-ph",
    reportNumber = "DO-TH-17-10, OUTP-17-07P, CERN-TH-2017-157",
    doi = "10.1007/JHEP11(2017)059",
    journal = "JHEP",
    volume = "11",
    pages = "059",
    year = "2017"
}

@article{Catena:2015uha,
    author = "Catena, Riccardo and Schwabe, Bodo",
    title = "{Form factors for dark matter capture by the Sun in effective theories}",
    eprint = "1501.03729",
    archivePrefix = "arXiv",
    primaryClass = "hep-ph",
    doi = "10.1088/1475-7516/2015/04/042",
    journal = "JCAP",
    volume = "04",
    pages = "042",
    year = "2015"
}

@article{Baxter:2021pqo,
    author = "Baxter, D. and others",
    title = "{Recommended conventions for reporting results from direct dark matter searches}",
    eprint = "2105.00599",
    archivePrefix = "arXiv",
    primaryClass = "hep-ex",
    doi = "10.1140/epjc/s10052-021-09655-y",
    journal = "Eur. Phys. J. C",
    volume = "81",
    number = "10",
    pages = "907",
    year = "2021"
}

@article{Mateu:2007tr,
    author = "Mateu, V. and Portoles, J.",
    title = "{Form-factors in radiative pion decay}",
    eprint = "0706.1039",
    archivePrefix = "arXiv",
    primaryClass = "hep-ph",
    reportNumber = "IFIC-07-29, FTUV-07-0607",
    doi = "10.1140/epjc/s10052-007-0393-5",
    journal = "Eur. Phys. J. C",
    volume = "52",
    pages = "325--338",
    year = "2007"
}

@article{Migdal:1941,
    author = "A. B. Migdal",
    title = "{Ionization of atoms accompanying $\alpha$-and $\beta$-decay}",
    journal = "J. Phys. 4, 449 (1941)",
    year = "1941"
}

@article{LZ:2023poo,
    author = "Aalbers, J. and others",
    collaboration = "LZ",
    title = "{Search for new physics in low-energy electron recoils from the first LZ exposure}",
    eprint = "2307.15753",
    archivePrefix = "arXiv",
    primaryClass = "hep-ex",
    reportNumber = "FERMILAB-PUB-23-397-PPD",
    doi = "10.1103/PhysRevD.108.072006",
    journal = "Phys. Rev. D",
    volume = "108",
    number = "7",
    pages = "072006",
    year = "2023"
}

@article{DelNobile:2018dfg,
    author = "Del Nobile, Eugenio",
    title = "{Complete Lorentz-to-Galileo dictionary for direct dark matter detection}",
    eprint = "1806.01291",
    archivePrefix = "arXiv",
    primaryClass = "hep-ph",
    doi = "10.1103/PhysRevD.98.123003",
    journal = "Phys. Rev. D",
    volume = "98",
    number = "12",
    pages = "123003",
    year = "2018"
}

@article{Baouche:2021wwa,
    author = "Baouche, Nabil and Ahriche, Amine and Faisel, Gaber and Nasri, Salah",
    title = "{Phenomenology of the hidden SU(2) vector dark matter model}",
    eprint = "2105.14387",
    archivePrefix = "arXiv",
    primaryClass = "hep-ph",
    doi = "10.1103/PhysRevD.104.075022",
    journal = "Phys. Rev. D",
    volume = "104",
    number = "7",
    pages = "075022",
    year = "2021"
}

@article{Hambye:2008bq,
    author = "Hambye, Thomas",
    title = "{Hidden vector dark matter}",
    eprint = "0811.0172",
    archivePrefix = "arXiv",
    primaryClass = "hep-ph",
    reportNumber = "ULB-TH-08-35",
    doi = "10.1088/1126-6708/2009/01/028",
    journal = "JHEP",
    volume = "01",
    pages = "028",
    year = "2009"
}

@article{Hagiwara:1986vm,
    author = "Hagiwara, Kaoru and Peccei, R. D. and Zeppenfeld, D. and Hikasa, K.",
    title = "{Probing the Weak Boson Sector in e+ e- ---\ensuremath{>} W+ W-}",
    reportNumber = "MAD/PH/279, DESY-86-058",
    doi = "10.1016/0550-3213(87)90685-7",
    journal = "Nucl. Phys. B",
    volume = "282",
    pages = "253--307",
    year = "1987"
}

@article{Gaemers:1978hg,
    author = "Gaemers, K. J. F. and Gounaris, G. J.",
    title = "{Polarization Amplitudes for e+ e- ---\ensuremath{>} W+ W- and e+ e- ---\ensuremath{>} Z Z}",
    reportNumber = "CERN-TH-2548",
    doi = "10.1007/BF01440226",
    journal = "Z. Phys. C",
    volume = "1",
    pages = "259",
    year = "1979"
}

@inproceedings{Gounaris:1996rz,
    author = "Gounaris, G. and others",
    title = "{Triple gauge boson couplings}",
    booktitle = "{AGS / RHIC Users Annual Meeting}",
    eprint = "hep-ph/9601233",
    archivePrefix = "arXiv",
    month = "1",
    year = "1996"
}

@article{Hisano:2020qkq,
    author = "Hisano, Junji and Ibarra, Alejandro and Nagai, Ryo",
    title = "{Direct detection of vector dark matter through electromagnetic multipoles}",
    eprint = "2007.03216",
    archivePrefix = "arXiv",
    primaryClass = "hep-ph",
    reportNumber = "IPMU20-0077, TUM-HEP 1269/20",
    doi = "10.1088/1475-7516/2020/10/015",
    journal = "JCAP",
    volume = "10",
    pages = "015",
    year = "2020"
}

@article{Chu:2023zbo,
    author = "Chu, Xiaoyong and Hisano, Junji and Ibarra, Alejandro and Kuo, Jui-Lin and Pradler, Josef",
    title = "{Multipole vector dark matter below the GeV scale}",
    eprint = "2303.13643",
    archivePrefix = "arXiv",
    primaryClass = "hep-ph",
    reportNumber = "UWThPh 2023-8, IPMU23-0006",
    doi = "10.1103/PhysRevD.108.015029",
    journal = "Phys. Rev. D",
    volume = "108",
    number = "1",
    pages = "015029",
    year = "2023"
}

@article{Kang:2018rad,
    author = "Kang, Sunghyun and Scopel, Stefano and Tomar, Gaurav and Yoon, Jong-Hyun",
    title = "{On the sensitivity of present direct detection experiments to WIMP\textendash{}quark and WIMP\textendash{}gluon effective interactions: A systematic assessment and new model\textendash{}independent approaches}",
    eprint = "1810.00607",
    archivePrefix = "arXiv",
    primaryClass = "hep-ph",
    doi = "10.1016/j.astropartphys.2019.07.001",
    journal = "Astropart. Phys.",
    volume = "114",
    pages = "80--91",
    year = "2020"
}

@article{Baek:2013dwa,
    author = "Baek, S. and Ko, P. and Park, Wan-Il",
    title = "{Hidden sector monopole, vector dark matter and dark radiation with Higgs portal}",
    eprint = "1311.1035",
    archivePrefix = "arXiv",
    primaryClass = "hep-ph",
    reportNumber = "KIAS-P13060",
    doi = "10.1088/1475-7516/2014/10/067",
    journal = "JCAP",
    volume = "10",
    pages = "067",
    year = "2014"
}

@article{Hu:2021pln,
    author = "Hu, Zexi and Cai, Chengfeng and Tang, Yi-Lei and Yu, Zhao-Huan and Zhang, Hong-Hao",
    title = "{Vector dark matter from split SU(2) gauge bosons}",
    eprint = "2103.00220",
    archivePrefix = "arXiv",
    primaryClass = "hep-ph",
    doi = "10.1007/JHEP07(2021)089",
    journal = "JHEP",
    volume = "07",
    pages = "089",
    year = "2021"
}

@article{Alexandrou:2020sml,
    author = "Alexandrou, C. and Bacchio, S. and Constantinou, M. and Finkenrath, J. and Hadjiyiannakou, K. and Jansen, K. and Koutsou, G. and Panagopoulos, H. and Spanoudes, G.",
    title = "{Complete flavor decomposition of the spin and momentum fraction of the proton using lattice QCD simulations at physical pion mass}",
    eprint = "2003.08486",
    archivePrefix = "arXiv",
    primaryClass = "hep-lat",
    doi = "10.1103/PhysRevD.101.094513",
    journal = "Phys. Rev. D",
    volume = "101",
    number = "9",
    pages = "094513",
    year = "2020"
}

@article{Barman:2021hhg,
    author = "Barman, Basabendu and Bhattacharya, Subhaditya and Girmohanta, Sudhakantha and Jahedi, Sahabub",
    title = "{Effective Leptophilic WIMPs at the ee collider}",
    eprint = "2109.10936",
    archivePrefix = "arXiv",
    primaryClass = "hep-ph",
    reportNumber = "Stony Brook preprint YITP-SB-2021-16, PI/UAN-2021-700FT",
    doi = "10.1007/JHEP04(2022)146",
    journal = "JHEP",
    volume = "04",
    pages = "146",
    year = "2022"
}

@article{Gorton:2022eed,
    author = "Gorton, Oliver C. and Johnson, Calvin W. and Jiao, Changfeng and Nikoleyczik, Jonathan",
    title = "{dmscatter: A fast program for WIMP-nucleus scattering}",
    eprint = "2209.09187",
    archivePrefix = "arXiv",
    primaryClass = "nucl-th",
    doi = "10.1016/j.cpc.2022.108597",
    journal = "Comput. Phys. Commun.",
    volume = "284",
    pages = "108597",
    year = "2023"
}

@article{Kundu:2021cmo,
    author = "Kundu, Saumyen and Guha, Atanu and Das, Prasanta Kumar and Dev, P. S. Bhupal",
    title = "{EFT analysis of leptophilic dark matter at future electron-positron colliders in the mono-photon and mono-Z channels}",
    eprint = "2110.06903",
    archivePrefix = "arXiv",
    primaryClass = "hep-ph",
    doi = "10.1103/PhysRevD.107.015003",
    journal = "Phys. Rev. D",
    volume = "107",
    number = "1",
    pages = "015003",
    year = "2023"
}

@article{Cheng:1988im,
    author = "Cheng, Hai-Yang",
    title = "{Low-energy interactions of scalar and pseudoscalar Higgs bosons with baryons}",
    reportNumber = "IP-ASTP-19-88",
    doi = "10.1016/0370-2693(89)90402-4",
    journal = "Phys. Lett. B",
    volume = "219",
    pages = "347--353",
    year = "1989"
}

@article{Lebedev:2011iq,
    author = "Lebedev, Oleg and Lee, Hyun Min and Mambrini, Yann",
    title = "{Vector Higgs-portal dark matter and the invisible Higgs}",
    eprint = "1111.4482",
    archivePrefix = "arXiv",
    primaryClass = "hep-ph",
    reportNumber = "CERN-PH-TH-2011-271, DESY-11-202, LPTORSAY-11-90",
    doi = "10.1016/j.physletb.2012.01.029",
    journal = "Phys. Lett. B",
    volume = "707",
    pages = "570--576",
    year = "2012"
}

@article{Abe:2012hb,
    author = "Abe, Tomohiro and Kakizaki, Mitsuru and Matsumoto, Shigeki and Seto, Osamu",
    title = "{Vector WIMP Miracle}",
    eprint = "1202.5902",
    archivePrefix = "arXiv",
    primaryClass = "hep-ph",
    reportNumber = "UT-HET-063, IPMU12-0029, HGU-CAP-14",
    doi = "10.1016/j.physletb.2012.05.051",
    journal = "Phys. Lett. B",
    volume = "713",
    pages = "211--215",
    year = "2012"
}

@article{Abe:2020mph,
    author = "Abe, Tomohiro and Fujiwara, Motoko and Hisano, Junji and Matsushita, Kohei",
    title = "{A model of electroweakly interacting non-abelian vector dark matter}",
    eprint = "2004.00884",
    archivePrefix = "arXiv",
    primaryClass = "hep-ph",
    reportNumber = "IPMU20-0032",
    doi = "10.1007/JHEP07(2020)136",
    journal = "JHEP",
    volume = "07",
    pages = "136",
    year = "2020"
}

@article{Kurylov:2003ra,
    author = "Kurylov, Andriy and Kamionkowski, Marc",
    title = "{Generalized analysis of weakly interacting massive particle searches}",
    eprint = "hep-ph/0307185",
    archivePrefix = "arXiv",
    doi = "10.1103/PhysRevD.69.063503",
    journal = "Phys. Rev. D",
    volume = "69",
    pages = "063503",
    year = "2004"
}

@article{Giuliani:2005my,
    author = "Giuliani, F.",
    title = "{Are direct search experiments sensitive to all spin-independent WIMP candidates?}",
    eprint = "hep-ph/0504157",
    archivePrefix = "arXiv",
    doi = "10.1103/PhysRevLett.95.101301",
    journal = "Phys. Rev. Lett.",
    volume = "95",
    pages = "101301",
    year = "2005"
}

@article{Chang:2010yk,
    author = "Chang, Spencer and Liu, Jia and Pierce, Aaron and Weiner, Neal and Yavin, Itay",
    title = "{CoGeNT Interpretations}",
    eprint = "1004.0697",
    archivePrefix = "arXiv",
    primaryClass = "hep-ph",
    reportNumber = "MCTP-10-16",
    doi = "10.1088/1475-7516/2010/08/018",
    journal = "JCAP",
    volume = "08",
    pages = "018",
    year = "2010"
}

@article{Carpenter:2015xaa,
    author = "Carpenter, Linda M. and Colburn, Russell and Goodman, Jessica",
    title = "{Indirect Detection Constraints on the Model Space of Dark Matter Effective Theories}",
    eprint = "1506.08841",
    archivePrefix = "arXiv",
    primaryClass = "hep-ph",
    doi = "10.1103/PhysRevD.92.095011",
    journal = "Phys. Rev. D",
    volume = "92",
    number = "9",
    pages = "095011",
    year = "2015"
}

@article{Belyaev:2018pqr,
    author = "Belyaev, Alexander and Bertuzzo, Enrico and Caniu Barros, Cristian and Eboli, Oscar and Grilli Di Cortona, Giovanni and Iocco, Fabio and Pukhov, Alexander",
    title = "{Interplay of the LHC and non-LHC Dark Matter searches in the Effective Field Theory approach}",
    eprint = "1807.03817",
    archivePrefix = "arXiv",
    primaryClass = "hep-ph",
    doi = "10.1103/PhysRevD.99.015006",
    journal = "Phys. Rev. D",
    volume = "99",
    number = "1",
    pages = "015006",
    year = "2019"
}

@article{Liu:2017kmx,
    author = "Liu, Zuowei and Su, Yushan and Sming Tsai, Yue-Lin and Yu, Bingrong and Yuan, Qiang",
    title = "{A combined analysis of PandaX, LUX, and XENON1T experiments within the framework of dark matter effective theory}",
    eprint = "1708.04630",
    archivePrefix = "arXiv",
    primaryClass = "hep-ph",
    reportNumber = "NCTS-PH-1717",
    doi = "10.1007/JHEP11(2017)024",
    journal = "JHEP",
    volume = "11",
    pages = "024",
    year = "2017"
}

@article{DarkSide:2021bnz,
    author = "Agnes, P. and others",
    collaboration = "DarkSide",
    title = "{Calibration of the liquid argon ionization response to low energy electronic and nuclear recoils with DarkSide-50}",
    eprint = "2107.08087",
    archivePrefix = "arXiv",
    primaryClass = "physics.ins-det",
    doi = "10.1103/PhysRevD.104.082005",
    journal = "Phys. Rev. D",
    volume = "104",
    number = "8",
    pages = "082005",
    year = "2021"
}

@article{Essig:2019xkx,
    author = "Essig, Rouven and Pradler, Josef and Sholapurkar, Mukul and Yu, Tien-Tien",
    title = "{Relation between the Migdal Effect and Dark Matter-Electron Scattering in Isolated Atoms and Semiconductors}",
    eprint = "1908.10881",
    archivePrefix = "arXiv",
    primaryClass = "hep-ph",
    reportNumber = "YITP-19-23",
    doi = "10.1103/PhysRevLett.124.021801",
    journal = "Phys. Rev. Lett.",
    volume = "124",
    number = "2",
    pages = "021801",
    year = "2020"
}

@article{Li:2022acp,
    author = "Li, Jiwei and Su, Liangliang and Wu, Lei and Zhu, Bin",
    title = "{Spin-dependent sub-GeV inelastic dark matter-electron scattering and Migdal effect. Part I. Velocity independent operator}",
    eprint = "2210.15474",
    archivePrefix = "arXiv",
    primaryClass = "hep-ph",
    doi = "10.1088/1475-7516/2023/04/020",
    journal = "JCAP",
    volume = "04",
    pages = "020",
    year = "2023"
}

@article{Kang:2024kec,
    author = "Kang, Sunghyun and Scopel, Stefano and Tomar, Gaurav",
    title = "{Low-mass constraints on WIMP effective models of inelastic scattering using the Migdal effect}",
    eprint = "2407.16187",
    archivePrefix = "arXiv",
    primaryClass = "hep-ph",
    doi = "10.1088/1475-7516/2025/01/035",
    journal = "JCAP",
    volume = "01",
    pages = "035",
    year = "2025"
}

@article{Bell:2019egg,
    author = "Bell, Nicole F. and Dent, James B. and Newstead, Jayden L. and Sabharwal, Subir and Weiler, Thomas J.",
    title = "{Migdal effect and photon bremsstrahlung in effective field theories of dark matter direct detection and coherent elastic neutrino-nucleus scattering}",
    eprint = "1905.00046",
    archivePrefix = "arXiv",
    primaryClass = "hep-ph",
    doi = "10.1103/PhysRevD.101.015012",
    journal = "Phys. Rev. D",
    volume = "101",
    number = "1",
    pages = "015012",
    year = "2020"
}

@article{XENON:2024xgd,
    author = "Aprile, E. and others",
    collaboration = "XENON",
    title = "{XENONnT WIMP Search: Signal \& Background Modeling and Statistical Inference}",
    eprint = "2406.13638",
    archivePrefix = "arXiv",
    primaryClass = "physics.data-an",
    month = "6",
    year = "2024"
}

@article{Dolan:2017xbu,
    author = "Dolan, Matthew J. and Kahlhoefer, Felix and McCabe, Christopher",
    title = "{Directly detecting sub-GeV dark matter with electrons from nuclear scattering}",
    eprint = "1711.09906",
    archivePrefix = "arXiv",
    primaryClass = "hep-ph",
    reportNumber = "KCL-PH-TH/2017-54, TTK-17-43, KCL-PH-TH-2017-54",
    doi = "10.1103/PhysRevLett.121.101801",
    journal = "Phys. Rev. Lett.",
    volume = "121",
    number = "10",
    pages = "101801",
    year = "2018"
}

@article{GrillidiCortona:2020owp,
    author = "Grilli di Cortona, Giovanni and Messina, Andrea and Piacentini, Stefano",
    title = "{Migdal effect and photon Bremsstrahlung: improving the sensitivity to light dark matter of liquid argon experiments}",
    eprint = "2006.02453",
    archivePrefix = "arXiv",
    primaryClass = "hep-ph",
    doi = "10.1007/JHEP11(2020)034",
    journal = "JHEP",
    volume = "11",
    pages = "034",
    year = "2020"
}

@article{Knapen:2020aky,
    author = "Knapen, Simon and Kozaczuk, Jonathan and Lin, Tongyan",
    title = "{Migdal Effect in Semiconductors}",
    eprint = "2011.09496",
    archivePrefix = "arXiv",
    primaryClass = "hep-ph",
    doi = "10.1103/PhysRevLett.127.081805",
    journal = "Phys. Rev. Lett.",
    volume = "127",
    number = "8",
    pages = "081805",
    year = "2021"
}

@article{Flambaum:2020xxo,
    author = "Flambaum, Victor V. and Su, Liangliang and Wu, Lei and Zhu, Bin",
    title = "{New strong bounds on sub-GeV dark matter from boosted and Migdal effects}",
    eprint = "2012.09751",
    archivePrefix = "arXiv",
    primaryClass = "hep-ph",
    doi = "10.1007/s11433-022-2090-7",
    journal = "Sci. China Phys. Mech. Astron.",
    volume = "66",
    number = "7",
    pages = "271011",
    year = "2023"
}
\bibliographystyle{JHEP}

\end{document}